\documentclass[aps,prd,longbibliography,reprint,twocolumn,amsmath,amssymb,amsfonts,showpacs,superscriptaddress]{revtex4-1}

\usepackage{ifpdf}
\usepackage{dcolumn}
\usepackage{enumitem}
\usepackage{bm}
\usepackage{float}

\floatstyle{plaintop}

\restylefloat{table}
\ifpdf

      \usepackage[pdftex]{graphicx}  
      \usepackage[pdftex]{epsfig}

      \usepackage[pdftex]{hyperref}
\hypersetup
	{
		colorlinks,%
		citecolor=blue,%
		linkcolor=red,%
		urlcolor=blue,%
	}

\else

      \usepackage[dvips]{graphicx}  
      \usepackage[dvips]{epsfig}

      \usepackage[dvipdfm]{hyperref}
      \hypersetup
	{
		colorlinks,%
		citecolor=blue,%
		linkcolor=red,%
		urlcolor=blue,%
	}

\fi

\usepackage{gensymb} 

\begin{document}

\title{Regge pole description of scattering of scalar and electromagnetic waves \\
 by a Schwarzschild black hole}

\author{Antoine Folacci}
\email{folacci@univ-corse.fr}
\affiliation{Equipe Physique
Th\'eorique, SPE, UMR 6134 du CNRS
et de l'Universit\'e de Corse,\\
Universit\'e de Corse, Facult\'e des Sciences, BP 52, F-20250 Corte,
France}

\author{Mohamed \surname{Ould~El~Hadj}}
\email{med.ouldelhadj@gmail.com}

\affiliation{Equipe Physique
Th\'eorique, SPE, UMR 6134 du CNRS
et de l'Universit\'e de Corse,\\
Universit\'e de Corse, Facult\'e des Sciences, BP 52, F-20250 Corte,
France}

\affiliation{Consortium for Fundamental Physics, School of Mathematics and Statistics, University of Sheffield,\\ Hicks Building, Hounsfield Road, Sheffield S3 7RH, United Kingdom}

\date{\today}

\begin{abstract}

We revisit the problem of scalar and electromagnetic waves impinging upon a Schwarzschild black hole from complex angular momentum techniques. We focus more particularly on the associated differential scattering cross sections.  We derive an exact representation of the corresponding scattering amplitudes by replacing the discrete sum over integer values of the angular momentum which defines their partial wave expansions by a background integral in the complex angular momentum plane plus a sum over the Regge poles of the $S$-matrix involving the associated residues. We show that, surprisingly, the background integral is numerically negligible for intermediate and high frequencies and, as a consequence, that the cross sections can be reconstructed in terms of Regge poles with very good agreement. We show in particular that, for large values of the scattering angle, a small number of Regge poles permits us to describe the black hole glory and that, by increasing the number of Regge poles, we can reconstruct very efficiently the differential scattering cross sections for small and intermediate scattering angles and therefore describe the orbiting oscillations. In fact, in the short-wavelength regime, the sum over Regge poles allows us to extract by resummation the information encoded in the partial wave expansion defining a scattering amplitude and, moreover, to overcome the difficulties linked to its lack of convergence due to the long-range nature of the fields propagating on the black hole. As a consequence, from asymptotic expressions for the lowest Regge poles and the associated residues based on the correspondence Regge poles -- ``surface waves'' propagating close to the photon sphere, we can provide an analytical approximation describing both the black hole glory and a large part of the orbiting oscillations. We finally discuss the role of the background integral for low frequencies.

\end{abstract}

\maketitle

\tableofcontents

\section{Introduction}
\label{Intro}

Studies concerning the scattering of waves by black holes (BHs) are mainly based on partial wave expansions (see, e.g., Ref.~\cite{Futterman:1988ni}).
This is due to the high degree of symmetry of the BH spacetimes usually considered and physically or astrophysically interesting.
For example, the Schwarzschild BH is a static spherically symmetric solution of the vacuum Einstein's equations while the Kerr BH is a stationary axisymmetric solution of these equations with, as a consequence, the separability of wave equations on these gravitational backgrounds (see, e.g., Ref.~\cite{Frolov:1998wf}).
Even if the approach based on partial wave expansions is natural and very effective in the context of scattering of waves by BHs, it presents some flaws. Due to the long-range nature of the fields propagating on a BH, some partial wave expansions encountered are formally divergent (see, e.g., Ref.~\cite{Futterman:1988ni}) and, moreover, it is in general rather difficult to interpret physically results described in terms of partial wave expansions. These problems can be overcome by using complex angular momentum (CAM) techniques (analytic continuation of partial wave expansions in the CAM plane, effective resummations involving the poles of the $S$-matrix in the CAM plane, i.e., the so-called Regge poles, and the associated residues, semiclassical interpretations of Regge pole expansions, etc.). Such techniques, which proved to be very helpful in quantum mechanics (see, e.g., Refs.~\cite{deAlfaro:1965zz,Newton:1982qc}), in electromagnetism and optics (see, e.g., Refs.~\cite{Watson18,Sommerfeld49,Newton:1982qc,Nus92,Grandy2000}), in acoustics and seismology (see, e.g., Refs.~\cite{Uberall1992,AkiRichards2002}) and in high-energy physics (see, e.g., Refs.~\cite{Gribov69,Collins77,BaronePredazzi2002,DonnachieETAL2005}) to describe and analyze resonant scattering are now also used in the context of BH physics (see, e.g., Refs.~\cite{Andersson:1994rk,Andersson:1994rm,Decanini:2002ha,Glampedakis:2003dn,Decanini:2009mu,Decanini:2009dn,Dolan:2009nk,Decanini:2010fz,Decanini:2011eh,Decanini:2011xi,
Decanini:2011xw,Dolan:2011ti,Macedo:2013afa,Nambu:2015aea,Folacci:2018sef,Benone:2018rtj}).

In this article we revisit the problem of plane monochromatic waves impinging upon a Schwarzschild BH from CAM techniques. More precisely, we focus on the differential scattering cross sections associated with scalar and electromagnetic waves. It should be recalled that the partial wave expansions of these cross sections have been obtained a long time ago by Matzner \cite{Matzner1968} for the scalar field and by Mashhoon \cite{Mashhoon:1973zz,Mashhoon:1974cq} and Fabbri \cite{Fabbri:1975sa} for the electromagnetic field and that many additional works have since been done which have theoretically and numerically completed these first investigations (see, e.g., Refs.~\cite{Sanchez:1976fcl,Sanchez:1977vz,Matzner:1985rjn,Anninos:1992ih,Andersson:1994rk,Andersson:1994rm,Andersson:1995vi,Glampedakis:2001cx,Dolan:2008kf,Crispino:2009xt} for some articles directly relevant to our own study). It is furthermore worth noting that the important case of the electromagnetic field is of fundamental interest with the emergence of multimessenger astronomy as well as with the possibility to produce experimentally BH images or, more precisely, to ``photograph'' with the Event Horizon Telescope the shadow of supermassive BHs \cite{Falcke:1999pj,EHT,EHTC2019I}. Here, we construct an exact representation of the scalar and electromagnetic scattering cross sections by replacing the discrete sum over integer values of the angular momentum which defines their partial wave expansions by a background integral in the CAM plane plus a sum over the Regge poles of the $S$-matrix which involves the associated residues. Surprisingly, we find that the background integral is numerically negligible for intermediate and high reduced frequencies (i.e., in the short-wavelength regime) and, as a consequence, that the differential scattering cross sections can be described in terms of Regge poles with very good agreement for arbitrary scattering angles.
In fact, in this wavelength regime, the sum over Regge poles allows us to extract by resummation the physical information encoded in the partial wave expansion defining a differential scattering cross section and, moreover, to overcome the difficulties linked to its lack of convergence due to the long-range nature of the fields propagating on the BH. We show in particular that, for large values of the scattering angle, i.e., in the backward direction, a small number of Regge poles permits us to describe the BH glory (see Refs.~\cite{Matzner:1985rjn,Andersson:1994rm} for semiclassical interpretations) and that, by increasing the number of Regge poles, we can reconstruct very efficiently the differential scattering cross sections for small and intermediate scattering angles and therefore describe the orbiting oscillations (see Refs.~\cite{Anninos:1992ih,Andersson:1994rm} for semiclassical interpretations). We then take advantage of these numerical results to derive an analytical approximation fitting both the BH glory and a large part of the orbiting oscillations. This is achieved by inserting in the Regge pole sums asymptotic approximations for the lowest Regge poles and the associated residues.

It is important to relate or compare our results with other results previously obtained:

 \begin{enumerate}[label=(\arabic*)]

\item Our work extends but also corrects the important studies by Andersson and Thylwe \cite{Andersson:1994rk,Andersson:1994rm} where we can find the first application of CAM techniques in BH physics. In Ref.~\cite{Andersson:1994rk}, Andersson and Thylwe have considered the scattering of scalar waves by a Schwarzschild BH from a theoretical point of view and adapted the CAM formalism to this problem. They have established some properties of the Regge poles and of the $S$-matrix in the CAM plane. In Ref.~\cite{Andersson:1994rm}, Andersson has used this formalism to interpret semiclassically the BH glory and the orbiting oscillations. He has, in particular, considered ``surface waves'' propagating close to the unstable circular photon (graviton) orbit at $r=3M$, i.e., near the so-called photon sphere, and associated them with the Regge poles. However, it should be noted that some of the analyses and comments we can find in Ref.~\cite{Andersson:1994rm} are invalidated by the fact that the residues numerically obtained are incorrect (see also Ref.~\cite{Glampedakis:2003dn}). In our article, we obtain precise values for the residues of a large number of Regge poles, thus permitting us to draw solid conclusions from our results. Furthermore, we show that the background integral in the CAM plane is numerically negligible for intermediate and high reduced frequencies. Here, it is important to recall that, in Refs.~\cite{Andersson:1994rk,Andersson:1994rm}, Andersson and Thylwe tried to extract some physical information of this background integral and that, in electromagnetism and optics \cite{Watson18,Sommerfeld49,Newton:1982qc,Nus92,Grandy2000} as well as in acoustics and seismology \cite{Uberall1992,AkiRichards2002}), it can be associated semiclassically (i.e., for high reduced frequencies) with incident and reflected rays. In the context of scattering by BHs (or, more precisely, if we focus on differential scattering cross sections in the short-wavelength regime), it plays only a minor role.

\item  We have developed Andersson's point of view concerning the surface waves propagating close to the photon sphere in a series of papers establishing, in particular, that the complex frequencies of the weakly damped quasinormal modes (QNMs) are Breit-Wigner-type resonances generated by these surface waves. We have then been able to construct semiclassically the spectrum of the QNM complex frequencies from the Regge trajectories, i.e., from the curves traced out in the CAM plane by the Regge poles as a function of the frequency \cite{Decanini:2002ha,Decanini:2009mu,Folacci:2018sef}, establishing on a ``rigorous'' basis the physically intuitive interpretation of the Schwarzschild BH QNMs suggested, as early as 1972, by Goebel \cite{Goebel} (see Refs.~\cite{Decanini:2009dn,Dolan:2009nk,Decanini:2010fz,Decanini:2011eh} for the extension of these results to other BHs and to massive fields). Moreover, from the Regge trajectories and the residues of the greybody factors, we have described analytically the high-energy absorption cross section for a wide class of BHs endowed with a photon sphere and explained its oscillations in terms of the geometrical characteristics (orbital period and Lyapunov exponent) of the null unstable geodesics lying on the photon sphere \cite{Decanini:2011xi,Decanini:2011xw,Decanini:2011eh}. All these results highlight the interpretive power of CAM techniques in BH physics.

\item  In order to derive analytical approximations which fit both the BH glory and a large part of the orbiting oscillations, we shall insert into the Regge pole sums asymptotic expansions for the Regge poles and the associated residues. In fact, such expansions which are valid in the short-wavelength regime are physically connected with the excitation of the surface waves previously mentioned and with diffractive effects due to the Schwarzschild photon sphere \cite{Andersson:1994rm,Decanini:2002ha,Decanini:2009mu,Dolan:2009nk,Decanini:2010fz,DFOEH2019}. It is moreover important to recall that glory scattering and orbiting scattering are usually considered as two different effects and are described analytically by two different semiclassical analytic formulas (see Refs.~\cite{Matzner:1985rjn,Anninos:1992ih} or Sec.~4.7.2 of Ref.~\cite{Frolov:1998wf} for a concise presentation). Here, we prove that it is possible from Regge pole sums to describe analytically both phenomena in a unique formula.

\end{enumerate}

Our paper is organized as follows. In Sec.~\ref{SecII}, by means of the Sommerfeld-Watson transform \cite{Watson18,Sommerfeld49,Newton:1982qc} and Cauchy's residue theorem, we construct exact CAM representations of the differential scattering cross sections for plane scalar and electromagnetic waves impinging upon a Schwarzschild BH from their partial wave expansions. These CAM representations are split into a background integral in the CAM plane and a sum over the Regge poles of the $S$-matrix involving the associated residues. In Sec.~\ref{SecIII}, we obtain numerically, for various reduced frequencies, the Regge poles of the $S$-matrix, the associated residues and the background integral. This permits us to reconstruct, for these particular frequencies of the impinging waves, the differential scattering cross sections of the BH and to show that, in the short-wavelength regime, they can be described from the Regge pole sum alone with very good agreement. We also discuss the role of the background integral for low reduced frequencies, i.e., in the long-wavelength regime. In Sec.~\ref{SecIV}, by inserting into the Regge pole sum asymptotic approximations for the lowest Regge poles and the associated residues, we derive a formula fitting both the BH glory and a large part of the orbiting oscillations. In the Conclusion, we summarize our main results and briefly consider possible extensions of our work. In the Appendix, we discuss the numerical evaluation of the background integrals. Due to the long-range nature of the fields propagating on a Schwarzschild BH, these integrals in the CAM plane (as the partial wave expansions) suffer a lack of convergence. We overcome this problem, i.e., we accelerate their convergence, by extending to integrals the iterative method developed in Ref.~\cite{Yennie:1954zz} for partial wave expansions.

Throughout this article, we adopt units such that $G = c = 1$. We furthermore consider that the exterior of the Schwarzschild BH is defined by the line element $ds^2= -f(r) dt^2+ f(r)^{-1}dr^2+ r^2 d\theta^2 + r^2 \sin^2\theta d\varphi^2$ where $f(r)=1-2M/r$ and $M$ is the mass of the BH while $t \in ]-\infty, +\infty[$, $r \in ]2M,+\infty[$, $\theta \in [0,\pi]$ and $\varphi \in [0,2\pi]$ are the usual Schwarzschild coordinates. We finally assume a time dependence $\exp(-i\omega t)$ for the plane monochromatic waves considered.

\section{Differential scattering cross sections for scalar and electromagnetic waves, their CAM representations and their Regge pole approximations}
\label{SecII}

In this section, we recall the partial wave expansions of the differential scattering cross sections for plane monochromatic scalar and electromagnetic waves impinging upon a Schwarzschild BH and we construct exact CAM representations of these cross sections by means of the Sommerfeld-Watson transform \cite{Watson18,Sommerfeld49,Newton:1982qc} and Cauchy's theorem. These CAM representations are split into a background integral in the CAM plane and a sum over the Regge poles of the $S$-matrix involving the associated residues.

\subsection{Partial wave expansions of differential scattering cross sections}
\label{SecIIa}

We recall that, for the scalar field, the differential scattering cross section is given by \cite{Matzner1968}
\begin{equation}\label{Scalar_Scattering_diff}
  \frac{d\sigma}{d\Omega} = |f(\omega,\theta)|^2
\end{equation}
where
\begin{equation}\label{Scalar_Scattering_amp}
  f(\omega,\theta) = \frac{1}{2 i \omega} \sum_{\ell = 0}^{\infty} (2\ell+1)[S_{\ell}(\omega)-1]P_{\ell}(\cos\theta)
\end{equation}
denotes the scattering amplitude and that, for the electromagnetic field, the differential scattering cross section can be written in the form \cite{Mashhoon:1973zz,Mashhoon:1974cq} (see also Refs.~\cite{Fabbri:1975sa,Crispino:2009xt})
\begin{equation}\label{EM_Scattering_diff}
  \frac{d\sigma}{d\Omega} = |A(\omega,\theta)|^2
\end{equation}
where the scattering amplitude is given by
\begin{equation}\label{EM_Scattering_amp}
  A(\omega,\theta) = \mathcal{D}_{\theta} B(\omega,\theta)
\end{equation}
with
\begin{equation}\label{EM_Scattering_amp2}
  B(\omega,\theta) =  \frac{1}{2 i \omega} \sum_{\ell = 1}^{\infty} \frac{(2\ell+1)}{\ell(\ell+1)}[S_{\ell}(\omega)-1]P_{\ell}(\cos\theta)
\end{equation}
and
\begin{subequations}\label{EM_Scattering_operator}
\begin{eqnarray}
\mathcal{D}_{\theta} &=& -(1 + \cos \theta)\frac{d \,\,\,}{d\cos\theta}\left[ (1-\cos\theta) \frac{d \,\,\,}{d\cos\theta}  \right] \\
              &=& - \left(\frac{d^2}{d\theta^2} + \frac{1}{\sin\theta} \frac{d}{d\theta}\right).
\end{eqnarray}
\end{subequations}
The expression (\ref{EM_Scattering_amp})-(\ref{EM_Scattering_operator}) takes into account the two polarizations of the electromagnetic field. In Eqs.~(\ref{Scalar_Scattering_amp}) and (\ref{EM_Scattering_amp2}), the functions $P_{\ell}(\cos\theta)$ are the Legendre polynomials \cite{AS65}.  We also recall that the $S$-matrix elements $S_{\ell}(\omega)$ appearing in Eqs.~(\ref{Scalar_Scattering_amp}) and (\ref{EM_Scattering_amp2}) can be defined from the modes $\phi_{\omega \ell}^{in}$ solutions of the homogenous Regge-Wheeler equation
\begin{equation}
\label{H_RW_equation}
\left[\frac{d^{2}}{dr_{\ast}^{2}}+\omega^{2}-V_{\ell}(r)\right]\phi_{\omega\ell}= 0
\end{equation}
[here $r^\ast = r+ 2M \ln [r/(2M) -1]+\mathrm{const}$ denotes the tortoise coordinate] where
\begin{equation}\label{Potentiel}
  V_{\ell}(r) =\left(1-\frac{2M}{r}\right)\left(\frac{\ell(\ell+1)}{r^2}+(1-s^2)\frac{2M}{r^3}\right)
\end{equation}
(here $s=0$ corresponds to the scalar field and $s=1$ to the electromagnetic field) which have a purely ingoing behavior at the event horizon $r=2M$ (i.e., for $r_\ast \to -\infty$)
\begin{subequations}
\label{bc_in}
\begin{equation}\label{bc_1_in}
\phi^\mathrm{in}_{\omega \ell} (r)\scriptstyle{\underset{r_\ast \to -\infty}{\sim}} \displaystyle{e^{-i\omega r_\ast}}
\end{equation}
and, at spatial infinity $r \to +\infty$ (i.e., for $r_\ast \to +\infty$), an
asymptotic behavior of the form
\begin{equation}\label{bc_2_in}
\phi^\mathrm{in}_{\omega  \ell}(r) \scriptstyle{\underset{r_\ast \to +\infty}{\sim}}
\displaystyle{ A^{(-)}_\ell (\omega) e^{-i\omega r_\ast} + A^{(+)}_\ell (\omega) e^{+i\omega r_\ast}}.
\end{equation}
\end{subequations}
In this last equation, the coefficients $A^{(-)}_\ell (\omega)$ and  $A^{(+)}_\ell (\omega)$ are complex amplitudes and we have
\begin{equation}\label{Matrix_S}
  S_{\ell}(\omega) =  e^{i(\ell+1)\pi} \, \frac{A_{\ell}^{(+)}(\omega)}{A_{\ell}^{(-)}(\omega)}.
\end{equation}

\subsection{CAM representation of the scattering amplitude for scalar waves}
\label{SecIIb}

\subsubsection{Sommerfeld-Watson representation of the scattering amplitude}
\label{SecIIb1}

\begin{figure}
 \includegraphics[scale=0.50]{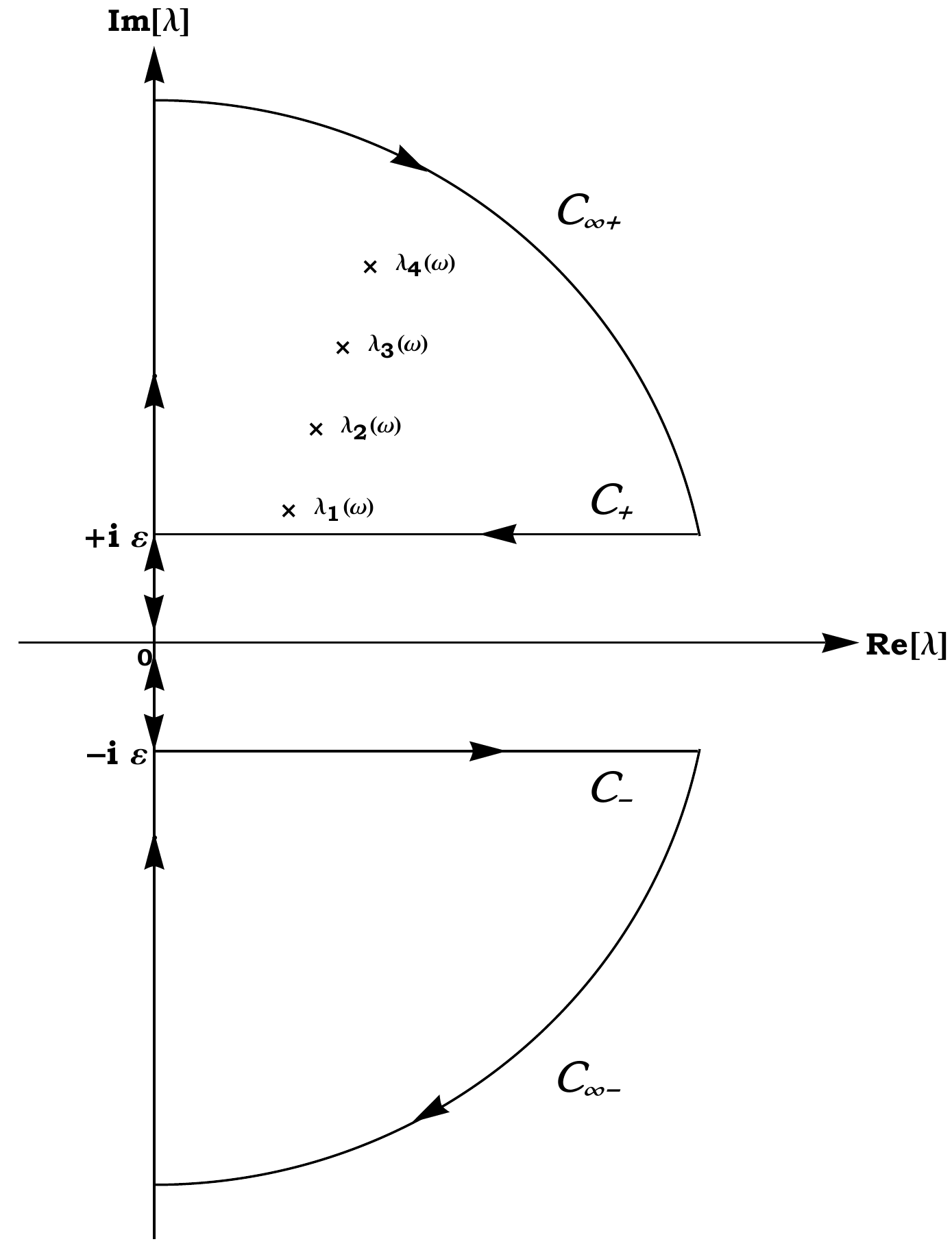}
\caption{\label{Contour_Deformation} The integration contour ${\cal C}={\cal C}_{+} \, \cup \, {\cal C}_{-}$ in the complex $\lambda $ plane. It defines the scattering amplitude (\ref{SW_Scalar_Scattering_amp}) and its deformations permit us to collect, by using Cauchy's theorem, the contributions of the Regge poles $\lambda_n(\omega)$.}
\end{figure}

By means of the Sommerfeld-Watson transformation \cite{Watson18,Sommerfeld49,Newton:1982qc} which permits us to write
\begin{equation}\label{SWT_gen}
\sum_{\ell=0}^{+\infty} (-1)^\ell F(\ell)= \frac{i}{2} \int_{\cal C} d\lambda \, \frac{F(\lambda -1/2)}{\cos (\pi \lambda)}
\end{equation}
for a function $F$ without any singularities on the real $\lambda$ axis, we can replace in Eq.~(\ref{Scalar_Scattering_amp}) the discrete sum over the ordinary angular momentum $\ell$ by a contour integral in the
complex $\lambda$ plane (i.e., in the complex $\ell$ plane with $\lambda = \ell +1/2$). By noting that $P_\ell (\cos \theta)=(-1)^\ell P_\ell (-\cos \theta)$, we obtain
\begin{eqnarray}\label{SW_Scalar_Scattering_amp}
& & f(\omega,\theta) = \frac{1}{2 \omega}  \int_{\cal C} d\lambda \, \frac{\lambda}{\cos (\pi \lambda)} \nonumber \\
&&  \qquad\qquad   \times \left[ S_{\lambda -1/2} (\omega) -1 \right]P_{\lambda -1/2} (-\cos \theta).
\end{eqnarray}
In Eqs.~(\ref{SWT_gen}) and (\ref{SW_Scalar_Scattering_amp}), the integration contour encircles counterclockwise the positive real axis of the complex $\lambda$ plane, i.e., we take ${\cal C}=]+\infty +i\epsilon,+i\epsilon] \cup
[+i\epsilon,-i\epsilon] \cup [-i\epsilon, +\infty -i\epsilon[$
with $\epsilon \to 0_+$ (see Fig.~\ref{Contour_Deformation}). We can recover (\ref{Scalar_Scattering_amp}) from (\ref{SW_Scalar_Scattering_amp}) by using Cauchy's theorem and by noting that the poles of the integrand in
(\ref{SW_Scalar_Scattering_amp}) that are enclosed into ${\cal C}$ are the zeros of $\cos (\pi \lambda)$, i.e., the semi-integers $\lambda = \ell + 1/2$ with $\ell \in \mathbb{N}$. It should be recalled that, in Eq.~(\ref{SW_Scalar_Scattering_amp}), the Legendre function of first kind $P_{\lambda -1/2} (z)$ denotes the analytic extension of the Legendre polynomials $P_\ell (z)$. It is defined in terms of hypergeometric functions by \cite{AS65}
\begin{equation}\label{Def_ext_LegendreP}
P_{\lambda -1/2} (z) = F[1/2-\lambda,1/2+\lambda;1;(1-z)/2].
\end{equation}
In Eq.~(\ref{SW_Scalar_Scattering_amp}), $S_{\lambda -1/2} (\omega)$ denotes ``the'' analytic extension of $S_\ell (\omega)$. It is given by [see Eq.~(\ref{Matrix_S})]
\begin{equation}\label{Matrix_S_CAM}
  S_{\lambda -1/2}(\omega) =  e^{i(\lambda + 1/2)\pi} \, \frac{A_{\lambda -1/2}^{(+)}(\omega)}{A_{\lambda -1/2}^{(-)}(\omega)}
\end{equation}
where the complex amplitudes $A^{(-)}_{\lambda -1/2} (\omega)$ and  $A^{(+)}_{\lambda -1/2} (\omega)$ are defined from the analytic extension of the modes $\phi_{\omega \ell}^{in}$, i.e., from the function $\phi_{\omega ,\lambda -1/2}^{in}$ solution of the problem (\ref{H_RW_equation})-(\ref{bc_in}) where we now replace $\ell$ by $\lambda -1/2$. With the deformation of the contour ${\cal C}$ in mind, it is important to note the symmetry property
\begin{equation}\label{Matrix_S_CAM_symm}
 e^{i \pi \lambda} \, S_{-\lambda -1/2}(\omega) =  e^{-i \pi \lambda} \, S_{\lambda -1/2}(\omega)
\end{equation}
of the $S$-matrix which can be easily obtained from its definition (see also Ref.~\cite{Andersson:1994rk}).

\subsubsection{Regge poles and associated residues}
\label{SecIIb2}

With the deformation of the contour ${\cal C}$ in mind, it is also important to recall that the poles of $S_{\lambda -1/2}(\omega)$ in the complex $\lambda$ plane (i.e., the Regge poles) lie in the first and third quadrants of this plane, symmetrically distributed with respect to the origin $O$. The poles lying in the first quadrant can be defined as the zeros $\lambda_n(\omega)$ with $n=1, 2, 3, \dots  $ of the coefficient $A^{(-)}_{\lambda-1/2} (\omega)$ [see Eq.~(\ref{Matrix_S_CAM})]. They therefore satisfy
\begin{equation}\label{PR_def_Am}
A^{(-)}_{\lambda_n(\omega)-1/2} (\omega)=0.
\end{equation}
In the following, the associated residues will play a central role. We note that the residue of the matrix $S_{\lambda-1/2}(\omega)$ at the pole $\lambda=\lambda_n(\omega)$ is defined by [see Eq.~(\ref{Matrix_S_CAM})]
\begin{equation}\label{residues_RP}
r_n(\omega)=e^{i\pi [\lambda_n(\omega)+1/2]} \left[ \frac{A_{\lambda -1/2}^{(+)}(\omega)}{\frac{d}{d \lambda}A_{\lambda -1/2}^{(-)}(\omega)}\right]_{\lambda=\lambda_n(\omega)}.
\end{equation}

\subsubsection{CAM representation of the scattering amplitude}
\label{SecIIb3}

We can now ``deform'' the contour ${\cal C}$ in Eq.~(\ref{SW_Scalar_Scattering_amp}) in order to collect, by using Cauchy's theorem, the contributions of the Regge poles lying in the first quadrant of the CAM plane (for more details, see, e.g., Ref.~\cite{Newton:1982qc}). We consider that ${\cal C}={\cal C}_{+} \cup {\cal C}_{-}$ with ${\cal C}_{+}=]+\infty +i\epsilon,+i\epsilon] \cup
[+i\epsilon,0] $ and ${\cal C}_{-}=[0,-i\epsilon]\cup [-i\epsilon, +\infty -i\epsilon[$ and we introduce the closed contours ${\cal C}_{+} \, \cup \, [0, +i\infty[ \, \cup \, {\cal C}_{\infty+}$ and ${\cal C}_{-} \, \cup \,  {\cal C}_{\infty-} \, \cup \, [-i\infty,0]$ (see Fig.~\ref{Contour_Deformation}). Here, the integration paths ${\cal C}_{\infty+}$ and ${\cal C}_{\infty-}$ are quarter circles at infinity lying respectively in the first and fourth quadrants of the complex $\lambda$ plane.

We first use Cauchy's residue theorem in connection with the closed contour ${\cal C}_{+} \, \cup \, [0, +i\infty[ \, \cup \, {\cal C}_{\infty+}$. We obtain
\begin{eqnarray}\label{Cauchy_Cplus}
& & \int_{{\cal C}_{+}} d\lambda \, \frac{\lambda}{\cos (\pi \lambda)} \left[ S_{\lambda -1/2} (\omega) -1 \right]P_{\lambda -1/2} (-\cos \theta) \nonumber \\
& &  \quad = -\int_0^{+i\infty} d\lambda \, \frac{\lambda}{\cos (\pi \lambda)} \left[ S_{\lambda -1/2} (\omega) -1 \right]P_{\lambda -1/2} (-\cos \theta) \nonumber \\
& &  \quad \phantom{=} -2i \pi    \sum_{n=1}^{+\infty}   \frac{ \lambda_n(\omega) r_n(\omega)}{\cos[\pi \lambda_n(\omega)]} P_{\lambda_n(\omega) -1/2} (-\cos \theta).
\end{eqnarray}
Here we have dropped the contribution coming from ${\cal C}_{\infty+}$ by noting that
\begin{equation}\label{for_quarter_circles_at_infinity_plus}
\frac{\lambda}{\cos (\pi \lambda)} \left[ S_{\lambda -1/2} (\omega) -1 \right]P_{\lambda -1/2} (-\cos \theta)
\end{equation}
vanishes faster than $1/\lambda$ for $|\lambda| \to +\infty$ and $\operatorname{Im} \lambda >0$.

We then use Cauchy's residue theorem in connection with the closed contour ${\cal C}_{-} \, \cup \,  {\cal C}_{\infty-} \, \cup \, ]-i\infty,0]$. This must be done with great caution because
\begin{equation}\label{for_quarter_circles_at_infinity_minus_b}
\frac{\lambda}{\cos (\pi \lambda)}  S_{\lambda -1/2} (\omega) P_{\lambda -1/2} (-\cos \theta)
\end{equation}
diverges for $|\lambda| \to +\infty$ and $\operatorname{Im} \lambda <0$ and, as a consequence, taking into account the contribution coming from the quarter circle ${\cal C}_{\infty-}$ is problematic. (It is interesting to note that, in addition, this result forbids us to collect, in a simple way, the Regge poles lying in the third quadrant of the complex $\lambda$ plane.) The difficulties encountered can be partially bypassed by using the relation \cite{AS65}
\begin{eqnarray}\label{Legendre_function_second_kind}
& & Q_{\lambda -1/2}(\cos \theta +i0) = \frac{\pi}{2 \cos(\pi \lambda)} \left[ P_{\lambda -1/2}(-\cos \theta) \phantom{e^{-i\pi \lambda}}\right. \nonumber \\
& & \qquad \qquad \left. - e^{-i\pi (\lambda-1/2)} P_{\lambda -1/2}(+\cos \theta )\right]
\end{eqnarray}
where $Q_{\lambda -1/2}(z)$ denotes the Legendre function of the second kind. Indeed, it permits us to replace (\ref{for_quarter_circles_at_infinity_minus_b}) by the equivalent expression
\begin{eqnarray}\label{for_quarter_circles_at_infinity_minus_c}
& & \frac{\lambda}{\cos (\pi \lambda)}  S_{\lambda -1/2} (\omega) e^{-i\pi (\lambda -1/2)}P_{\lambda -1/2} (+\cos \theta)\nonumber \\
 & & \qquad + \frac{2}{\pi } \lambda S_{\lambda -1/2} (\omega) Q_{\lambda -1/2} (\cos \theta +i0)
\end{eqnarray}
where the first term vanishes faster than $1/\lambda$ for $|\lambda| \to +\infty$ and $\operatorname{Im} \lambda <0$. By moreover noting that \begin{equation}\label{for_quarter_circles_at_infinity_minus_a}
\frac{\lambda}{\cos (\pi \lambda)} P_{\lambda -1/2} (-\cos \theta)
\end{equation}
vanishes faster than $1/\lambda$ for $|\lambda| \to +\infty$ and $\operatorname{Im} \lambda <0$, we can write
\begin{eqnarray}\label{Cauchy_Cmoins}
& & \int_{{\cal C}_{-}} d\lambda \, \frac{\lambda}{\cos (\pi \lambda)} \left[ S_{\lambda -1/2} (\omega) -1 \right]P_{\lambda -1/2} (-\cos \theta) \nonumber \\
& &  \quad = \int_{-i\infty}^0 d\lambda \, \frac{\lambda}{\cos (\pi \lambda)} P_{\lambda -1/2} (-\cos \theta) \nonumber \\
& &  \quad \phantom{=}
+ \frac{2}{\pi}  \int_{{\cal C}_{-}} d\lambda \, \lambda S_{\lambda -1/2} (\omega) Q_{\lambda -1/2} (\cos \theta +i0)
 \nonumber \\
& &  \quad \phantom{=}
-\int_{-i\infty}^0 d\lambda \, \frac{\lambda}{\cos (\pi \lambda)} S_{\lambda -1/2} (\omega) \nonumber \\
& & \qquad \qquad \qquad \qquad \times e^{-i\pi (\lambda-1/2)} P_{\lambda -1/2} (+\cos \theta) .
\end{eqnarray}

We finally insert the results (\ref{Cauchy_Cplus}) and (\ref{Cauchy_Cmoins}) into (\ref{SW_Scalar_Scattering_amp}) and by using (\ref{Matrix_S_CAM_symm}) as well as the relation \cite{AS65}
\begin{equation}\label{prop_ext_LegendreP}
P_{-\lambda -1/2} (z) = P_{\lambda -1/2} (z)
\end{equation}
we obtain
\begin{equation}\label{CAM_Scalar_Scattering_amp_tot}
f (\omega, \theta) =  f^\text{\tiny{B}} (\omega, \theta) +  f^\text{\tiny{RP}} (\omega, \theta)
\end{equation}
where
\begin{subequations}\label{CAM_Scalar_Scattering_amp_decomp}
\begin{equation}\label{CAM_Scalar_Scattering_amp_decomp_Background}
f^\text{\tiny{B}} (\omega, \theta) = f^\text{\tiny{B},\tiny{Re}} (\omega, \theta)+f^\text{\tiny{B},\tiny{Im}} (\omega, \theta)
\end{equation}
with
\begin{equation}\label{CAM_Scalar_Scattering_amp_decomp_Background_a}
f^\text{\tiny{B},\tiny{Re}} (\omega, \theta) = \frac{1}{\pi \omega} \int_{{\cal C}_{-}} d\lambda \, \lambda S_{\lambda -1/2}(\omega) Q_{\lambda -1/2}(\cos \theta +i0)
\end{equation}
and
\begin{equation}\label{CAM_Scalar_Scattering_amp_decomp_Background_b}
f^\text{\tiny{B},\tiny{Im}} (\omega, \theta) = \frac{1}{\pi \omega} \int_{+i\infty}^0 d\lambda \, \lambda S_{\lambda -1/2}(\omega) Q_{\lambda -1/2}(\cos \theta +i0)
\end{equation}
\end{subequations}
is a background integral contribution and where
\begin{eqnarray}\label{CAM_Scalar_Scattering_amp_decomp_RP}
& & f^\text{\tiny{RP}} (\omega, \theta) = -\frac{i \pi}{\omega}    \sum_{n=1}^{+\infty}   \frac{ \lambda_n(\omega) r_n(\omega)}{\cos[\pi \lambda_n(\omega)]}  \nonumber \\
&&  \qquad\qquad \qquad\qquad \times  P_{\lambda_n(\omega) -1/2} (-\cos \theta)
\end{eqnarray}
is a sum over the Regge poles lying in the first quadrant of the CAM plane. Of course, Eqs.~(\ref{CAM_Scalar_Scattering_amp_tot}), (\ref{CAM_Scalar_Scattering_amp_decomp}) and (\ref{CAM_Scalar_Scattering_amp_decomp_RP}) provide an exact representation of the scattering amplitude $f (\omega, \theta)$ for the scalar field, equivalent to the initial partial wave expansion (\ref{Scalar_Scattering_amp}). From this CAM representation, we can extract the contribution $f^\text{\tiny{RP}} (\omega, \theta)$ given by (\ref{CAM_Scalar_Scattering_amp_decomp_RP}) which, as a sum over Regge poles, is only an approximation of $f (\omega, \theta)$, and which provides us with an approximation of the differential scattering cross section (\ref{Scalar_Scattering_diff}).

\subsubsection{Remarks concerning the background integral}
\label{SecIIb4}

It is important to note that the path of integration defining the background integral (\ref{CAM_Scalar_Scattering_amp_decomp}) is a continuous one running down first the positive imaginary axis and then running along ${\cal C}_{-}$, i.e., slightly below the positive real axis. But, in fact, because the integrand in the right-hand side of (\ref{CAM_Scalar_Scattering_amp_decomp_Background_a}) is regular, this second branch can be deformed in order to coincide exactly with the positive real axis. As a consequence, we can take for the path of integration defining the background integral (\ref{CAM_Scalar_Scattering_amp_decomp}) the path $]+i \infty,0] \, \cup \, [0,+\infty[$ (see Fig.~\ref{Contour_integ_de_fond}) and we can replace (\ref{CAM_Scalar_Scattering_amp_decomp_Background_a}) by
\begin{eqnarray}\label{CAM_Scalar_Scattering_amp_decomp_Background_a_bis}
& & f^\text{\tiny{B},\tiny{Re}} (\omega, \theta) = \frac{1}{\pi \omega} \int_0^{+\infty} d\lambda \, \lambda S_{\lambda -1/2}(\omega) \nonumber \\
& & \qquad\qquad\qquad\qquad\qquad \times Q_{\lambda -1/2}(\cos \theta +i0).
\end{eqnarray}

\begin{figure}
 \includegraphics[scale=0.50]{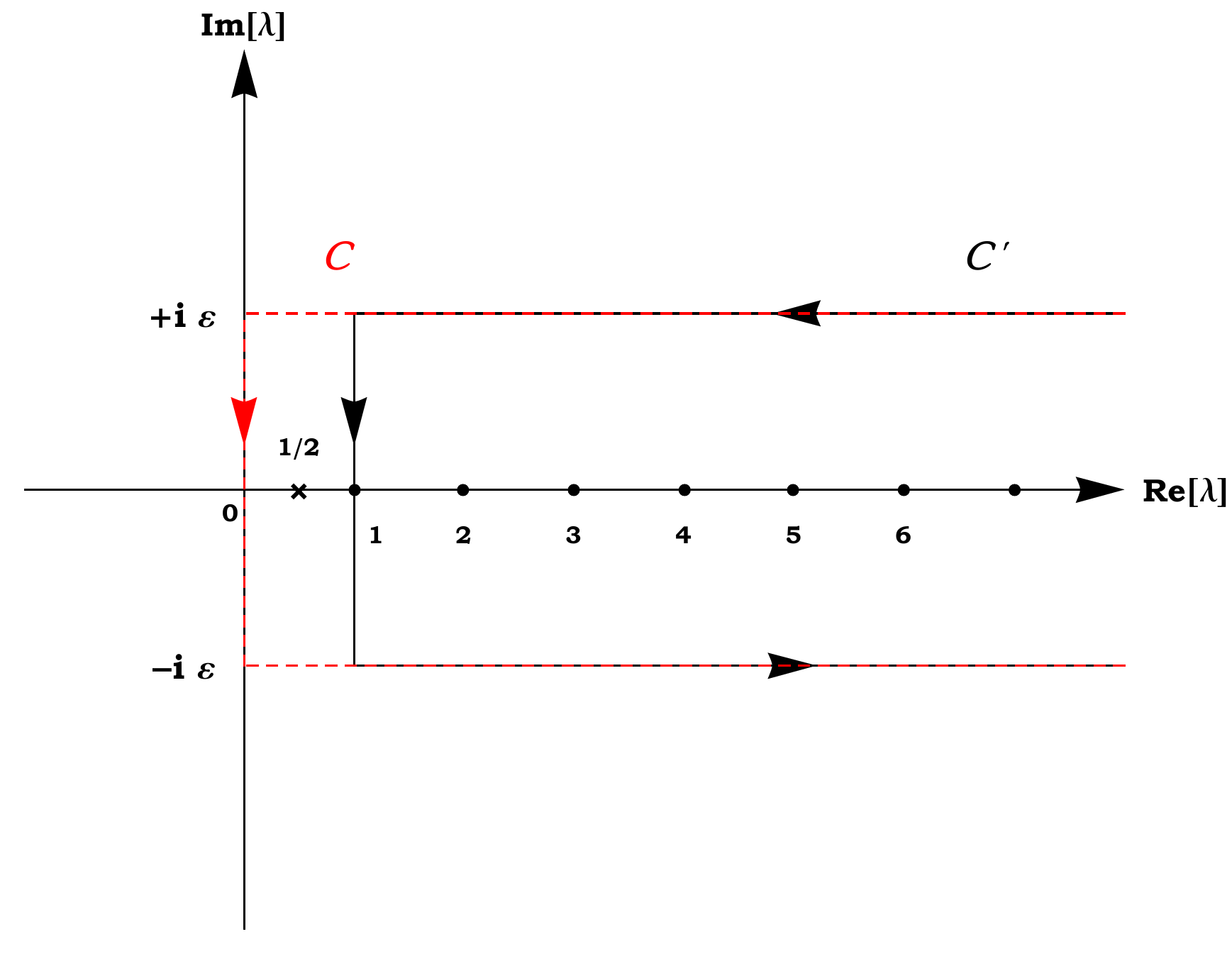}
\caption{\label{Contour_Shift_s1} Integration contours in the CAM plane: ${\cal C'}$ is associated with the scattering amplitude (\ref{SW_EM_Scattering_amp}) and ${\cal C}$  with the scattering amplitude (\ref{SW_EM_Scattering_amp_prov}).}
\end{figure}

\begin{figure}
 \includegraphics[scale=0.50]{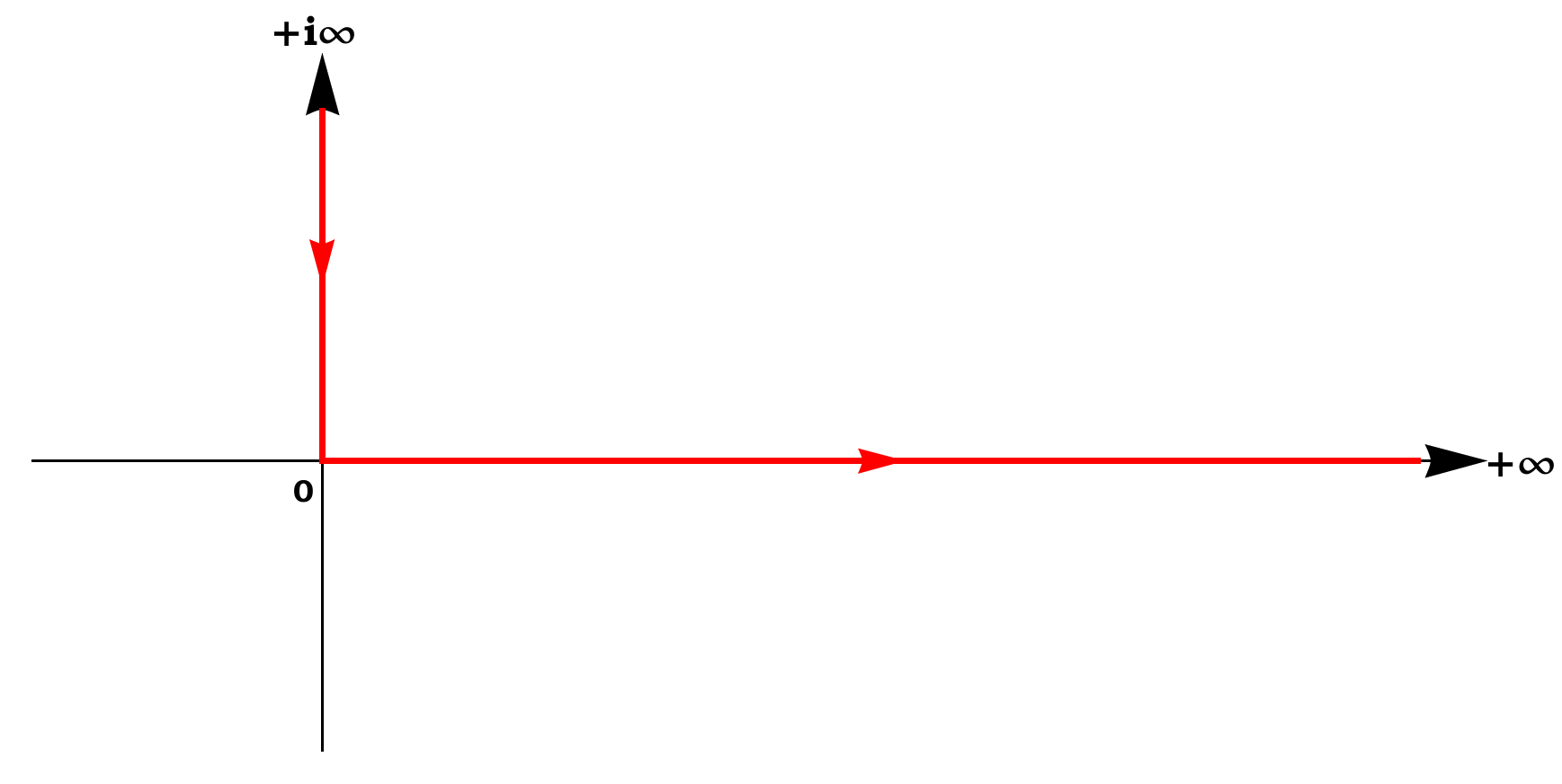}
\caption{\label{Contour_integ_de_fond} The path of integration in the complex $\lambda $ plane which defines the background integrals (\ref{CAM_Scalar_Scattering_amp_decomp}) and (\ref{CAM_EM_Scattering_amp_decomp}).}
\end{figure}

Moreover, with numerical calculations in mind, it is interesting to remark that the function $Q_{\lambda -1/2}(\cos \theta +i0)$ appearing in the integrand of the background integral can be expressed in the alternative forms \cite{AS65}
\begin{subequations} \label{Legendre_function_second_kind_bis}
\begin{eqnarray}\label{Legendre_function_second_kind_bis_a}
& & Q_{\lambda -1/2}(\cos \theta +i0) = Q_{\lambda -1/2}(\cos \theta) \nonumber \\
& & \qquad\qquad\qquad\qquad\qquad - i\frac{\pi}{2} P_{\lambda -1/2}(\cos \theta)
\end{eqnarray}
and
\begin{eqnarray}\label{Legendre_function_second_kind_bis_b}
& & Q_{\lambda -1/2}(\cos \theta +i0) = \frac{\Gamma (\lambda + 1/2)}{\Gamma (\lambda +1)} \sqrt{\frac{\pi}{2 \sin \theta}} e^{-i \left[\lambda \theta + \pi /4 \right]}  \nonumber \\
& & \qquad\qquad\qquad \times F[1/2,1/2;\lambda +1; i e^{-i \theta}/ (2 \sin \theta)].
\end{eqnarray}
\end{subequations}

\subsection{CAM representation of the scattering amplitude for electromagnetic waves}
\label{SecIIc}

\subsubsection{Sommerfeld-Watson representation of the scattering amplitude}
\label{SecIIc1}

In order to obtain the CAM representation of the scattering amplitudes (\ref{EM_Scattering_amp}) and (\ref{EM_Scattering_amp2}), we use the Sommerfeld-Watson transformation in the form
\begin{equation}\label{SWT_gen_mod1}
\sum_{\ell=1}^{+\infty} (-1)^\ell F(\ell)= \frac{i}{2} \int_{\cal C'} d\lambda \, \frac{F(\lambda -1/2)}{\cos (\pi \lambda)}.
\end{equation}
Here ${\cal C'}=]+\infty +i\epsilon,1+i\epsilon] \cup
[1+i\epsilon,1-i\epsilon] \cup [1-i\epsilon, +\infty -i\epsilon[$
with $\epsilon \to 0_+$ because we must take into account that the sum over $\ell$ defining the scattering amplitude (\ref{EM_Scattering_amp2}) begins at $\ell=1$. We then obtain
\begin{eqnarray}\label{SW_EM_Scattering_amp}
& & B(\omega,\theta) = \frac{1}{2 \omega}  \int_{\cal C'} d\lambda \, \frac{\lambda}{(\lambda^2-1/4)\cos (\pi \lambda)} \nonumber \\
&&  \qquad\qquad   \times \left[ S_{\lambda -1/2} (\omega) -1 \right]P_{\lambda -1/2} (-\cos \theta).
\end{eqnarray}
With the collect of Regge poles in mind, the contour ${\cal C'}$ is not as convenient as the contour ${\cal C}$ used in Sec.~\ref{SecIIb}. However, it is possible to move ${\cal C'}$ to the left so that it coincides with ${\cal C}$ but we then introduce a spurious double pole at $\lambda=1/2$ (i.e., at $\ell=0$) corresponding to the term $1/[(\lambda-1/2)\cos (\pi \lambda)]$ (see Fig.~\ref{Contour_Shift_s1}). It is necessary to remove the associated residue contribution and we obtain
\begin{widetext}
\begin{eqnarray}\label{SW_EM_Scattering_amp_prov}
& & B(\omega,\theta) = \frac{1}{2 \omega}  \left( \int_{\cal C} d\lambda \, \frac{\lambda}{(\lambda^2-1/4)\cos (\pi \lambda)} \left[ S_{\lambda -1/2} (\omega) -1 \right]P_{\lambda -1/2} (-\cos \theta) \right. \nonumber \\
&&  \qquad \left. - 2i\pi \lim_{\lambda \to 1/2} \frac{d}{d\lambda} \left[ (\lambda-1/2)^2 \times  \frac{\lambda}{(\lambda^2-1/4)\cos (\pi \lambda)} \left[ S_{\lambda -1/2} (\omega) -1 \right]P_{\lambda -1/2} (-\cos \theta)\right] \right).
\end{eqnarray}
In this equation, the residue contribution can be explicitly derived by using
\begin{equation}\label{der_FuncLeg_1}
\left[\frac{d}{d\lambda} P_{\lambda-1/2}(-x)\right]_{\lambda=1/2} = \ln \left( \frac{1-x}{2} \right)
\end{equation}
which is a consequence of the definition (\ref{Def_ext_LegendreP}) and by noting that, for the electromagnetic field, we have formally $S_0(\omega)=0$ [see Eqs.~(\ref{H_RW_equation})--(\ref{bc_in})]. We can then write
\begin{eqnarray}\label{SW_EM_Scattering_amp_def1}
& & B(\omega,\theta) = \frac{1}{2 \omega}  \int_{\cal C} d\lambda \, \frac{\lambda}{(\lambda^2-1/4)\cos (\pi \lambda)} \left[ S_{\lambda -1/2} (\omega) -1 \right]P_{\lambda -1/2} (-\cos \theta) \nonumber \\
&& \qquad \qquad -\frac{i}{2\omega} \ln\left[ \frac{1}{2} (1- \cos \theta)\right] + \mathrm{terms \,\, independent \,\, of \,\,} \theta.
\end{eqnarray}
\end{widetext}
By now noting that
\begin{eqnarray}
\mathcal{D}_{\theta}\, \left\lbrace \ln\left[ \frac{1}{2} (1- \cos \theta)\right] \right\rbrace =0
\end{eqnarray}
we finally obtain
\begin{eqnarray}\label{SW_EM_Scattering_amp_def2}
& & B(\omega,\theta) = \frac{1}{2 \omega}  \int_{\cal C} d\lambda \, \frac{\lambda}{(\lambda^2-1/4)\cos (\pi \lambda)} \nonumber \\
& & \qquad \times \left[ S_{\lambda -1/2} (\omega) -1 \right] P_{\lambda -1/2} (-\cos \theta).
\end{eqnarray}
Here we have dropped terms which do not contribute to the scattering amplitude $A(\omega,\theta)$.

\subsubsection{CAM representation of the scattering amplitude}
\label{SecIIc2}

We now deform the contour ${\cal C}$ in Eq.~(\ref{SW_EM_Scattering_amp_def2}) in order to collect, by using Cauchy's residue theorem, the Regge pole contributions. This is achieved by following, {\it mutatis mutandis}, the approach of Sec.~\ref{SecIIb}. We then obtain
\begin{equation}\label{CAM_EM_Scattering_amp_tot}
A (\omega, \theta) =  A^\text{\tiny{B}} (\omega, \theta) +  A^\text{\tiny{RP}} (\omega, \theta)
\end{equation}
where
\begin{subequations}\label{CAM_EM_Scattering_amp_decomp}
\begin{equation}\label{CAM_EM_Scattering_amp_decomp_Background}
A^\text{\tiny{B}} (\omega, \theta) = A^\text{\tiny{B},\tiny{Re}} (\omega, \theta)+A^\text{\tiny{B},\tiny{Im}} (\omega, \theta)
\end{equation}
with
\begin{eqnarray}\label{CAM_EM_Scattering_amp_decomp_Background_a}
& & A^\text{\tiny{B},\tiny{Re}} (\omega, \theta) = \mathcal{D}_{\theta} \left\lbrace \frac{1}{\pi \omega} \int_{{\cal C}_{-}} d\lambda \, \frac{\lambda}{\lambda^2-1/4} S_{\lambda -1/2}(\omega) \right. \nonumber \\
& &  \qquad\qquad \left. \phantom{\int_{{\cal C}_{-}}} \times Q_{\lambda -1/2}(\cos \theta +i0) \right\rbrace
\end{eqnarray}
and
\begin{eqnarray}\label{CAM_EM_Scattering_amp_decomp_Background_b}
& & A^\text{\tiny{B},\tiny{Im}} (\omega, \theta) = \mathcal{D}_{\theta} \left\lbrace \frac{1}{\pi \omega} \int_{+i\infty}^0 d\lambda \, \frac{\lambda}{\lambda^2-1/4} S_{\lambda -1/2}(\omega) \right. \nonumber \\
& &  \qquad\qquad \left. \phantom{\int_{+i\infty}^0} \times Q_{\lambda -1/2}(\cos \theta +i0) \right\rbrace
\end{eqnarray}
\end{subequations}
is a background integral contribution and where
\begin{eqnarray}\label{CAM_EM_Scattering_amp_decomp_RP}
& & A^\text{\tiny{RP}} (\omega, \theta) = \mathcal{D}_{\theta} \left\lbrace -\frac{i \pi}{\omega}    \sum_{n=1}^{+\infty}   \frac{ \lambda_n(\omega) r_n(\omega)}{[\lambda_n(\omega)^2-1/4] \cos[\pi \lambda_n(\omega)]} \right. \nonumber \\
&&  \qquad\qquad\qquad \left. \phantom{\sum_{n=1}^{+\infty}} \times  P_{\lambda_n(\omega) -1/2} (-\cos \theta) \right\rbrace
\end{eqnarray}
is a sum over the Regge poles of the $S$-matrix lying in the first quadrant of the CAM plane involving the associated residues.

It is worth pointing out that, as for the scalar theory, we can take for the path of integration defining the background integral (\ref{CAM_EM_Scattering_amp_decomp}) the path $]+i \infty,0] \, \cup \, [0,+\infty[$ depicted in Fig.~\ref{Contour_integ_de_fond}. Indeed, because the integrand in the right-hand side of (\ref{CAM_EM_Scattering_amp_decomp_Background_a}) is regular [note that the divergence of $1/(\lambda-1/2)$ for $\lambda=1/2$ is compensated by the vanishing of $S_{\lambda-1/2}(\omega)$], the path ${\cal C}_{-}$ can be deformed in order to coincide exactly with the positive real axis.

\section{Reconstruction of differential scattering cross sections from Regge pole sums}
\label{SecIII}

In this section, we compare numerically the partial wave expansions of the differential scattering cross sections with their CAM representations or, more precisely, with their Regge pole approximations in order to highlight the benefits of working with Regge pole sums.

\subsection{Computational methods}
\label{SecIIIa}

In order to construct numerically the scattering amplitudes (\ref{Scalar_Scattering_amp}) and (\ref{EM_Scattering_amp}), the background integrals  (\ref{CAM_Scalar_Scattering_amp_decomp_Background_a}), (\ref{CAM_Scalar_Scattering_amp_decomp_Background_b}), (\ref{CAM_EM_Scattering_amp_decomp_Background_a}) and (\ref{CAM_EM_Scattering_amp_decomp_Background_b}) as well as the Regge pole sums (\ref{CAM_Scalar_Scattering_amp_decomp_RP}) and (\ref{CAM_EM_Scattering_amp_decomp_RP}), it is necessary:
\begin{enumerate}[label=(\arabic*)]

\item  To solve the problem (\ref{H_RW_equation})-(\ref{bc_in}) permitting us to obtain the function $\phi_{\omega, \ell}^{\mathrm {in}}(r)$, the coefficients $A^{(-)}_\ell (\omega)$ and  $A^{(+)}_\ell (\omega)$ and the $S$-matrix elements $S_\ell (\omega)$. This must be achieved (i) for $\ell \in \mathbb{N}$ and $\omega >0$ as well as (ii) for $\ell=\lambda-1/2 \in \mathbb{C}$ and $\omega >0$.

\item To determine for $\omega>0$ the Regge poles $\lambda_n(\omega)$, i.e., the solutions of (\ref{PR_def_Am}) and to obtain the corresponding residues (\ref{residues_RP}).

\end{enumerate}
All these numerical results can be obtained by using, {\it mutatis mutandis}, the methods that have permitted us to provide in Ref.~\cite{Folacci:2018sef} a description of gravitational radiation from BHs based on CAM techniques (see, in particular, Secs.~IIIB and IVA of this previous paper). It is moreover important to note that, due to the long-range nature of the fields propagating on a Schwarzschild BH, the scattering amplitudes (\ref{Scalar_Scattering_amp}) and (\ref{EM_Scattering_amp}) and the background integrals  (\ref{CAM_Scalar_Scattering_amp_decomp_Background_a}) and (\ref{CAM_EM_Scattering_amp_decomp_Background_a}) suffer a lack of convergence [this is not the case for the background integrals (\ref{CAM_Scalar_Scattering_amp_decomp_Background_b}) and (\ref{CAM_EM_Scattering_amp_decomp_Background_b}) because their integrands vanish exponentially as $\lambda \to +i\infty$]. In the Appendix, we explain how to overcome this problem, i.e., how to accelerate their convergence by employing an iterative method, the number of iterations being chosen to obtain stable numerical results. It should be noted that we have performed all the numerical calculations by using {\it Mathematica} \cite{Mathematica}.

\subsection{Results and comments}
\label{SecIIIb}

\begin{figure*}
 \includegraphics[scale=0.50]{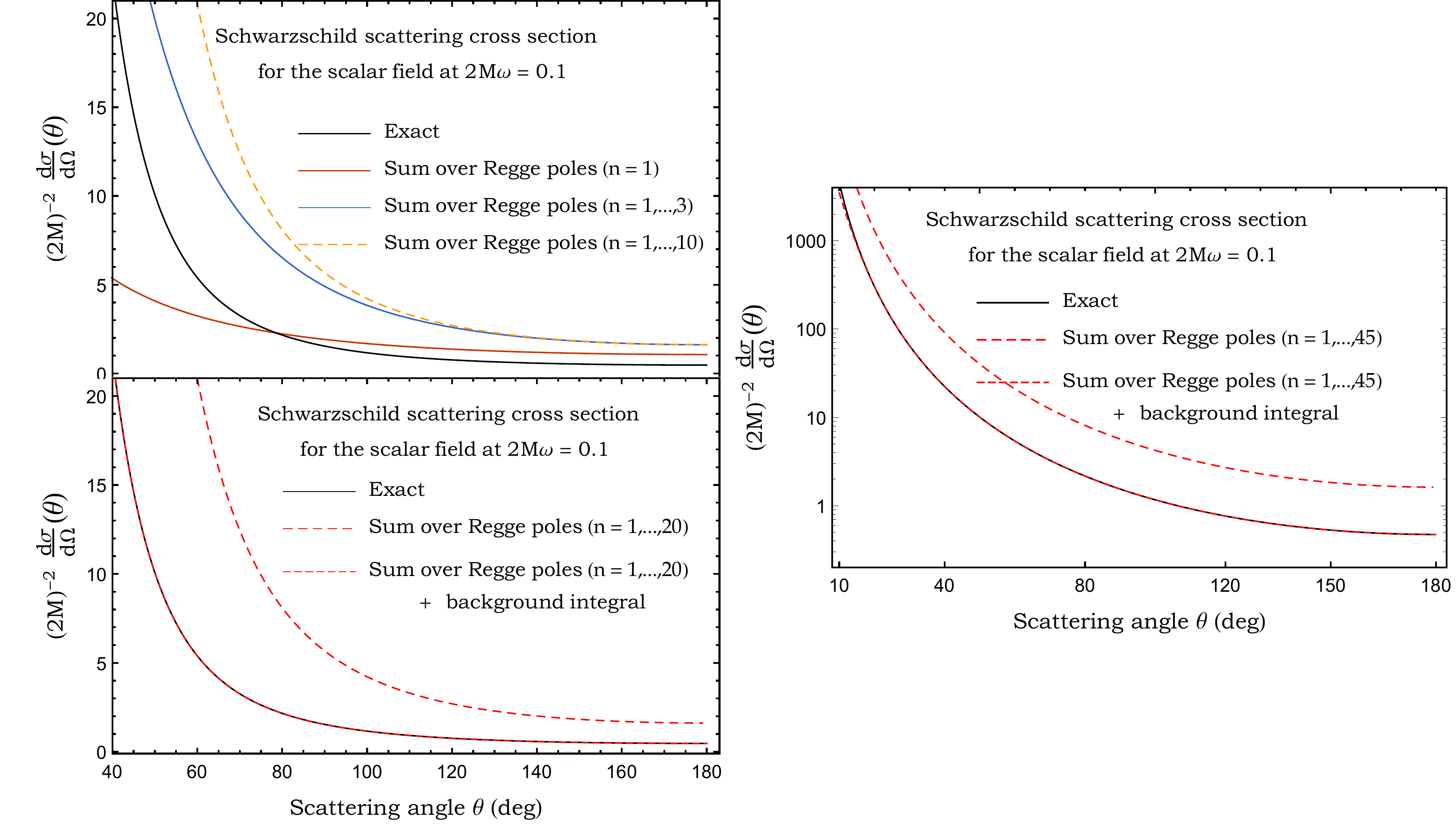}
\caption{\label{S_0_2Mw_01_Exact_vs_CAM} The scalar cross section of a Schwarzschild BH for $2M\omega=0.1$, its Regge pole approximation and the background integral contribution.}
\end{figure*}

\begin{figure*}
 \includegraphics[scale=0.50]{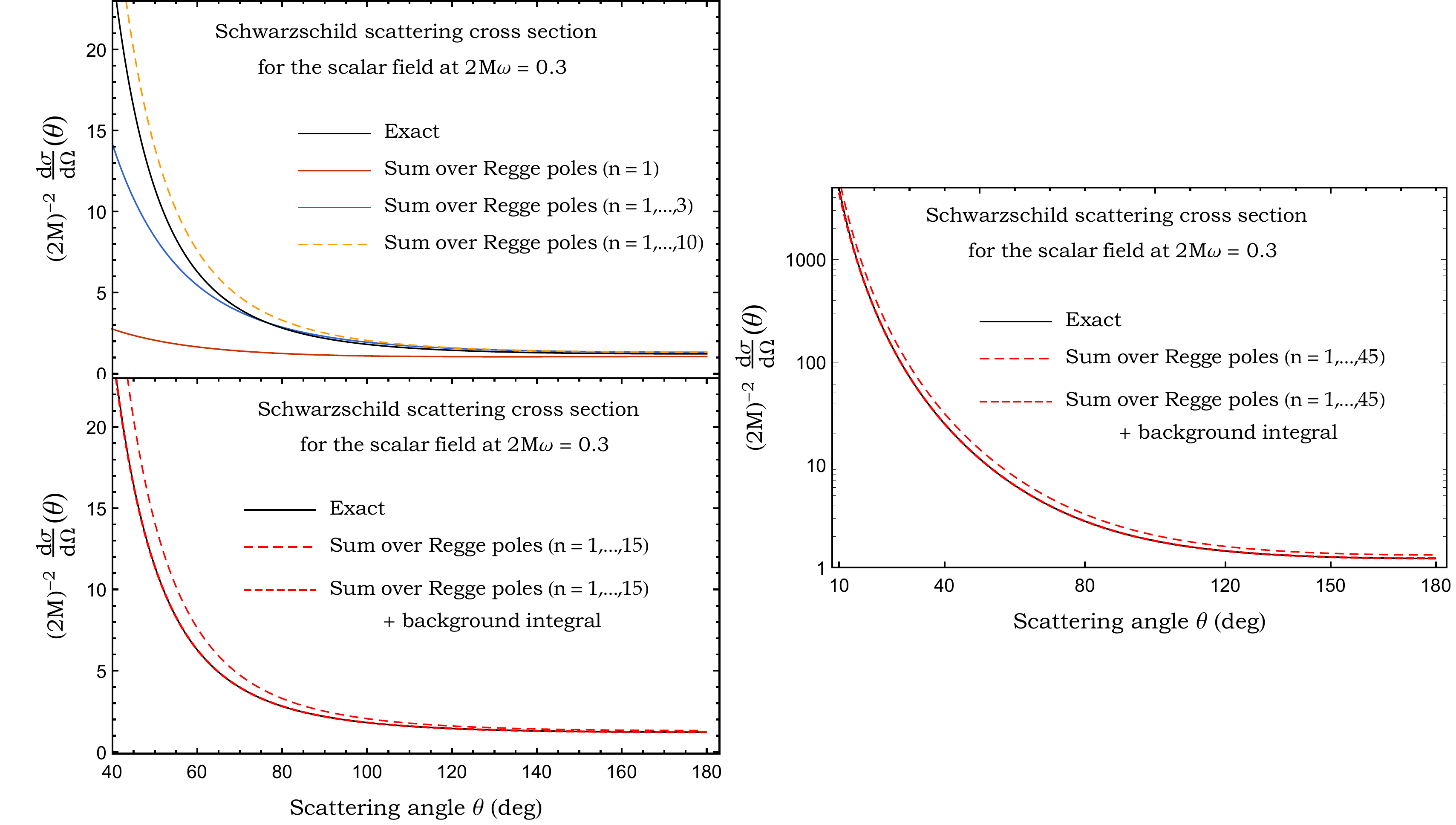}
\caption{\label{S_0_2Mw_03_Exact_vs_CAM} The scalar cross section of a Schwarzschild BH for $2M\omega=0.3$, its Regge pole approximation and the background integral contribution.}
\end{figure*}

\begin{figure*}
\centering
 \includegraphics[scale=0.50]{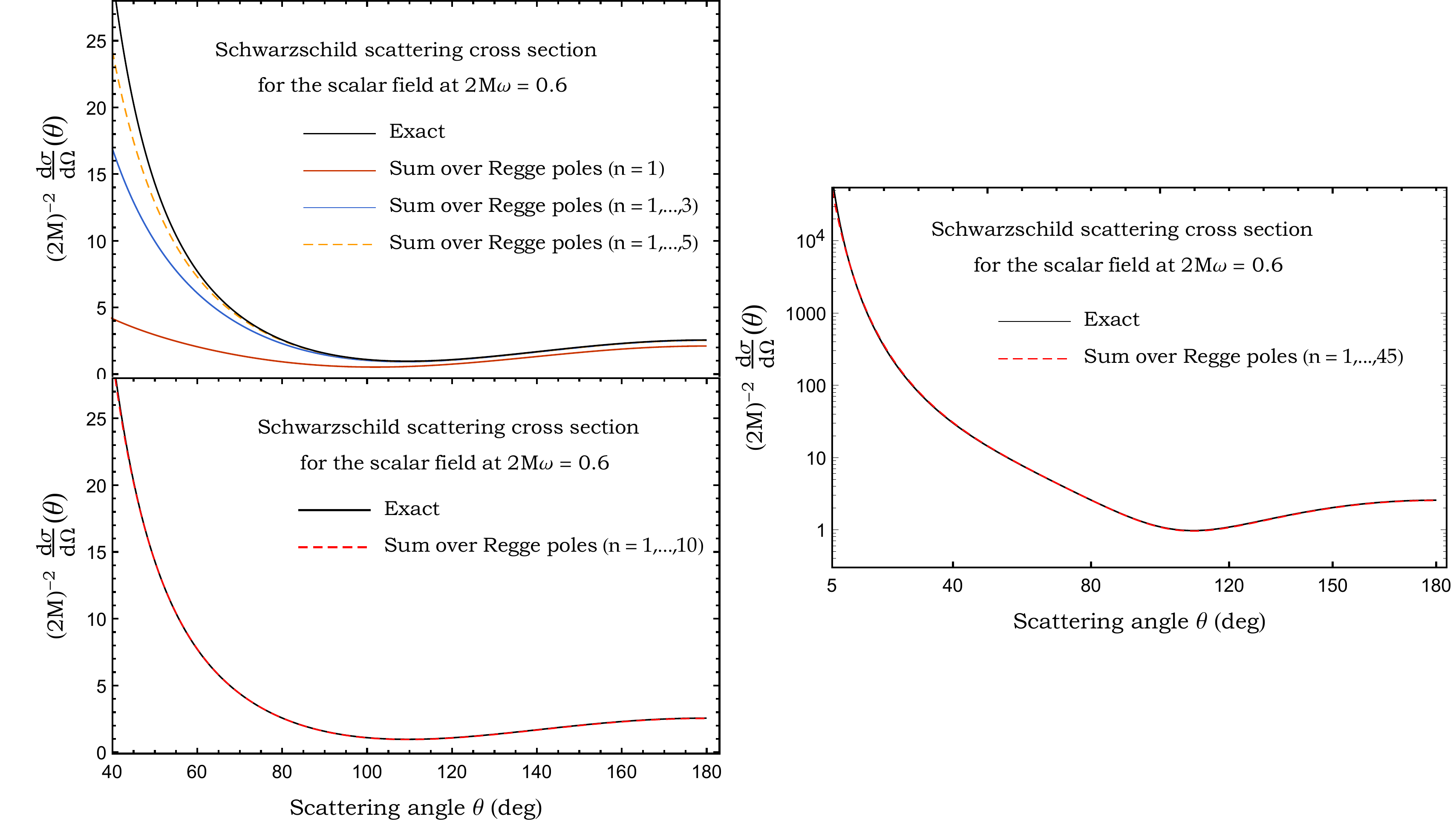}
\caption{\label{S_0_2Mw_06_Exact_vs_CAM} The scalar cross section of a Schwarzschild BH for $2M\omega=0.6$ and its Regge pole approximation.}
\end{figure*}

\begin{figure*}
\centering
 \includegraphics[scale=0.50]{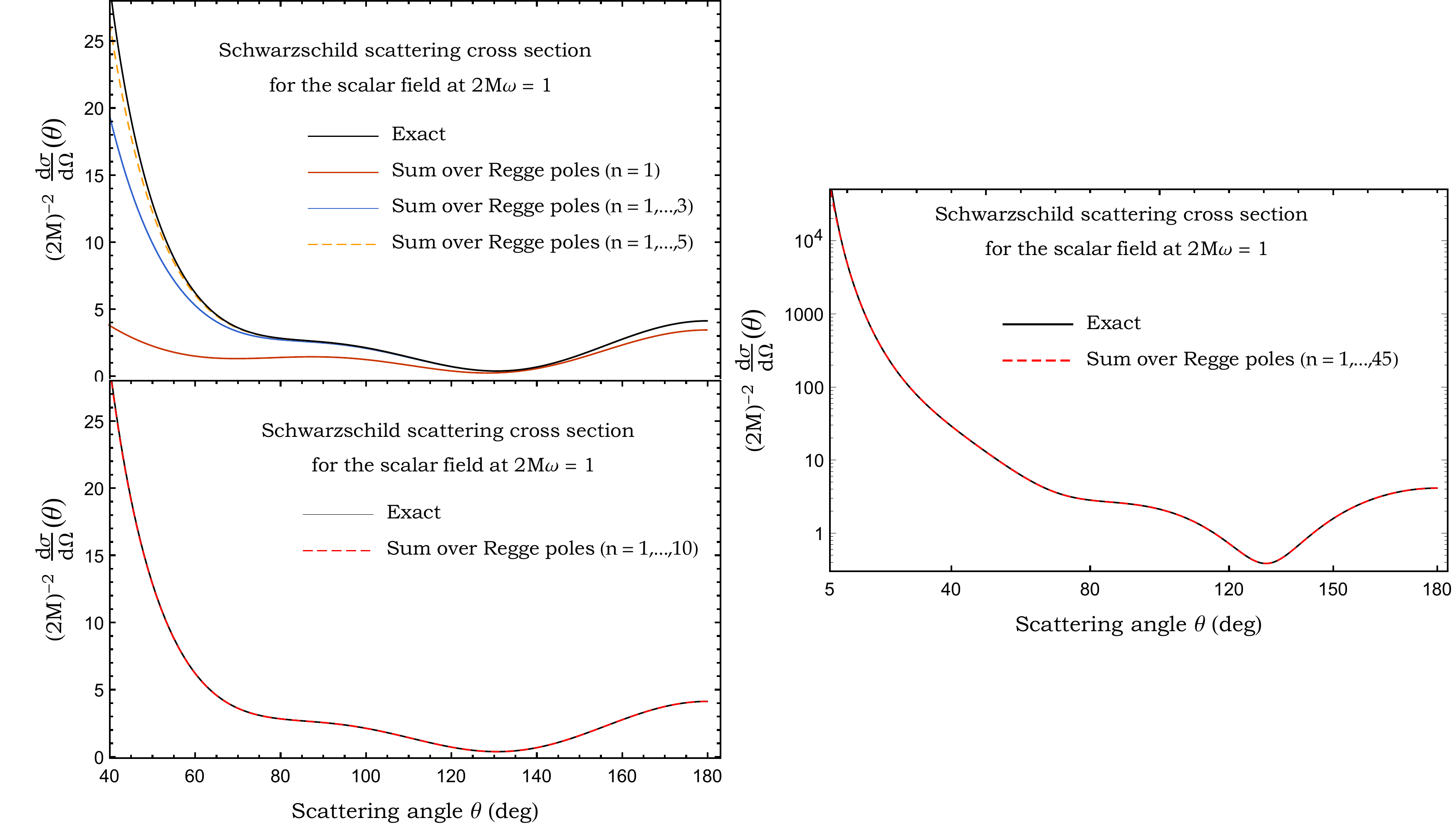}
\caption{\label{S_0_2Mw_1_Exact_vs_CAM} The scalar cross section of a Schwarzschild BH for $2M\omega=1$ and its Regge pole approximation.}
\end{figure*}

\begin{figure*}
\centering
 \includegraphics[scale=0.50]{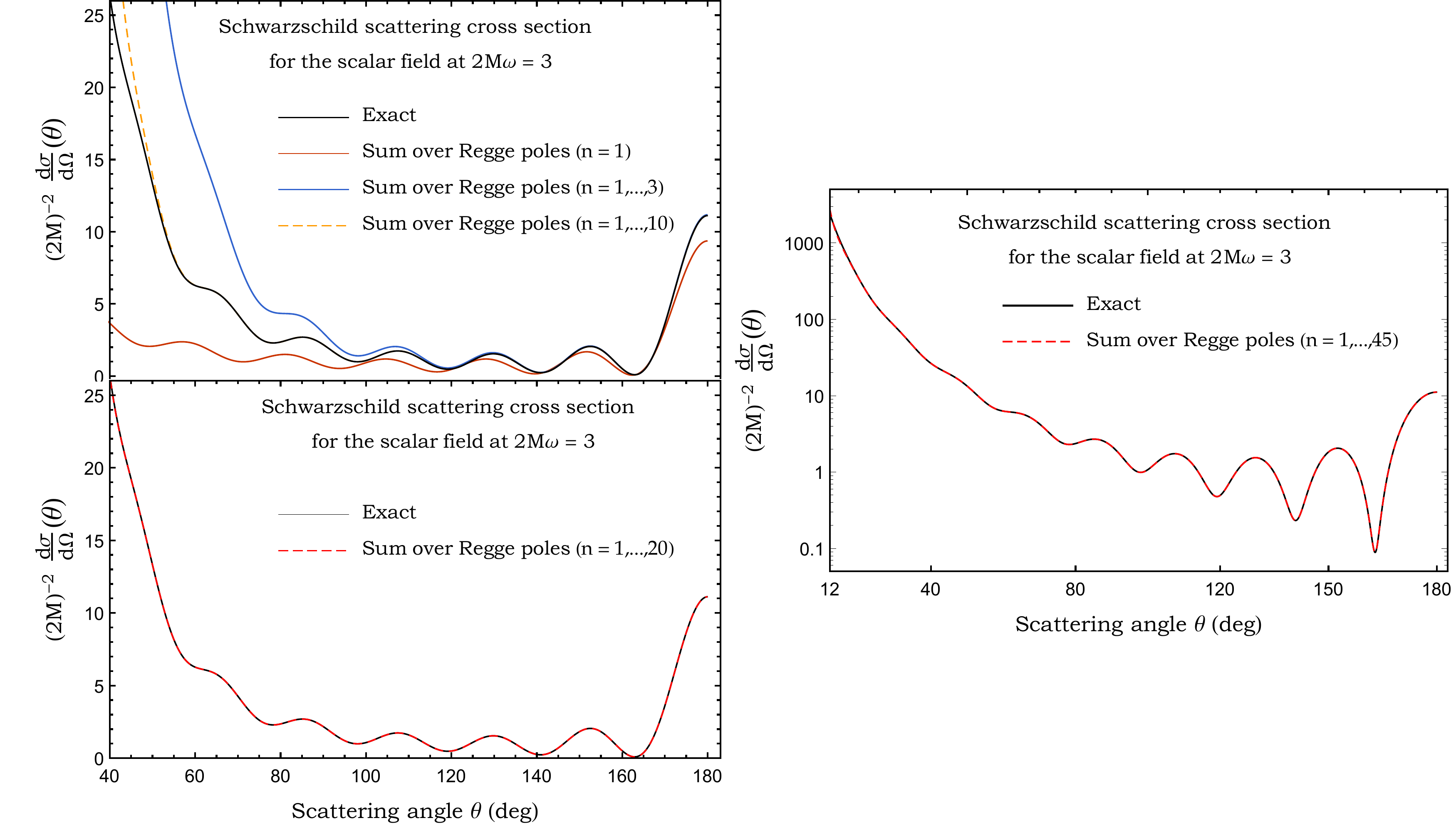}
\caption{\label{S_0_2Mw_3_Exact_vs_CAM} The scalar cross section of a Schwarzschild BH for $2M\omega=3$ and its Regge pole approximation.}
\end{figure*}

\begin{figure*}
\centering
 \includegraphics[scale=0.50]{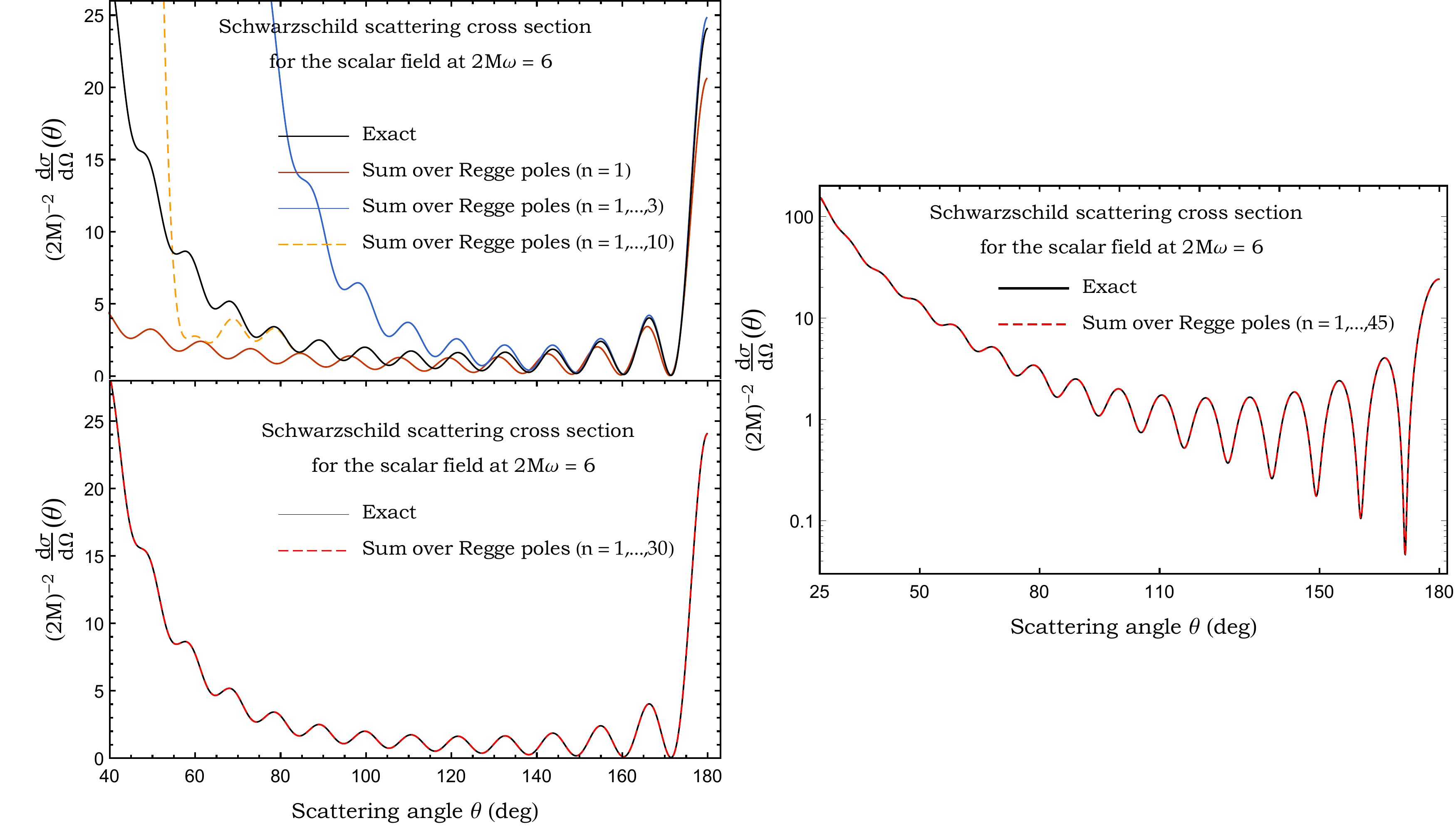}
\caption{\label{S_0_2Mw_6_Exact_vs_CAM} The scalar cross section of a Schwarzschild BH for $2M\omega=6$ and its Regge pole approximation.}
\end{figure*}

\begin{figure*}
\centering
 \includegraphics[scale=0.50]{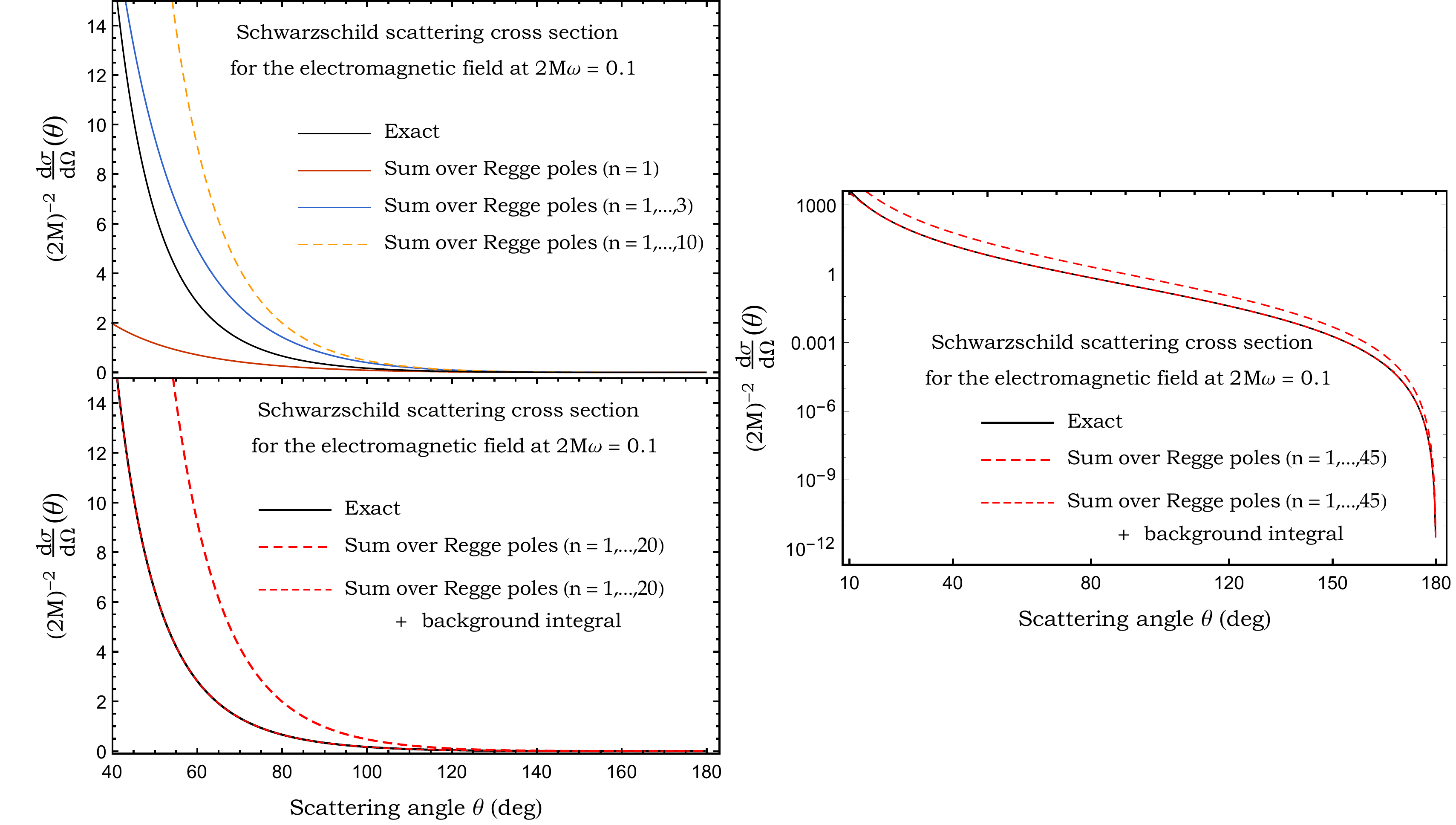}
\caption{\label{S_1_2Mw_01_Exact_vs_CAM} The electromagnetic cross section of a Schwarzschild BH for $2M\omega=0.1$, its Regge pole approximation and the background integral contribution.}
\end{figure*}

\begin{figure*}
\centering
 \includegraphics[scale=0.50]{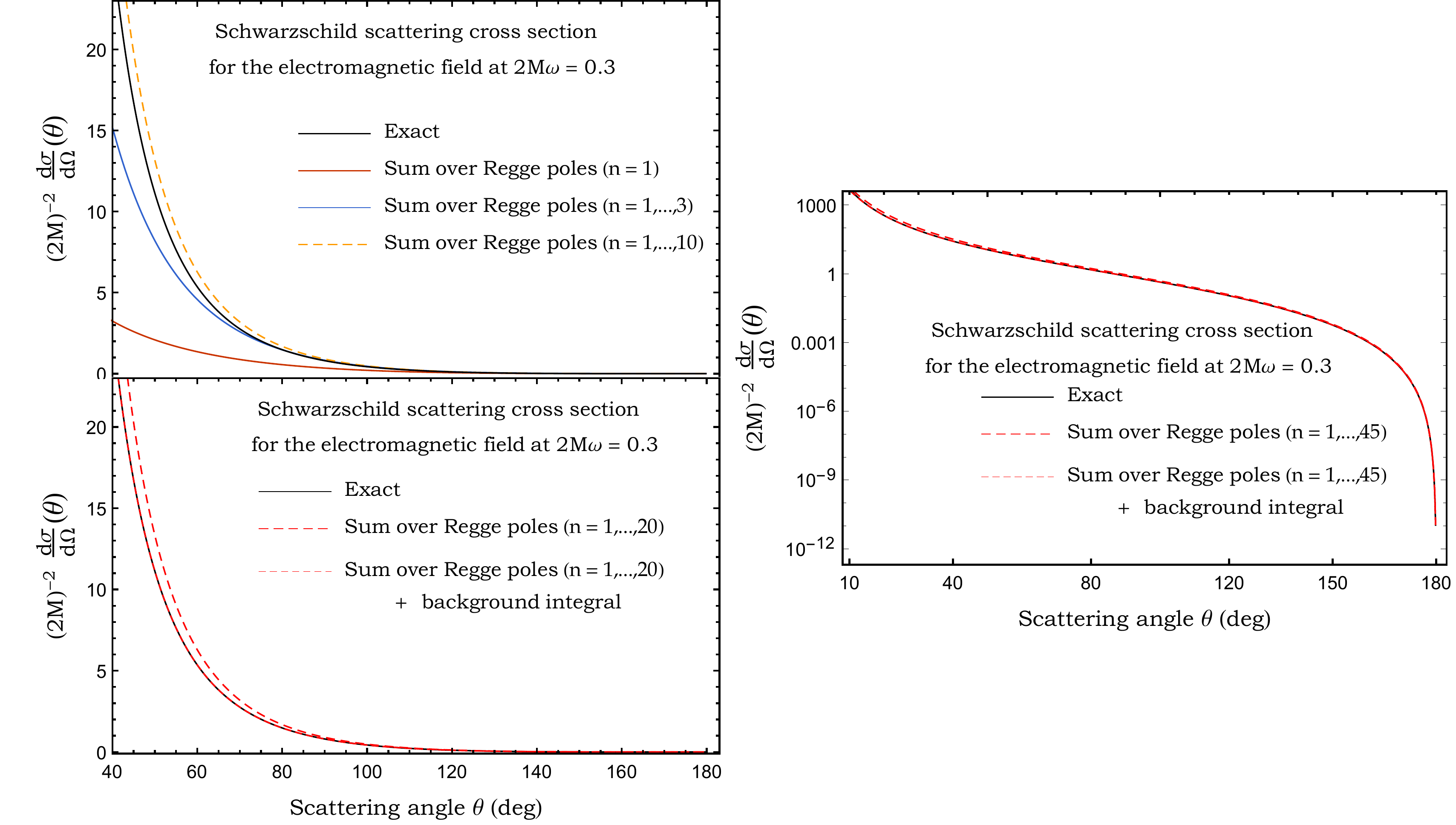}
\caption{\label{S_1_2Mw_03_Exact_vs_CAM} The electromagnetic cross section of a Schwarzschild BH for $2M\omega=0.3$, its Regge pole approximation and the background integral contribution.}
\end{figure*}

\begin{figure*}
\centering
 \includegraphics[scale=0.50]{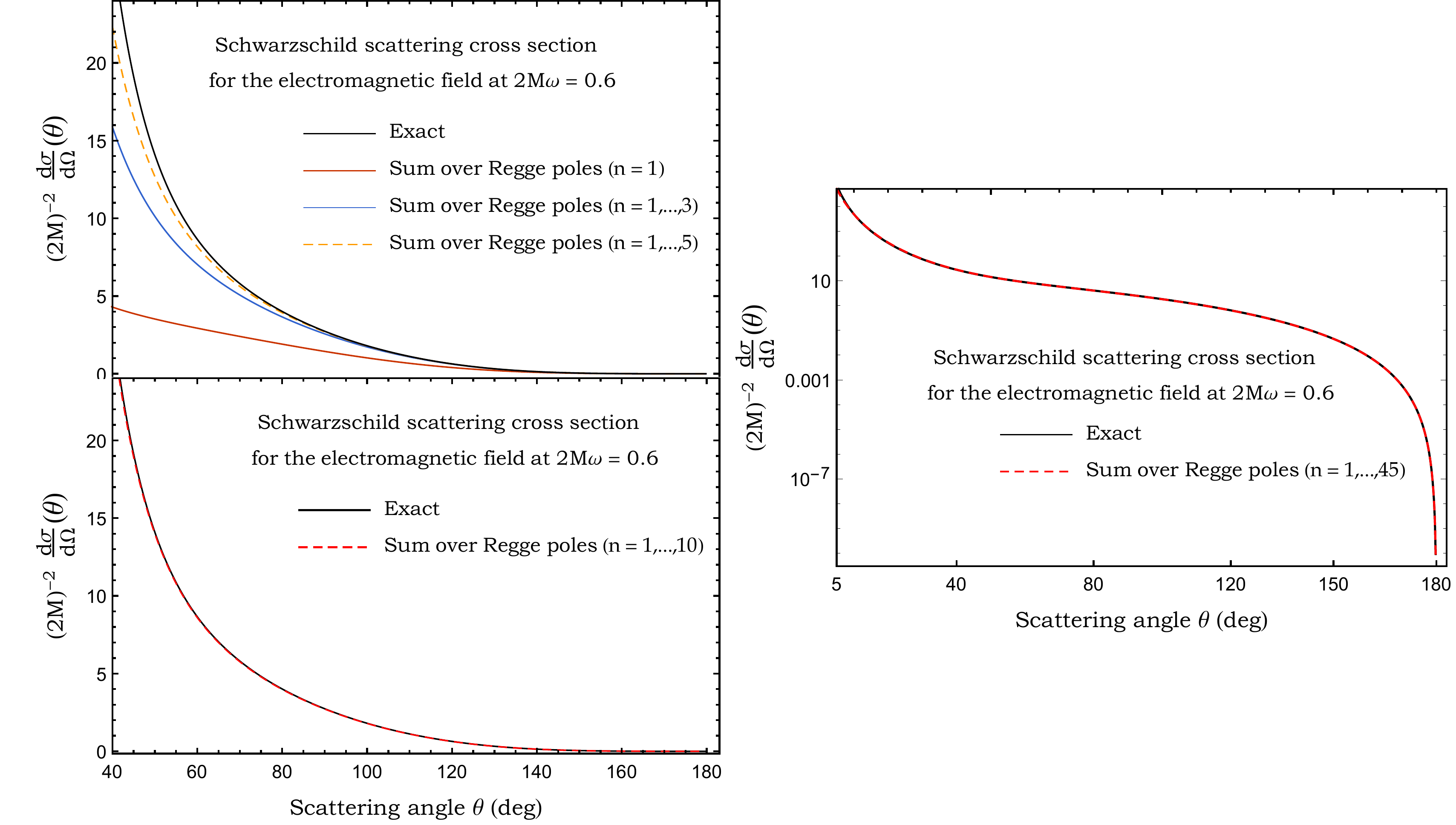}
\caption{\label{S_1_2Mw_06_Exact_vs_CAM} The electromagnetic cross section of a Schwarzschild BH for $2M\omega=0.6$ and its Regge pole approximation.}
\end{figure*}

\begin{figure*}
\centering
 \includegraphics[scale=0.50]{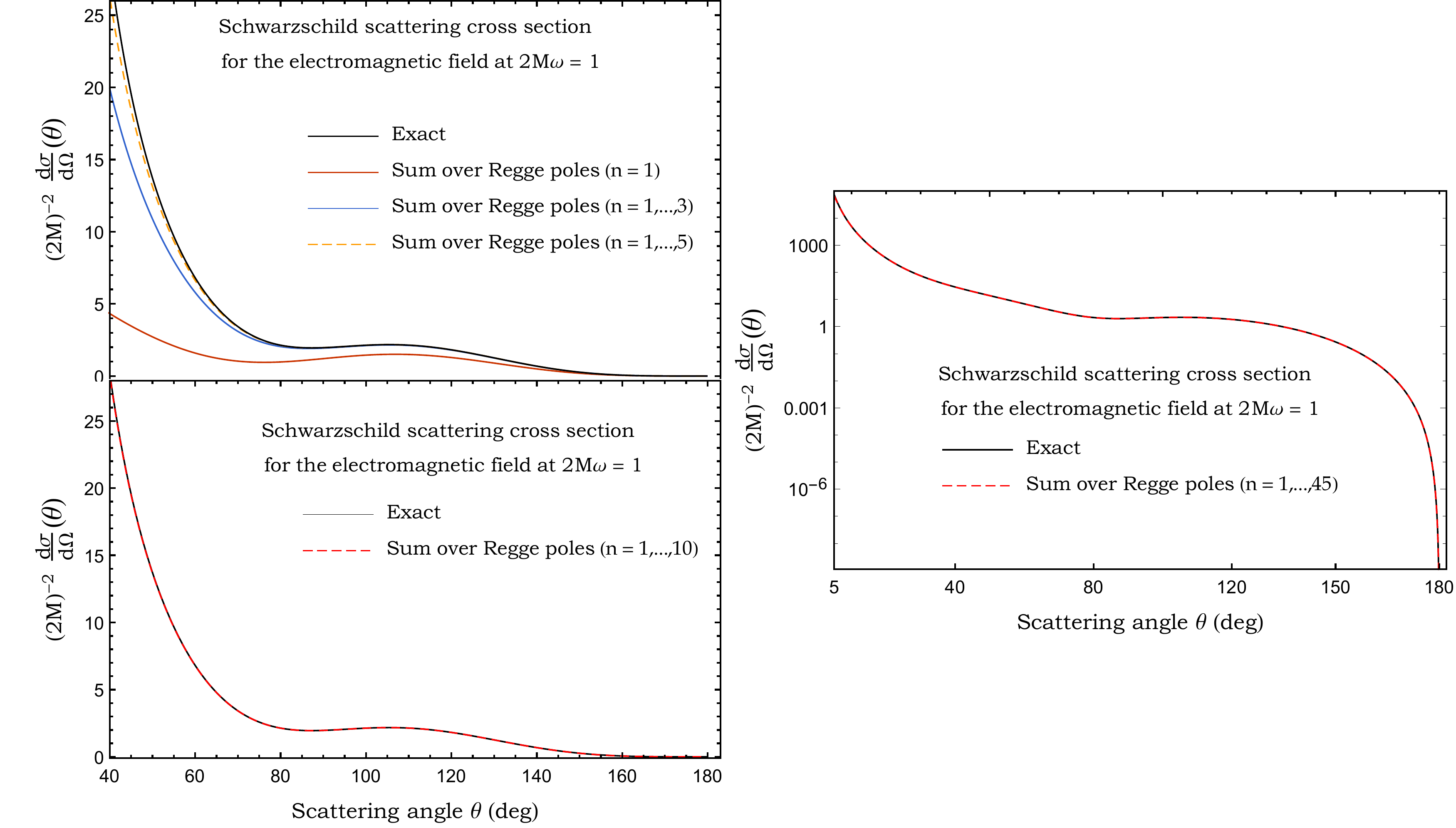}
\caption{\label{S_1_2Mw_1_Exact_vs_CAM} The electromagnetic cross section of a Schwarzschild BH for $2M\omega=1$ and its Regge pole approximation.}
\end{figure*}

\begin{figure*}
\centering
 \includegraphics[scale=0.50]{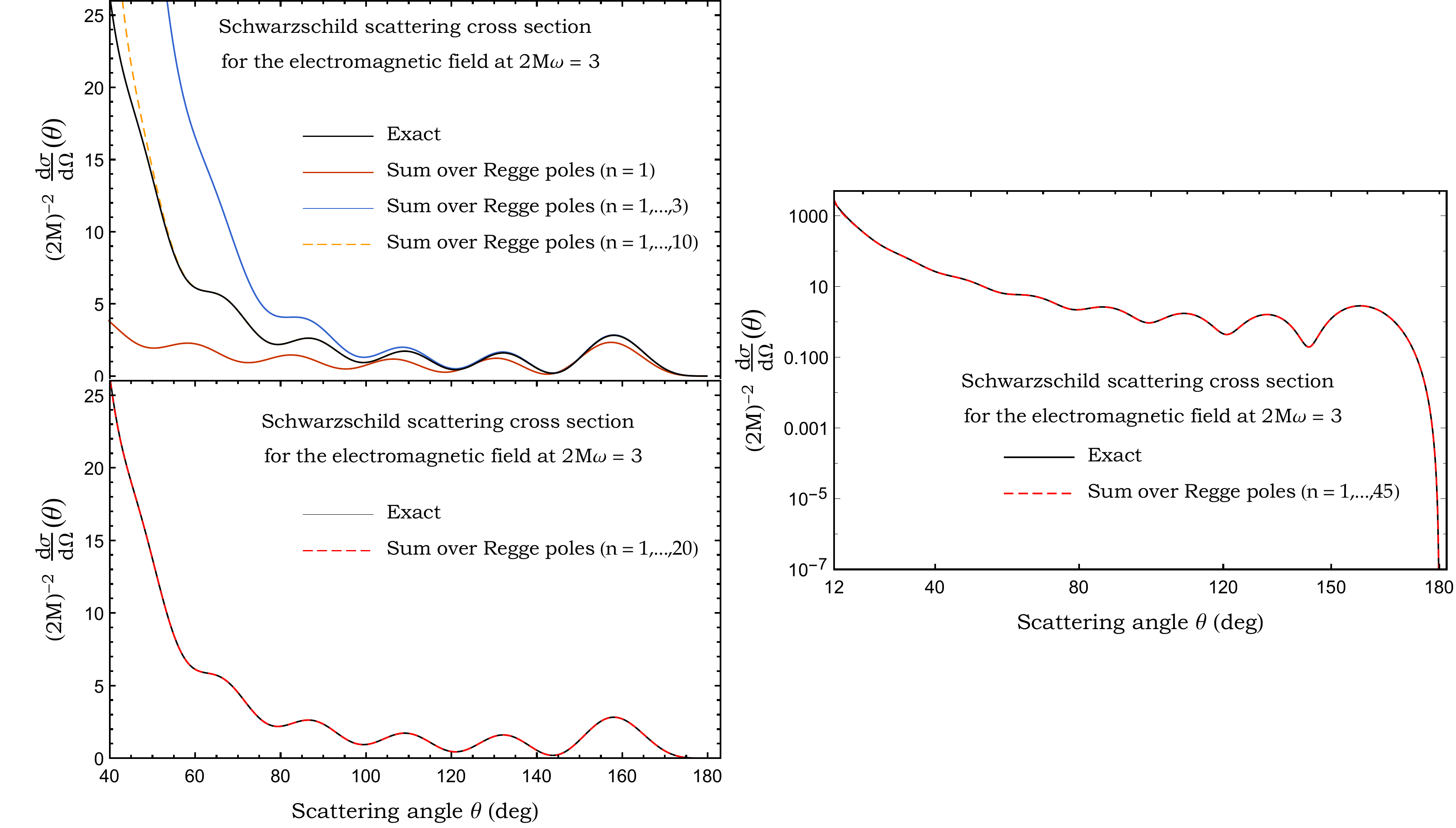}
\caption{\label{S_1_2Mw_3_Exact_vs_CAM} The electromagnetic cross section of a Schwarzschild BH for $2M\omega=3$ and its Regge pole approximation.}
\end{figure*}

\begin{figure*}
\centering
 \includegraphics[scale=0.50]{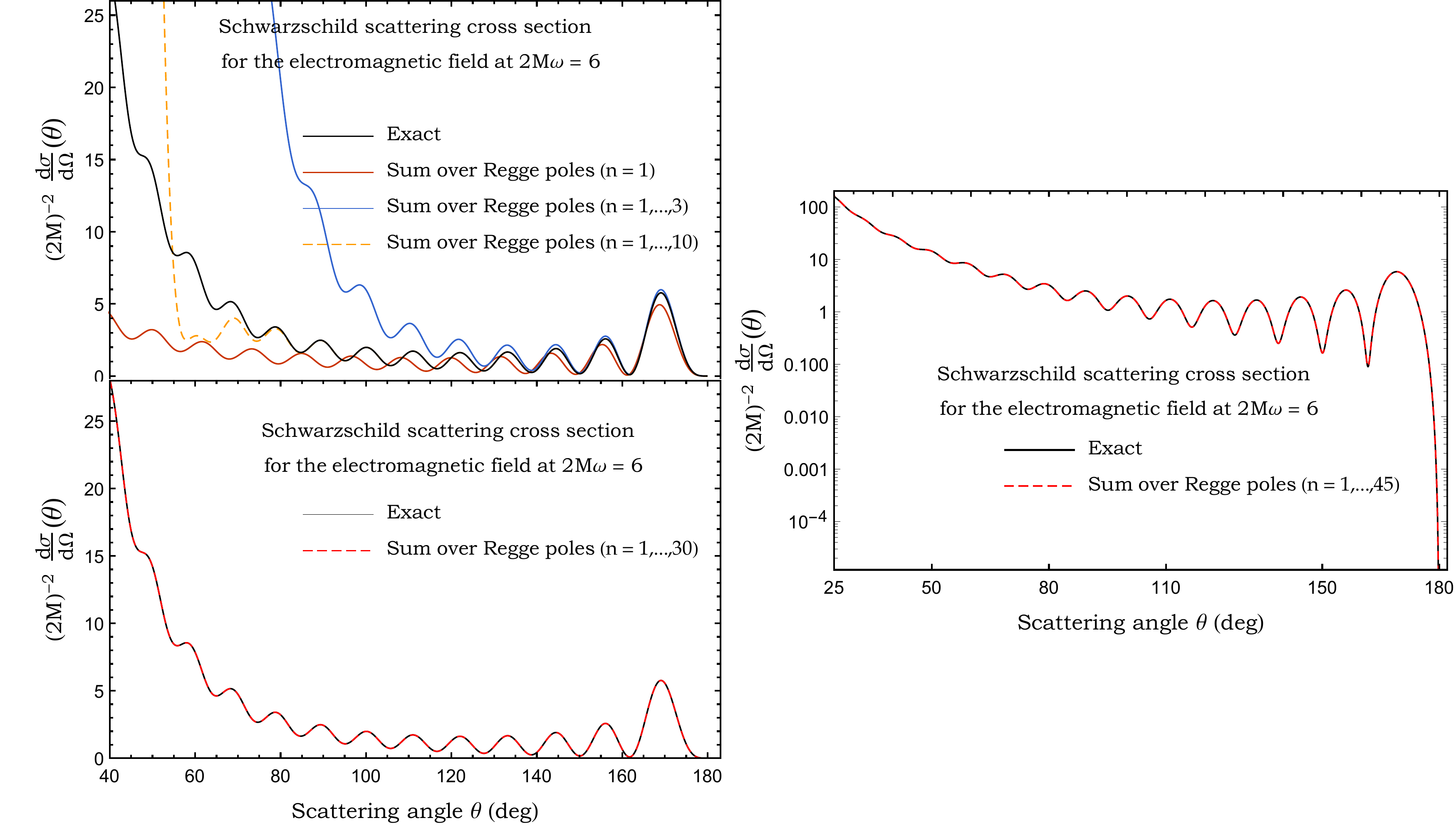}
\caption{\label{S_1_2Mw_6_Exact_vs_CAM} The electromagnetic cross section of a Schwarzschild BH for $2M\omega=6$ and its Regge pole approximation.}
\end{figure*}

In Figs.~\ref{S_0_2Mw_01_Exact_vs_CAM}-\ref{S_0_2Mw_6_Exact_vs_CAM}, we focus on the scalar field and we compare the differential scattering cross section (\ref{Scalar_Scattering_diff}) constructed from the partial wave expansion (\ref{Scalar_Scattering_amp}) with its Regge pole approximation obtained from the Regge pole sum (\ref{CAM_Scalar_Scattering_amp_decomp_RP}). In Figs.~\ref{S_1_2Mw_01_Exact_vs_CAM}-\ref{S_1_2Mw_6_Exact_vs_CAM} we focus on the electromagnetic field and we compare the differential scattering cross section (\ref{EM_Scattering_diff}) constructed from the partial wave expansion (\ref{EM_Scattering_amp})-(\ref{EM_Scattering_operator}) with its Regge pole approximation obtained from the Regge pole sum (\ref{CAM_EM_Scattering_amp_decomp_RP}). The comparisons are achieved for the reduced frequencies $2M\omega=$ 0.1, 0.3, 0.6, 1, 3 and 6 and, for these frequencies, we have displayed the lowest Regge poles and the associated residues in Table~\ref{tab:table1} (for the scalar field ) and in Table~\ref{tab:table2} (for the electromagnetic field). The higher Regge poles and their residues that have been necessary to obtain some of the results displayed in Figs.~\ref{S_0_2Mw_01_Exact_vs_CAM}-\ref{S_1_2Mw_6_Exact_vs_CAM} are available upon request from the authors.

In Figs.~\ref{S_0_2Mw_06_Exact_vs_CAM}-\ref{S_0_2Mw_6_Exact_vs_CAM} and \ref{S_1_2Mw_06_Exact_vs_CAM}-\ref{S_1_2Mw_6_Exact_vs_CAM}, we display the results obtained for intermediate and high reduced frequencies (here, we consider the reduced frequencies $2M\omega=$ 0.6, 1, 3 and 6). We can observe that, in the ``short''-wavelength regime, the Regge pole approximation involving a small number of Regge poles permits us to describe very well the cross sections for intermediate and large values of the scattering angle and, in particular, the BH glory. Taking into account additional Regge poles improves the Regge pole approximation and we can see that, by summing over a large number of Regge poles, the whole scattering cross section is impressively described, this being valid even for small scattering angles.

It is important to note that, for ``high'' reduced frequencies, it is not necessary to take into account the background integrals in order to reproduce the differential scattering cross sections. In fact, we can numerically obtain these contributions and observe that they are completely negligible for intermediate and large scattering angles. It seems they begin to play a role only for small angles. In Table~\ref{tab:table3}, we have considered, for the electromagnetic field at $2M\omega =1$, the various contributions to the CAM representation (\ref{CAM_EM_Scattering_amp_tot}). We can see that, for the scattering angles $\theta = 15\degree$ and $\theta = 20\degree$, the background integrals, although not completely negligible, play a minor role. It would be interesting to check if this remains valid even for scattering angles $\theta \ll 1/(2M\omega)$ but, due to numerical instabilities when $\theta \to 0$, we are not currently able to provide such a result.

In Figs.~\ref{S_0_2Mw_01_Exact_vs_CAM}, \ref{S_0_2Mw_03_Exact_vs_CAM}, \ref{S_1_2Mw_01_Exact_vs_CAM} and \ref{S_1_2Mw_03_Exact_vs_CAM}, we focus on the results obtained for low reduced frequencies (here, we consider the reduced frequencies $2M\omega=$ 0.1 and 0.3). We can observe that, in the long-wavelength regime, the Regge pole approximations (\ref{CAM_Scalar_Scattering_amp_decomp_RP}) and (\ref{CAM_EM_Scattering_amp_decomp_RP}) alone do not permit us to reconstruct the differential scattering cross sections but that this can be achieved by taking into account the background integral contributions (\ref{CAM_Scalar_Scattering_amp_decomp}) and (\ref{CAM_EM_Scattering_amp_decomp}).

\begingroup
\squeezetable
\begin{table}[H]
\caption{\label{tab:table1} The lowest Regge poles $\lambda_{n}(\omega)$ for the scalar field and the associated residues $r_{n}(\omega)$. We assume $2M=1$.}
\smallskip
\centering
\begin{ruledtabular}
\begin{tabular}{cccc}
 $n$ & $\omega$  & $\lambda_n(\omega)$ & $r_{n}(\omega)$
 \\ \hline
$1$  & $0.1$ &$\phantom{1}0.299705 + 0.532035 i$  & $\phantom{-}0.110788 - 0.088226 i$   \\
     & $0.3$ &$\phantom{1}0.768159 + 0.531039 i $  & $\phantom{-} 0.229999 - 0.159051 i$   \\
     & $0.6$ &$\phantom{1}1.543385 + 0.511888 i $  & $\phantom{-} 0.389142 - 0.125974 i $   \\
     & $1$ &$\phantom{1}2.586845 + 0.504736 i$  & $\phantom{-} 0.527572 + 0.019732 i$   \\
     & $3$ &$\phantom{1}7.790118 + 0.500553 i$  & $-0.136670 + 0.901886 i$   \\
     & $6$ &$15.586384 + 0.500139i$  & $-0.594258 - 1.144512 i$   \\

$2$  & $0.1$ &$\phantom{1}0.495889 + 1.176250 i $  & $\phantom{-}0.099116 - 0.087234 i $   \\
     & $0.3$ &$\phantom{1}0.976977 + 1.364123 i$  & $\phantom{-}0.119559 - 0.228260 i $   \\
     & $0.6$ &$\phantom{1}1.695146 + 1.448184 i $  & $\phantom{-} 0.181312 - 0.487058 i $   \\
     & $1$ &$\phantom{1}2.688355 + 1.479587 i$  & $\phantom{-} 0.447998 - 0.888968 i$\\
     & $3$ &$\phantom{1}7.825611 + 1.497769 i$  & $\phantom{-} 4.596565 + 1.422003 i$   \\
     & $6$ &$15.604187 + 1.499447i $  & $-12.419829 + 5.305460 i$ \\

$3$  & $0.1$ &$\phantom{1}0.650192 + 1.778513 i $  & $\phantom{-} 0.087664 - 0.087739 i $   \\
     & $0.3$ &$\phantom{1}1.164769 + 2.098578 i $  & $\phantom{-} 0.056734 - 0.236550 i $   \\
     & $0.6$ &$\phantom{1}1.878969 + 2.288202 i $  & $-0.078148 - 0.527337 i  $   \\
     & $1$ &$\phantom{1}2.843093 + 2.391435 i$  & $-0.357084 - 1.170498 i$\\
     & $3$ &$\phantom{1}7.893803 + 2.484084 i$  & $\phantom{-} 7.751902 - 10.715411 i$\\
     & $6$ &$15.639426 + 2.495896i$  & $\phantom{-} 16.758993 + 69.494273 i$ \\

$4$  & $0.1$ &$\phantom{1}0.788665 + 2.349868 i $  & $\phantom{-} 0.079894 - 0.087791 i $   \\
     & $0.3$ &$\phantom{1}1.339372 + 2.783917 i $  & $\phantom{-} 0.016812 - 0.232342 i$   \\
     & $0.6$ &$\phantom{1}2.065320 + 3.068702 i $  & $-0.245770 - 0.466956 i  $   \\
     & $1$ &$\phantom{1}3.018870 + 3.248630 i$  & $-0.987253 - 0.88251 i$\\
     & $3$ &$\phantom{1}7.990039 + 3.454396 i$  & $-11.471634 - 23.211359 i$\\
     & $6$ &$15.691407 + 3.487740 i$ & $\phantom{-} 256.923913 + 5.083667 i$ \\

$5$  & $0.1$ &$\phantom{1}0.917562 + 2.899702 i $  & $\phantom{-} 0.074199 - 0.087641 i $   \\
     & $0.3$ &$\phantom{1}1.504128 + 3.437384 i $  & $           -0.011069 - 0.224974 i$   \\
     & $0.6$ &$\phantom{1}2.248056 + 3.809074 i $  & $           -0.348103 - 0.384562 i $   \\
     & $1$ &$\phantom{1}3.201835 + 4.063591 i $    & $           -1.301089 - 0.421317 i $\\
     & $3$ &$\phantom{1}8.109029 + 4.405790 i $    & $           -40.72701 - 6.755946 i $\\
     & $6$ &$15.759163 + 4.473447 i$               & $\phantom{-}251.611779-664.675235 i $ \\

$6$  & $0.1$ &$\phantom{1}1.039655 + 3.433425 i $  & $\phantom{-} 0.069763 - 0.087418 i$   \\
     & $0.3$ &$\phantom{1}1.661166 + 4.067794 i $  & $           -0.031853 - 0.216897 i $   \\
     & $0.6$ &$\phantom{1}2.425852 + 4.520191 i $  & $-0.410128 - 0.302558 i  $   \\
     & $1$ &$\phantom{1}3.386161 + 4.845810 i $  & $-1.384667 + 0.028934 i $\\
     & $3$ &$\phantom{1}8.245770 + 5.337166 i $  & $-45.341516 + 35.668449 i $\\
     & $6$ &$15.841536 + 5.451746 i$ & $-1147.818563 - 1147.131306 i $ \\

$7$  & $0.1$ &$\phantom{1}1.156528 + 3.954439 i $  & $\phantom{-}0.066156 - 0.087170 i $   \\
     & $0.3$ &$\phantom{1}1.811951 + 4.680400 i $  & $          -0.048060 - 0.208882 i $   \\
     & $0.6$ &$\phantom{1}2.598670 + 5.208747 i $  & $          -0.446908 - 0.227138 i $   \\
     & $1$ &$\phantom{1}3.569316 + 5.602015 i $ & $-1.328185 + 0.409299 i $\\
     & $3$ &$\phantom{1}8.396000 + 6.248658 i $ & $-12.517767 + 72.751731 i $\\
     & $6$ &$15.937250 + 6.421641 i$  & $-3071.847422 + 904.87469 i $ \\

$8$  & $0.1$ &$\phantom{1}1.269207 + 4.465047 i $  & $\phantom{-}0.063133 - 0.086918 i  $   \\
     & $0.3$ &$\phantom{1}1.957532 + 5.278644 i $  & $          -0.061116 - 0.201188 i  $   \\
     & $0.6$ &$\phantom{1}2.766835 + 5.879184 i  $  & $         -0.467343 - 0.159566 i   $   \\
     & $1$ &$\phantom{1}3.750196 + 6.337011 i $  & $-1.193204 + 0.709887 i $\\
     & $3$ &$\phantom{1}8.556293 + 7.141111 i $  & $\phantom{-} 41.280909 + 78.754435 i $\\
     & $6$ &$16.044990 + 7.382406 i $  & $-1576.690323 + 5414.570984 i $ \\

$9$  & $0.1$ &$\phantom{1}1.378403 + 4.966897 i$  & $\phantom{-}0.060539 - 0.086669 i $   \\
     & $0.3$ &$\phantom{1}2.098692 + 5.864936 i$  & $          -0.071892 - 0.193894 i $   \\
     & $0.6$ &$\phantom{1}2.930753 + 6.534615 i $ & $          -0.476862 - 0.099550 i $   \\
     & $1$ &$\phantom{1}3.928352 + 7.054311 i $  & $-1.017928 + 0.937403 i $\\
     & $3$ &$\phantom{1}8.723992 + 8.015714 i $  & $\phantom{-} 89.972545 + 49.380751 i $\\
     & $6$ &$16.163453 + 8.333555 i $ & $\phantom{-} 5509.049968 + 7209.509940 i $ \\

\end{tabular}
\end{ruledtabular}
\end{table}
\endgroup

\vfill

\vfill

\vfill

\vfill

\vfill

\vfill

\begingroup
\squeezetable
\begin{table}[H]
\caption{\label{tab:table2} The lowest Regge poles $\lambda_{n}(\omega)$ for the electromagnetic field and the associated residues $r_{n}(\omega)$. We assume $2M=1$.}
\smallskip
\centering
\begin{ruledtabular}
\begin{tabular}{cccc}
 $n$ & $\omega$  & $\lambda_n(\omega)$ & $r_{n}(\omega)$
 \\ \hline
$1$  & $0.1$ &$\phantom{1}0.610235 + 0.1857157 i $  & $0.141533 + 0.042574 i $   \\
     & $0.3$ &$\phantom{1}1.022046 + 0.3598438 i $  & $0.261310 - 0.042212 i $   \\
     & $0.6$ &$\phantom{1}1.719903 + 0.4428220 i $  & $0.383610 - 0.048683 i $   \\
     & $1$ &$\phantom{1}2.705358 + 0.475454 i $  & $0.508191 + 0.067329 i $   \\
     & $3$ &$\phantom{1}7.832474 + 0.496948 i $  & $-0.159996 + 0.894283 i $   \\
     & $6$ &$15.607716 + 0.499228 i $  & $-0.578675 - 1.150971 i $   \\

$2$  & $0.1$ &$\phantom{1}0.525188 + 1.002194 i $  & $0.125937 - 0.097912 i $   \\
     & $0.3$ &$\phantom{1}1.041088 + 1.218701 i $  & $0.159661 - 0.245504 i $   \\
     & $0.6$ &$\phantom{1}1.778902 + 1.346966 i $  & $0.242164 - 0.499689 i $   \\
     & $1$ &$\phantom{1}2.769262 + 1.419171 i $  & $0.526456 - 0.877988 i $\\
     & $3$ &$\phantom{1}7.865654 + 1.487542 i$  & $4.568657 + 1.545802 i $   \\
     & $6$ &$15.62521 + 1.496753 i $  & $-12.494555 + 5.146258 i$ \\

$3$  & $0.1$ &$\phantom{1}0.662252 + 1.674293 i $  & $0.095812 - 0.093134 i $   \\
     & $0.3$ &$\phantom{1}1.190599 + 2.002822 i $  & $0.065892 - 0.251733 i $   \\
     & $0.6$ &$\phantom{1}1.919344 + 2.206835 i $  & $-0.064979- 0.560444 i $   \\
     & $1$ &$\phantom{1}2.892291 + 2.330093 i $  & $-0.325379 - 1.231546 i $\\
     & $3$ &$\phantom{1}7.929898 + 2.468721 i $  & $8.065346 - 10.612212 i $\\
     & $6$ &$15.65986 + 2.491532 i $  & $15.910509 + 69.844271 i $ \\

$4$  & $0.1$ &$\phantom{1}0.795653 + 2.274956 i $  & $\phantom{-}0.083712 - 0.091011 i $   \\
     & $0.3$ &$\phantom{1}1.353709 + 2.713393 i $  & $\phantom{-}0.018874 - 0.241698 i $   \\
     & $0.6$ &$\phantom{1}2.088487 + 3.004734 i $  & $-0.250327- 0.488373 i $   \\
     & $1$ &$\phantom{1}3.049989 + 3.194709 i $  & $-1.005893 - 0.931457 i $\\
     & $3$ &$\phantom{1}8.021481 + 3.435669 i $  & $-11.138536 - 23.703501 i $\\
     & $6$ &$15.711016 + 3.481878 i $ & $257.725106 + 8.206397 i $ \\

$5$  & $0.1$ &$\phantom{1}0.922279 + 2.841097 i $  & $\phantom{-}0.076365 - 0.089785 i $   \\
     & $0.3$ &$\phantom{1}1.513455 + 3.381624 i $  & $          -0.011075 - 0.231005 i $   \\
     & $0.6$ &$\phantom{1}2.263145 + 3.757064 i $  & $          -0.356562 - 0.396595 i $   \\
     & $1$ &$\phantom{1}3.22292 + 4.017259 i $ & $-1.333866 - 0.443471 i $\\
     & $3$ &$\phantom{1}8.135861 + 4.385262 i  $  & $ -41.163827 - 7.491406 i $\\
     & $6$ &$15.77777 + 4.466296 i $ & $260.251292 - 664.857185 i  $ \\

$6$  & $0.1$ &$\phantom{1}1.043127 + 3.385241 i $  & $\phantom{-}0.071135 - 0.088951 i $   \\
     & $0.3$ &$\phantom{1}1.667826 + 4.021677 i $  & $          -0.032532 - 0.221005 i $   \\
     & $0.6$ &$\phantom{1}2.436547 + 4.476543 i $  & $          -0.418506 - 0.308999 i $   \\
     & $1$ &$\phantom{1} 3.401342 + 4.805736 i $  & $-1.41496 + 0.025197 i $\\
     & $3$ &$\phantom{1}8.268455 + 5.315989 i  $  & $-46.408075 + 35.620183 i $\\
     & $6$ &$15.859025 + 5.443533 i $ & $-1142.290033 - 1165.297141 i $ \\

$7$  & $0.1$ &$\phantom{1}1.159231 + 3.913502 i $  & $\phantom{-}0.067092 - 0.088324 i $   \\
     & $0.3$ &$\phantom{1}1.817004 + 4.641067 i $  & $          -0.048946 - 0.211809 i $   \\
     & $0.6$ &$\phantom{1}2.606707 + 5.171198 i $  & $          -0.454210 - 0.230350 i $   \\
     & $1$ &$\phantom{1}3.580784 + 5.566912 i $ & $-1.351547 + 0.415443 i $\\
     & $3$ &$\phantom{1}8.41515 + 6.227581 i $ & $-13.342348 + 73.643709 i  $\\
     & $6$ &$15.953564 + 6.412592 i $  & $-3099.407958 + 881.950573 i $ \\

$8$  & $0.1$ &$\phantom{1}1.271395 + 4.429445 i $  & $\phantom{-}0.063805 - 0.087819 i $   \\
     & $0.3$ &$\phantom{1}1.961534 + 5.244345 i $  & $          -0.062033 - 0.203348 i $   \\
     & $0.6$ &$\phantom{1}2.773137 + 5.846253 i $  & $          -0.473449 - 0.160898 i $   \\
     & $1$ &$\phantom{1}3.759191 + 6.305865 i  $  & $-1.209644 + 0.720509 i  $\\
     & $3$ &$\phantom{1}8.572514 + 7.120576 i  $  & $41.311379 + 80.054461 i  $\\
     & $6$ &$16.060119 + 7.372728 i $  & $-1632.012756 + 5438.076537 i  $ \\

$9$  & $0.1$ &$\phantom{1}1.380226 + 4.935390 i $  & $\phantom{-}0.061041 - 0.087394 i $   \\
     & $0.3$ &$\phantom{1}2.101964 + 5.834520 i $  & $          -0.072774 - 0.195533 i $   \\
     & $0.6$ &$\phantom{1}2.935855 + 6.505296 i $  & $          -0.481896 - 0.099774 i$   \\
     & $1$ &$\phantom{1}3.935619 + 7.026361 i $  & $-1.028681 + 0.949498 i $\\
     & $3$ &$\phantom{1}8.737816 + 7.995955 i  $  & $90.843271 + 50.395101 i  $\\
     & $6$ &$16.177423 + 8.323434 i  $ & $5495.804514 + 7300.293176 i $ \\

\end{tabular}
\end{ruledtabular}
\end{table}
\endgroup

\begingroup
\squeezetable
\begin{table}[H]
\caption{\label{tab:table3} We compare, for the electromagnetic field at $\omega =1$ ($2M=1$), the exact value of the scattering amplitude (\ref{EM_Scattering_amp}) with the sum over Regge poles (\ref{CAM_EM_Scattering_amp_decomp_RP}) and we highlight the minor role of the background integral contributions (\ref{CAM_EM_Scattering_amp_decomp_Background_a}) and (\ref{CAM_EM_Scattering_amp_decomp_Background}).}
\centering
\begin{ruledtabular}
\begin{tabular}{lcc}
    Electromagnetic field at $\omega =1$ ($2M=1$) & $\theta = 15\degree$ &$\theta = 20\degree$ \tabularnewline
    \hline
    $|A (\omega, \theta)|^2$ & $ 1023.0681$ &$335.2717$  \tabularnewline
    $|A^\text{\tiny{RP}} (\omega, \theta)|^2$ (n=1,\dots, 60) & $1025.5432$ &$335.8305$ \tabularnewline
    $|A^\text{\tiny{RP}} (\omega, \theta)|^2$ (n=1,\dots, 90) & $1025.4676$ &$335.8343$  \tabularnewline
    $|A^\text{\tiny{B},\tiny{Re}} (\omega, \theta)+A^\text{\tiny{RP}} (\omega, \theta)|^2$ (n=1,\dots, 90)  & $1023.0754$ &$335.2743$\tabularnewline
    $|A^\text{\tiny{B}} (\omega, \theta)+A^\text{\tiny{RP}} (\omega, \theta)|^2$ (n=1,\dots, 90) & $1023.0762$ &$335.2718$\tabularnewline
\end{tabular}
\end{ruledtabular}
\end{table}
\endgroup

\section{BH glory and orbiting oscillations}
\label{SecIV}

In this section, we provide an analytical description of both the glory and orbiting cross sections based on the Regge pole sums (\ref{CAM_Scalar_Scattering_amp_decomp_RP}) and (\ref{CAM_EM_Scattering_amp_decomp_RP}). This is achieved by inserting in these sums analytical approximations for the lowest Regge poles and the associated residues. These approximations are obtained by assuming that, at large reduced frequencies $2M\omega$, the lowest Regge poles are in a one-to-one correspondence with ``surface waves'' propagating close to the unstable circular photon (graviton) orbit at $r=3M$, i.e., near the Schwarzschild photon sphere \cite{Decanini:2002ha,Decanini:2009mu,Dolan:2009nk,DFOEH2019}.

\subsection{Remarks concerning backward glory scattering and orbiting scattering}
\label{SecIVa}

It is well known that, for intermediate and high reduced frequencies, two interesting structures which are associated ``classically/semiclassically'' with backward glory scattering and orbiting scattering can be observed in the plots of BH cross sections (see, e.g., Ref.~\cite{Futterman:1988ni} or Chap.~4 of Ref.~\cite{Frolov:1998wf} and references therein). Roughly speaking, we can consider that glory scattering explains the behavior of the cross section in the backward direction, i.e., for scattering angles $\theta \approx \pi$ and that orbiting scattering permits us to understand the oscillations arising for ``intermediate'' scattering angles. A pioneering analysis, in the context of BH physics, of these two phenomena, can be found in an article by Handler and Matzner \cite{Handler:1980un}. It is based on ideas developed by Ford and Wheeler in their semiclassically study of quantum mechanical scattering \cite{Ford1959259}. A deeper analysis based on path integration has been carried out by DeWitt-Morette and co-workers \cite{Matzner:1985rjn,Anninos:1992ih} leading to semiclassical analytic formulas.

In Ref.~\cite{Matzner:1985rjn}, the authors have established that, for a Schwarzschild BH, the glory scattering cross section of massless waves of spin $s$ can be approximated by
\begin{equation}\label{Glory_scattering_formula}
  \left. \frac{d\sigma}{d\Omega}\right|_\mathrm{glory} = 30.752 M^3 \omega \, \left[J_{2s} \left(5.357 M\omega \sin \theta  \right)\right]^2,
\end{equation}
a formula which is formally valid for $2M\omega \gg 1$ and $|\theta -\pi| \ll 1$. Here, $J_{2s}$ is a Bessel function of the first kind. It should be noted that (\ref{Glory_scattering_formula}) encodes only the contribution of the first backward glory and that the numerical factors appearing in this approximation are those obtained in Ref.~\cite{Crispino:2009ki}. It is moreover interesting to recall that, in Ref.~\cite{Anninos:1992ih}, the authors have proposed a formula which roughly describes the orbiting oscillations of the scalar cross section. In our opinion, we cannot be satisfied with that formula. Indeed, it involves two free parameters which must be adjusted and, even if this job is done effectively, it does not permit us to reproduce with good agreement the exact result (see Fig.~9 of Ref.~\cite{Anninos:1992ih}).

\subsection{Analytical Regge pole approximations of scattering amplitudes}
\label{SecIVb}

As we shall see, it is possible from Regge pole sums to describe analytically both phenomena, i.e., to establish a unique formula which fits the glory and orbiting cross sections. At first sight, this is rather surprising because, in the geometrical-optics limit (i.e., if we only focus on the concepts of geodesic and bundle of
geometrical rays), the two phenomena are very different in that they are associated with very different mathematical behavior of the deflection function \cite{Ford1959259,Handler:1980un,Matzner:1985rjn,Anninos:1992ih}. But, here, we must not be too obsessed by classical results. Indeed, we are working in the framework of wave physics and it is tempting to consider that the glory and orbiting effects are not fundamentally different insofar as they are both generated by the excitation of surface waves propagating close to the BH photon sphere. By using the Regge pole approach of BH physics, we can take into account these diffractive effects due to the Schwarzschild photon sphere \cite{Andersson:1994rm,Decanini:2002ha,Decanini:2009mu,Dolan:2009nk,Decanini:2010fz} and derive asymptotic expansions for the Regge poles and the associated residues.

Analytical expressions for the lowest Regge
poles of the Schwarzschild BH  have been obtained in Ref.~\cite{Decanini:2009mu}. This has been achieved by using
a third-order WKB approximation to solve the Regge-Wheeler equation (\ref{H_RW_equation})-(\ref{Potentiel}) or, more precisely, by extending to Regge poles the approach developed in the context of the determination
of the QNMs by Schutz and Will \cite{Schutz:1985zz} and by Iyer and Will \cite{Iyer:1986vv,Iyer:1986nq}. It should be noted that it is possible to go beyond the third-order WKB approximation (see Refs.~\cite{Dolan:2009nk,Decanini:2011eh}) but here we will not need these improved results. In fact, we shall only consider that, at large $2M\omega$, the lowest Regge poles closely adhere to the asymptotic form
\begin{equation}\label{PRapproxWKB}
\lambda_n(\omega) \approx 3\sqrt{3} \, M\omega + i \,(n-1/2) + \frac{\sqrt{3} \,
a_n}{18 M\omega}
\end{equation}
[here we have neglected  the terms of order $1/(2M\omega)^2$ derived in Ref.~\cite{Decanini:2009mu}] where
\begin{equation}\label{PRapproxWKB_denote1}
a_n =\frac{2}{3}\left[\frac{5}{12}(n-1/2)^2 + \frac{115}{144} -1
+s^2 \right].
\end{equation}

It is moreover possible to obtain an analytical expression for the residues associated with the lowest Regge
poles of the Schwarzschild BH. By extending to Regge poles the calculations which have permitted to Dolan and Ottewill to derive an analytical expression for the QNM excitation factors \cite{Dolan:2011fh}, we have obtained \cite{DFOEH2019}
\begin{eqnarray}\label{Residues_approx}
&& r_n(\omega) \approx  \frac{\left[-i \, 216 (3\sqrt{3} M\omega) / \xi \right]^{n-1/2}}{\sqrt{2\pi} \, (n-1)!} e^{i 2M\omega y} e^{i\pi \lambda_n(\omega)} \nonumber \\
& &
\end{eqnarray}
where
\begin{equation}\label{Residues_approx_xi_y}
\xi=7+4\sqrt{3} \quad \mathrm{and} \quad y=3-3\sqrt{3} + 4 \ln 2 - 3 \ln \xi.
\end{equation}

\begin{figure}[htb]
\centering
 \includegraphics[scale=0.50]{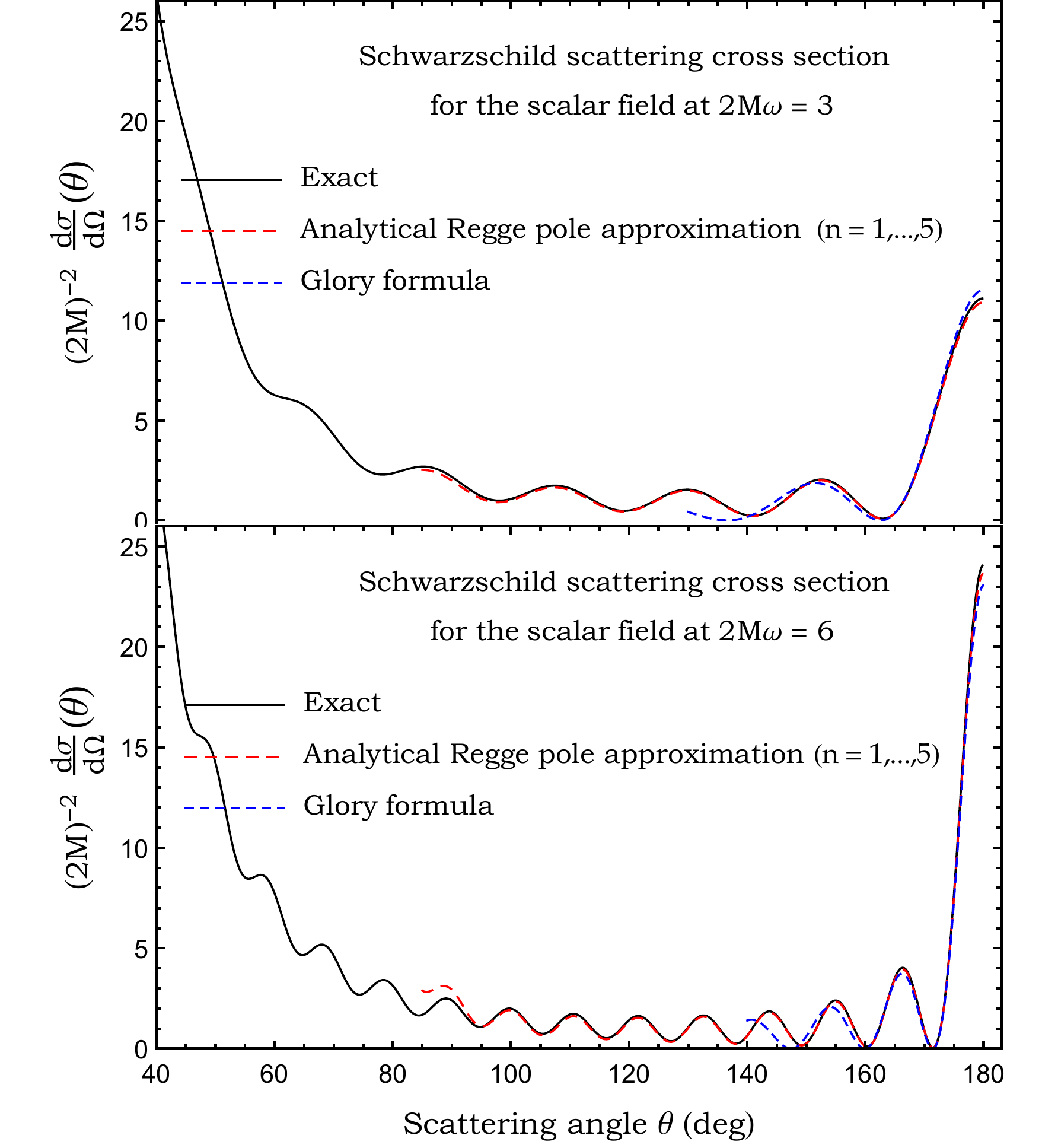}
\caption{\label{RP_approx_2Mw_3_6_s_0} The scalar cross section of a Schwarzschild BH for $2M\omega=3$ and $2M\omega=6$. We compare the exact result with the glory formula and with that obtained from the analytical Regge pole approximation.}
\end{figure}

 \begin{figure}[htb]
\centering
 \includegraphics[scale=0.50]{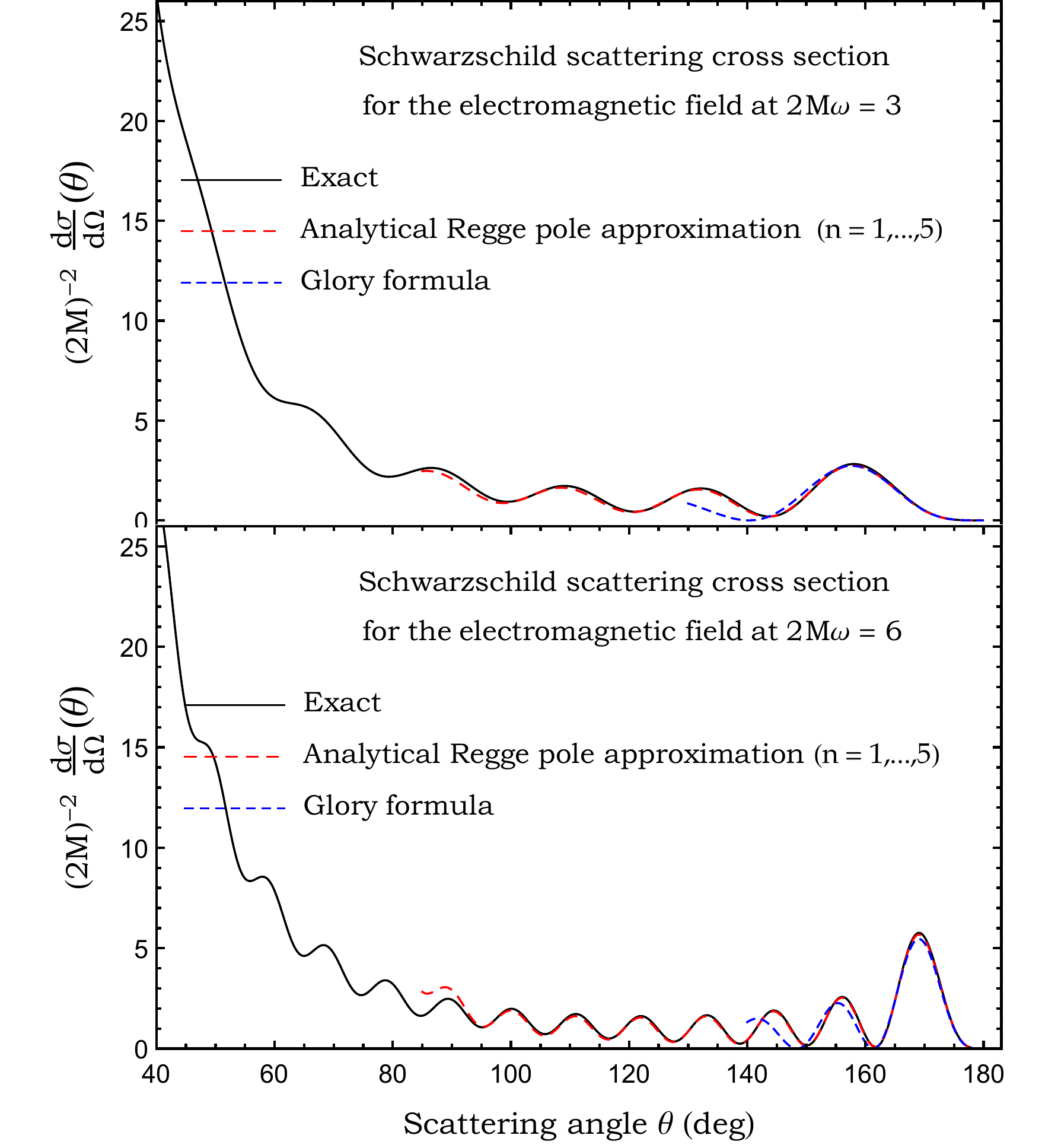}
\caption{\label{RP_approx_2Mw_3_6_s_1} The electromagnetic cross section of a Schwarzschild BH for $2M\omega=3$ and $2M\omega=6$. We compare the exact result with the glory formula and with that obtained from the analytical Regge pole approximation.}
\end{figure}
By inserting now the approximations (\ref{PRapproxWKB})-(\ref{PRapproxWKB_denote1}) for the Regge poles and (\ref{Residues_approx})-(\ref{Residues_approx_xi_y}) for the residues into the Regge pole sum (\ref{CAM_Scalar_Scattering_amp_decomp_RP}) for the scalar field and into the Regge pole sum (\ref{CAM_EM_Scattering_amp_decomp_RP}) for the electromagnetic field, we have at our disposal analytical approximations for the scattering cross sections associated with these two fields which are formally valid for $2M\omega \to +\infty$. It should be however noted that the Regge pole sums considered can only involve a small number of terms because the approximations (\ref{PRapproxWKB}) and (\ref{Residues_approx}) are accurate only for the lowest Regge poles. As a consequence, from a theoretical point of view, our analytical Regge pole approximations cannot describe the cross sections for small scattering angles.

\subsection{Results and comments}
\label{SecIVc}

\begin{figure}
\centering
 \includegraphics[scale=0.50]{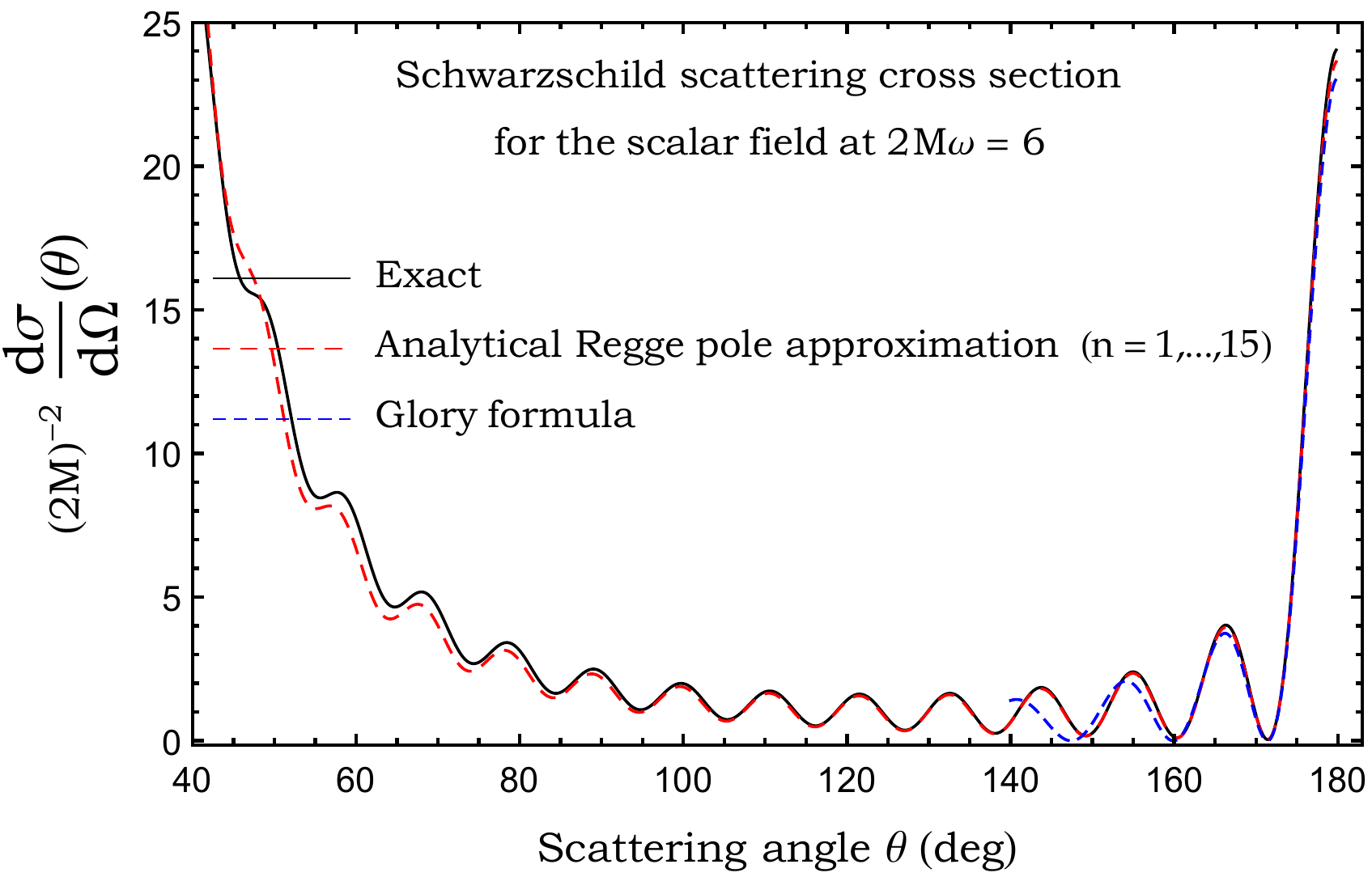}
\caption{\label{RP_approx_2Mw_6_s_0_n15} The scalar cross section of a Schwarzschild BH for $2M\omega=6$. We compare the exact result with the glory formula and with that obtained from the analytical Regge pole approximation where we sum over a large number of terms despite the inaccuracy of the approximations (\ref{PRapproxWKB}) and (\ref{Residues_approx}) for the higher Regge poles.}
\end{figure}

In Figs.~\ref{RP_approx_2Mw_3_6_s_0} and \ref{RP_approx_2Mw_3_6_s_1}, we complete our study of Sec.~\ref{SecIIIb} by now comparing the exact scattering cross sections with their analytical approximations constructed in Sec.~\ref{SecIVb}. The comparisons are achieved for the reduced frequencies $2M\omega=3$ and 6 and the summations are over the first five Regge poles. We can observe that the analytical Regge pole approximations permit us to reproduce with very good agreement both the glory cross section and a large part of the orbiting cross section. It is moreover surprising to note that, by considering analytical Regge pole sums involving a large number of terms, we are able to describe the cross sections in a wide range of scattering angles despite the inaccuracy of the approximations (\ref{PRapproxWKB}) and (\ref{Residues_approx}) for the higher Regge poles (see Fig.~\ref{RP_approx_2Mw_6_s_0_n15} where we have considered the case of the scalar field at $2M\omega=6$).

\section{Conclusion}
\label{Conc}

In this article, we have considered the scattering of scalar and electromagnetic waves by a Schwarzschild BH by focusing on the associated differential scattering cross sections and we have shown that, for intermediate and high reduced frequencies, these cross sections can be reconstructed in terms of Regge poles with great precision. This is really surprising and certainly due to the fact that BHs are very particular physical objects. Indeed, in quantum mechanics \cite{deAlfaro:1965zz,Newton:1982qc}, electromagnetism and optics \cite{Watson18,Sommerfeld49,Newton:1982qc,Nus92,Grandy2000} and acoustics \cite{Uberall1992}, background integral contributions are never negligible (this remains true regardless of the frequency) and, as a consequence, a Regge pole sum alone does not permit us to reproduce a differential scattering cross section. In the context of scattering of scalar and electromagnetic waves by a Schwarzschild BH, we have observed that it is necessary to take into account background integral contributions only for low reduced frequencies.

In the short-wavelength regime, from Regge pole sums we have also been able to describe numerically, with an impressive agreement, the BH glory occurring in the backward direction as well as the orbiting oscillations appearing on the differential scattering cross sections for small and intermediate scattering angles. Moreover, it is important to note that working with Regge pole sums has permitted us to overcome the difficulties linked to the lack of convergence of the partial wave expansions defining the cross sections which are due to the long-range nature of the fields propagating on a Schwarzschild background.

Finally, taking into account the fact that, in the short-wavelength regime, a Regge pole sum encodes the physical information hidden into the partial wave expansion defining a differential scattering cross section, we have derived an analytical approximation fitting both the BH glory and a large part of the orbiting oscillations. This has been achieved by inserting into the Regge pole sum asymptotic approximations for the lowest Regge poles and the associated residues. In our opinion, our analytical approximation is far superior to existing formulas \cite{Matzner:1985rjn,Anninos:1992ih}.

We hope in next works to extend our study to scattering of gravitational waves by a Schwarzschild BH as well as to scattering of waves by a Kerr BH (see Refs.~\cite{Matzner:1977dn,MatznerRyan1978,Handler:1980un,Glampedakis:2001cx,Dolan:2007ut,Dolan:2008kf} for articles concerning these two topics and which could serve as departure points for such works).

\begin{acknowledgments}

M.O.E.H. wishes to thank Sam Dolan for various discussions and for his kind invitation to the University of Sheffield where this work was completed. The authors are grateful to Yves Decanini for discussions concerning the excitation of Regge modes.

\end{acknowledgments}

\appendix*

\section{Iterative method to accelerate the convergence of background integrals}
\label{appen}

Due to the long-range nature of the fields propagating on a Schwarzschild BH, the background integrals (\ref{CAM_Scalar_Scattering_amp_decomp_Background_a}) and (\ref{CAM_EM_Scattering_amp_decomp_Background_a}) suffer a lack of convergence. In this Appendix we explain how to overcome this problem, i.e., how to accelerate their convergence. In fact, the same problem occurs for the partial wave expansions (\ref{Scalar_Scattering_amp}), (\ref{EM_Scattering_amp2}) and  (\ref{EM_Scattering_amp}) (see, e.g., Refs.~\cite{Anninos:1992ih} where the case of the scalar field is discussed). For these partial wave expansions, the convergence can be obtained by employing an iterative method introduced a long time ago in the context of Coulomb scattering by Yennie, Ravenhall and Wilson \cite{Yennie:1954zz} [see also Refs.~\cite{Dolan:2008kf,Crispino:2009xt} where this method is used in the context of scattering by BHs and see, e.g., Ref.~\cite{Anninos:1992ih} for another approach based on the truncation of partial wave expansions and on the matching of the Schwarzschild $S$-matrix elements $S_\ell (\omega)$ with the Newtonian ones]. In this Appendix, we briefly recall this iterative method because we use it in Sec.~\ref{SecIII} in order to accelerate the convergence of the partial wave expansions (\ref{Scalar_Scattering_amp}) and (\ref{EM_Scattering_amp2}) and we generalize it in order to deal with the background integrals (\ref{CAM_Scalar_Scattering_amp_decomp_Background_a}) and (\ref{CAM_EM_Scattering_amp_decomp_Background_a}).

\subsection{Acceleration of the convergence of partial wave expansions}

In order to improve the convergence of the partial wave expansion
\begin{equation}\label{PWE_div}
\alpha (\theta)= \sum_{\ell = 0}^{\infty} a_\ell \, P_{\ell}(\cos\theta)
\end{equation}
[see Eqs.~(\ref{Scalar_Scattering_amp}) and (\ref{EM_Scattering_amp2})] we introduce the associated ``reduced'' series \cite{Yennie:1954zz}
\begin{equation}\label{PWE_div_tilde}
{\widetilde \alpha}^{(n)} (\theta)= \sum_{\ell = 0}^{\infty} {\widetilde a}^{(n)}_\ell \, P_{\ell}(\cos\theta)
\end{equation}
defined by
\begin{equation}\label{PWE_div_conv}
\alpha (\theta)=  (1-\cos \theta)^{-n} \, {\widetilde \alpha} (\theta).
\end{equation}
It should be noted that, by reexpressing the series (\ref{PWE_div}) in the form (\ref{PWE_div_conv}), we isolate the pathological behavior of the partial wave expansions (\ref{Scalar_Scattering_amp}) and (\ref{EM_Scattering_amp2}) occurring for $\theta \to 0$. By using now the recursion relation \cite{AS65}
\begin{eqnarray}\label{PolLegendre_RR}
& & (\ell +1) P_{\ell+1} (\cos \theta) -(2\ell+1) \cos \theta P_\ell (\cos \theta) \nonumber \\
& & \qquad \qquad+ \ell P_{\ell-1} (\cos \theta)=0
\end{eqnarray}
in the form
\begin{eqnarray}\label{PolLegendre_RR_bis}
& & (1-\cos \theta) P_\ell (\cos \theta)= P_\ell (\cos \theta) \nonumber \\
& & \quad - \frac{\ell+1}{2\ell+1}P_{\ell+1} (\cos \theta)- \frac{\ell}{2\ell+1}P_{\ell-1} (\cos \theta)
\end{eqnarray}
we can show from (\ref{PWE_div})--(\ref{PWE_div_conv}) that the coefficients ${\widetilde a}^{(n)}_\ell$ can be expressed in terms of the coefficients ${\widetilde a}^{(n-1)}_\ell$. We have, for $\ell \in \mathbb{N}$,
\begin{equation}\label{RR_an_anmun}
{\widetilde a}^{(n)}_\ell = {\widetilde a}^{(n-1)}_\ell -\frac{\ell +1}{2\ell+3} {\widetilde a}^{(n-1)}_{\ell+1} -\frac{\ell}{2\ell-1} {\widetilde a}^{(n-1)}_{\ell-1}
\end{equation}
(here we take ${\widetilde a}^{(n-1)}_{-1} = 0$). As noted in Ref.~\cite{Yennie:1954zz}, we have for large values of $\ell$,
\begin{equation}\label{CONV_series}
{\widetilde a}^{(n)}_\ell = {\cal O} \left( {\widetilde a}^{(n-1)}_\ell /\ell^2 \right).
\end{equation}
This explicitly shows that using the reduced series (\ref{PWE_div_tilde}) greatly improves the convergence of the initial partial wave expansion (\ref{PWE_div}).

\subsection{Acceleration of the convergence of background integrals}

In order to improve the convergence of the background integral
\begin{equation}\label{BackGroundInt_div}
\alpha (\theta)= \int_0^{+\infty} d\lambda  \, a(\lambda)  \, Q_{\lambda -1/2}(\cos\theta +i0)
\end{equation}
[see Eqs.~(\ref{CAM_Scalar_Scattering_amp_decomp_Background_a}) and (\ref{CAM_EM_Scattering_amp_decomp_Background_a})], we first split it in the form
\begin{equation}\label{BackGroundInt_div_bis}
\alpha (\theta)= \int_0^{\lambda_0} d\lambda \,  a(\lambda)  \, Q_{\lambda -1/2}(\cos\theta +i0) + \alpha_{\lambda_0} (\theta)
\end{equation}
where
\begin{equation}\label{BackGroundInt_div_inf}
\alpha_{\lambda_0} (\theta) = \int_{\lambda_0}^{+\infty} d\lambda  \, a(\lambda)  \, Q_{\lambda -1/2}(\cos\theta +i0).
\end{equation}
The choice of the truncation parameter ${\lambda_0}$ will be discussed later. We then introduce the ``reduced'' integrals ${\widetilde \alpha}_{\lambda_0}^{(n)} (\theta)$ defined by
\begin{equation}\label{BackGroundInt_div_conv_prov}
\alpha_{\lambda_0} (\theta)=  (1-\cos \theta)^{-n} \, {\widetilde \alpha}_{\lambda_0}^{(n)} (\theta).
\end{equation}
By using the relation
\begin{eqnarray}\label{FuncLegendre_RR_bis}
& & (1-\cos \theta) Q_{\lambda-1/2} (\cos \theta+i0)= Q_{\lambda-1/2} (\cos \theta+i0) \nonumber \\
& & \qquad - \left(\frac{\lambda+1/2}{2\lambda} \right) Q_{\lambda+1/2} (\cos \theta+i0) \nonumber \\
& & \qquad- \left(\frac{\lambda-1/2}{2\lambda}\right) Q_{\lambda-3/2} (\cos \theta+i0)
\end{eqnarray}
which is a consequence of the definition (\ref{Legendre_function_second_kind}) and of the relation \cite{AS65}
\begin{eqnarray}\label{FuncLegendre_RR}
& & (\nu +1) P_{\nu+1} (\cos \theta) -(2\nu+1) \cos \theta P_\nu (\cos \theta) \nonumber \\
& & \qquad \qquad + \nu P_{\nu-1} (\cos \theta)=0,
\end{eqnarray}
we can show that these reduced integrals can be written in the form
\begin{equation}\label{BackGroundInt_div_conv}
{\widetilde \alpha}_{\lambda_0}^{(n)} (\theta) =  \int_{\lambda_0}^{+\infty} d\lambda  \, {\widetilde a}^{(n)}(\lambda)  \, Q_{\lambda -1/2}(\cos\theta +i0) + {\widetilde R}^{(n)}(\theta)
\end{equation}
where ${\widetilde R}^{(n)}(\theta)$ is an integral over a finite integration domain which can be expressed in terms of the integral ${\widetilde R}^{(n-1)}(\theta)$ and where the function ${\widetilde a}^{(n)}(\lambda)$ can be expressed in terms of the function ${\widetilde a}^{(n-1)}(\lambda)$. We have
\begin{eqnarray}\label{RR_an_anmun_lambda}
& & {\widetilde a}^{(n)} (\lambda) = {\widetilde a}^{(n-1)} (\lambda)   - \left(\frac{\lambda+1/2}{2(\lambda+1)} \right) {\widetilde a}^{(n-1)} (\lambda+1) \nonumber \\
& & \qquad\qquad -\left(\frac{\lambda-1/2}{2(\lambda-1)}\right) {\widetilde a}^{(n-1)} (\lambda-1)
\end{eqnarray}
and
\begin{eqnarray}\label{RR_Rn_anmun_lambda}
& & {\widetilde R}^{(n)}(\theta) = (1-\cos \theta) {\widetilde R}^{(n-1)}(\theta) \nonumber \\
& & \quad + \int_{\lambda_0-1}^{\lambda_0} d\lambda \left[ \left(\frac{\lambda+1/2}{2\lambda}\right)  {\widetilde a}^{(n-1)}(\lambda)  \, Q_{\lambda +1/2}(\cos\theta +i0) \right. \nonumber \\
& & \qquad  \left.
- \left(\frac{\lambda+1/2}{2(\lambda+1)}\right)  {\widetilde a}^{(n-1)}(\lambda+1)  \, Q_{\lambda -1/2}(\cos\theta +i0)\right]. \nonumber \\
& &
\end{eqnarray}
From the relation (\ref{RR_an_anmun_lambda}) we can see that, for large values of $\lambda$,
\begin{equation}\label{CONV_Integ}
{\widetilde a}^{(n)} (\lambda) = {\cal O} \left( {\widetilde a}^{(n-1)} (\lambda) /\lambda^2 \right).
\end{equation}
This explicitly shows that using the reduced integral (\ref{BackGroundInt_div_conv}) greatly improves the convergence of the initial background integral (\ref{BackGroundInt_div}). As far as the choice of the truncation parameter ${\lambda_0}$ is concerned, it should be noted that it depends on the number $n$ of iterations performed. Indeed, due to the shift of the variable $\lambda$ induced by the relation (\ref{RR_an_anmun_lambda}), we can observe that ${\widetilde a}^{(n)} (\lambda)$ is not defined for $\lambda \in [0,n[$ and, in order to have the integrals (\ref{RR_Rn_anmun_lambda}) and (\ref{BackGroundInt_div_conv}) well defined, it is necessary to take $\lambda_0 \geq n$.

\bibliography{BH_scattering_and_RP}

\begin{thebibliography}{60}%
\makeatletter
\providecommand \@ifxundefined [1]{%
 \@ifx{#1\undefined}
}%
\providecommand \@ifnum [1]{%
 \ifnum #1\expandafter \@firstoftwo
 \else \expandafter \@secondoftwo
 \fi
}%
\providecommand \@ifx [1]{%
 \ifx #1\expandafter \@firstoftwo
 \else \expandafter \@secondoftwo
 \fi
}%
\providecommand \natexlab [1]{#1}%
\providecommand \enquote  [1]{``#1''}%
\providecommand \bibnamefont  [1]{#1}%
\providecommand \bibfnamefont [1]{#1}%
\providecommand \citenamefont [1]{#1}%
\providecommand \href@noop [0]{\@secondoftwo}%
\providecommand \href [0]{\begingroup \@sanitize@url \@href}%
\providecommand \@href[1]{\@@startlink{#1}\@@href}%
\providecommand \@@href[1]{\endgroup#1\@@endlink}%
\providecommand \@sanitize@url [0]{\catcode `\\12\catcode `\$12\catcode
  `\&12\catcode `\#12\catcode `\^12\catcode `\_12\catcode `\%12\relax}%
\providecommand \@@startlink[1]{}%
\providecommand \@@endlink[0]{}%
\providecommand \url  [0]{\begingroup\@sanitize@url \@url }%
\providecommand \@url [1]{\endgroup\@href {#1}{\urlprefix }}%
\providecommand \urlprefix  [0]{URL }%
\providecommand \Eprint [0]{\href }%
\providecommand \doibase [0]{http://dx.doi.org/}%
\providecommand \selectlanguage [0]{\@gobble}%
\providecommand \bibinfo  [0]{\@secondoftwo}%
\providecommand \bibfield  [0]{\@secondoftwo}%
\providecommand \translation [1]{[#1]}%
\providecommand \BibitemOpen [0]{}%
\providecommand \bibitemStop [0]{}%
\providecommand \bibitemNoStop [0]{.\EOS\space}%
\providecommand \EOS [0]{\spacefactor3000\relax}%
\providecommand \BibitemShut  [1]{\csname bibitem#1\endcsname}%
\let\auto@bib@innerbib\@empty
\bibitem [{\citenamefont {Futterman}\ \emph {et~al.}(2012)\citenamefont
  {Futterman}, \citenamefont {Handler},\ and\ \citenamefont
  {Matzner}}]{Futterman:1988ni}%
  \BibitemOpen
  \bibfield  {author} {\bibinfo {author} {\bibfnamefont {J.~A.~H.}\
  \bibnamefont {Futterman}}, \bibinfo {author} {\bibfnamefont {F.~A.}\
  \bibnamefont {Handler}}, \ and\ \bibinfo {author} {\bibfnamefont {R.~A.}\
  \bibnamefont {Matzner}},\ }\href {\doibase 10.1017/CBO9780511735615} {\emph
  {\bibinfo {title} {{Scattering from Black Holes}}}},\ Cambridge Monographs on
  Mathematical Physics\ (\bibinfo  {publisher} {Cambridge University Press},\
  \bibinfo {year} {2012})\BibitemShut {NoStop}%
\bibitem [{\citenamefont {Frolov}\ and\ \citenamefont
  {Novikov}(1998)}]{Frolov:1998wf}%
  \BibitemOpen
  \bibfield  {author} {\bibinfo {author} {\bibfnamefont {V.~P.}\ \bibnamefont
  {Frolov}}\ and\ \bibinfo {author} {\bibfnamefont {I.~D.}\ \bibnamefont
  {Novikov}},\ }\href {\doibase 10.1007/978-94-011-5139-9} {\emph {\bibinfo
  {title} {{Black Hole Physics: Basic Concepts and New Developments}}}},\
  \bibinfo {series} {Fundamental Theories of Physics}, Vol.~\bibinfo {volume}
  {96}\ (\bibinfo  {publisher} {Kluwer Academic Publishers},\ \bibinfo {year}
  {1998})\BibitemShut {NoStop}%
\bibitem [{\citenamefont {de~Alfaro}\ and\ \citenamefont
  {Regge}(1965)}]{deAlfaro:1965zz}%
  \BibitemOpen
  \bibfield  {author} {\bibinfo {author} {\bibfnamefont {V.}~\bibnamefont
  {de~Alfaro}}\ and\ \bibinfo {author} {\bibfnamefont {T.}~\bibnamefont
  {Regge}},\ }\href@noop {} {\emph {\bibinfo {title} {{Potential
  Scattering}}}}\ (\bibinfo  {publisher} {North-Holland Publishing Company,
  Amsterdam},\ \bibinfo {year} {1965})\BibitemShut {NoStop}%
\bibitem [{\citenamefont {Newton}(1982)}]{Newton:1982qc}%
  \BibitemOpen
  \bibfield  {author} {\bibinfo {author} {\bibfnamefont {R.~G.}\ \bibnamefont
  {Newton}},\ }\href@noop {} {\emph {\bibinfo {title} {{Scattering Theory of
  Waves and Particles}}}},\ \bibinfo {edition} {2nd}\ ed.\ (\bibinfo
  {publisher} {Springer-Verlag, New York},\ \bibinfo {year} {1982})\BibitemShut
  {NoStop}%
\bibitem [{\citenamefont {Watson}(1918)}]{Watson18}%
  \BibitemOpen
  \bibfield  {author} {\bibinfo {author} {\bibfnamefont {G.~N.}\ \bibnamefont
  {Watson}},\ }\bibfield  {title} {\enquote {\bibinfo {title} {{The diffraction
  of electric waves by the Earth}},}\ }\href@noop {} {\bibfield  {journal}
  {\bibinfo  {journal} {Proc.\ R.\ Soc.\ A}\ }\textbf {\bibinfo {volume}
  {95}},\ \bibinfo {pages} {83} (\bibinfo {year} {1918})}\BibitemShut {NoStop}%
\bibitem [{\citenamefont {Sommerfeld}(1949)}]{Sommerfeld49}%
  \BibitemOpen
  \bibfield  {author} {\bibinfo {author} {\bibfnamefont {A.}~\bibnamefont
  {Sommerfeld}},\ }\href@noop {} {\emph {\bibinfo {title} {Partial Differential
  Equations of Physics}}}\ (\bibinfo  {publisher} {Academic Press, New York},\
  \bibinfo {year} {1949})\BibitemShut {NoStop}%
\bibitem [{\citenamefont {Nussenzveig}(1992)}]{Nus92}%
  \BibitemOpen
  \bibfield  {author} {\bibinfo {author} {\bibfnamefont {H.~M.}\ \bibnamefont
  {Nussenzveig}},\ }\href@noop {} {\emph {\bibinfo {title} {Diffraction Effects
  in Semiclassical Scattering}}}\ (\bibinfo  {publisher} {Cambridge University
  Press, Cambridge},\ \bibinfo {year} {1992})\BibitemShut {NoStop}%
\bibitem [{\citenamefont {Grandy}(2000)}]{Grandy2000}%
  \BibitemOpen
  \bibfield  {author} {\bibinfo {author} {\bibfnamefont {W.~T.}\ \bibnamefont
  {Grandy}},\ }\href@noop {} {\emph {\bibinfo {title} {Scattering of Waves from
  Large Spheres}}}\ (\bibinfo  {publisher} {Cambridge University Press,
  Cambridge},\ \bibinfo {year} {2000})\BibitemShut {NoStop}%
\bibitem [{\citenamefont {{\"U}berall}(1992)}]{Uberall1992}%
  \BibitemOpen
  \bibfield  {author} {\bibinfo {author} {\bibfnamefont {H.}~\bibnamefont
  {{\"U}berall}},\ }\href@noop {} {\emph {\bibinfo {title} {Acoustic Resonance
  Scattering}}}\ (\bibinfo  {publisher} {Gordon and Breach, New York},\
  \bibinfo {year} {1992})\BibitemShut {NoStop}%
\bibitem [{\citenamefont {Aki}\ and\ \citenamefont
  {Richards}(2002)}]{AkiRichards2002}%
  \BibitemOpen
  \bibfield  {author} {\bibinfo {author} {\bibfnamefont {K.}~\bibnamefont
  {Aki}}\ and\ \bibinfo {author} {\bibfnamefont {P.}~\bibnamefont {Richards}},\
  }\href@noop {} {\emph {\bibinfo {title} {Quantitative Seismology}}},\
  \bibinfo {edition} {2nd}\ ed.\ (\bibinfo  {publisher} {University Science
  Book, Sausalito},\ \bibinfo {year} {2002})\BibitemShut {NoStop}%
\bibitem [{\citenamefont {Gribov}(2003)}]{Gribov69}%
  \BibitemOpen
  \bibfield  {author} {\bibinfo {author} {\bibfnamefont {V.~N.}\ \bibnamefont
  {Gribov}},\ }\href@noop {} {\emph {\bibinfo {title} {The Theory of Complex
  Angular Momenta: Gribov Lectures on Theoretical Physics}}}\ (\bibinfo
  {publisher} {Cambridge University Press, Cambridge},\ \bibinfo {year}
  {2003})\BibitemShut {NoStop}%
\bibitem [{\citenamefont {Collins}(1977)}]{Collins77}%
  \BibitemOpen
  \bibfield  {author} {\bibinfo {author} {\bibfnamefont {P.~D.~B.}\
  \bibnamefont {Collins}},\ }\href@noop {} {\emph {\bibinfo {title} {An
  Introduction to Regge Theory and High-Energy Physics}}}\ (\bibinfo
  {publisher} {Cambridge University Press, Cambridge},\ \bibinfo {year}
  {1977})\BibitemShut {NoStop}%
\bibitem [{\citenamefont {Barone}\ and\ \citenamefont
  {Predazzi}(2002)}]{BaronePredazzi2002}%
  \BibitemOpen
  \bibfield  {author} {\bibinfo {author} {\bibfnamefont {V.}~\bibnamefont
  {Barone}}\ and\ \bibinfo {author} {\bibfnamefont {E.}~\bibnamefont
  {Predazzi}},\ }\href@noop {} {\emph {\bibinfo {title} {High-Energy Particle
  Diffraction}}}\ (\bibinfo  {publisher} {Springer-Verlag, Berlin},\ \bibinfo
  {year} {2002})\BibitemShut {NoStop}%
\bibitem [{\citenamefont {Donnachie}\ \emph {et~al.}(2005)\citenamefont
  {Donnachie}, \citenamefont {Dosch}, \citenamefont {Landshoff},\ and\
  \citenamefont {Nachtmann}}]{DonnachieETAL2005}%
  \BibitemOpen
  \bibfield  {author} {\bibinfo {author} {\bibfnamefont {S.}~\bibnamefont
  {Donnachie}}, \bibinfo {author} {\bibfnamefont {G.}~\bibnamefont {Dosch}},
  \bibinfo {author} {\bibfnamefont {P.~V.}\ \bibnamefont {Landshoff}}, \ and\
  \bibinfo {author} {\bibfnamefont {O.}~\bibnamefont {Nachtmann}},\ }\href@noop
  {} {\emph {\bibinfo {title} {Pomeron Physics and QCD}}}\ (\bibinfo
  {publisher} {Cambridge University Press, Cambridge},\ \bibinfo {year}
  {2005})\BibitemShut {NoStop}%
\bibitem [{\citenamefont {Andersson}\ and\ \citenamefont
  {Thylwe}(1994)}]{Andersson:1994rk}%
  \BibitemOpen
  \bibfield  {author} {\bibinfo {author} {\bibfnamefont {N.}~\bibnamefont
  {Andersson}}\ and\ \bibinfo {author} {\bibfnamefont {K.~E.}\ \bibnamefont
  {Thylwe}},\ }\bibfield  {title} {\enquote {\bibinfo {title} {{Complex angular
  momentum approach to black hole scattering}},}\ }\href {\doibase
  10.1088/0264-9381/11/12/013} {\bibfield  {journal} {\bibinfo  {journal}
  {Class.\ Quant.\ Grav.}\ }\textbf {\bibinfo {volume} {11}},\ \bibinfo {pages}
  {2991} (\bibinfo {year} {1994})}\BibitemShut {NoStop}%
\bibitem [{\citenamefont {Andersson}(1994)}]{Andersson:1994rm}%
  \BibitemOpen
  \bibfield  {author} {\bibinfo {author} {\bibfnamefont {N.}~\bibnamefont
  {Andersson}},\ }\bibfield  {title} {\enquote {\bibinfo {title} {{Complex
  angular momenta and the black hole glory}},}\ }\href {\doibase
  10.1088/0264-9381/11/12/014} {\bibfield  {journal} {\bibinfo  {journal}
  {Class.\ Quant.\ Grav.}\ }\textbf {\bibinfo {volume} {11}},\ \bibinfo {pages}
  {3003} (\bibinfo {year} {1994})}\BibitemShut {NoStop}%
\bibitem [{\citenamefont {Decanini}\ \emph {et~al.}(2003)\citenamefont
  {Decanini}, \citenamefont {Folacci},\ and\ \citenamefont
  {Jensen}}]{Decanini:2002ha}%
  \BibitemOpen
  \bibfield  {author} {\bibinfo {author} {\bibfnamefont {Y.}~\bibnamefont
  {Decanini}}, \bibinfo {author} {\bibfnamefont {A.}~\bibnamefont {Folacci}}, \
  and\ \bibinfo {author} {\bibfnamefont {B.}~\bibnamefont {Jensen}},\
  }\bibfield  {title} {\enquote {\bibinfo {title} {{Complex angular momentum in
  black hole physics and the quasinormal modes}},}\ }\href {\doibase
  10.1103/PhysRevD.67.124017} {\bibfield  {journal} {\bibinfo  {journal}
  {Phys.\ Rev.\ D}\ }\textbf {\bibinfo {volume} {67}},\ \bibinfo {pages}
  {124017} (\bibinfo {year} {2003})},\ \Eprint
  {http://arxiv.org/abs/gr-qc/0212093} {arXiv:gr-qc/0212093} \BibitemShut
  {NoStop}%
\bibitem [{\citenamefont {Glampedakis}\ and\ \citenamefont
  {Andersson}(2003)}]{Glampedakis:2003dn}%
  \BibitemOpen
  \bibfield  {author} {\bibinfo {author} {\bibfnamefont {K.}~\bibnamefont
  {Glampedakis}}\ and\ \bibinfo {author} {\bibfnamefont {N.}~\bibnamefont
  {Andersson}},\ }\bibfield  {title} {\enquote {\bibinfo {title} {{Quick and
  dirty methods for studying black hole resonances}},}\ }\href {\doibase
  10.1088/0264-9381/20/15/312} {\bibfield  {journal} {\bibinfo  {journal}
  {Class.\ Quant.\ Grav.}\ }\textbf {\bibinfo {volume} {20}},\ \bibinfo {pages}
  {3441} (\bibinfo {year} {2003})},\ \Eprint
  {http://arxiv.org/abs/gr-qc/0304030} {arXiv:gr-qc/0304030} \BibitemShut
  {NoStop}%
\bibitem [{\citenamefont {Decanini}\ and\ \citenamefont
  {Folacci}(2010)}]{Decanini:2009mu}%
  \BibitemOpen
  \bibfield  {author} {\bibinfo {author} {\bibfnamefont {Y.}~\bibnamefont
  {Decanini}}\ and\ \bibinfo {author} {\bibfnamefont {A.}~\bibnamefont
  {Folacci}},\ }\bibfield  {title} {\enquote {\bibinfo {title} {{Regge poles of
  the Schwarzschild black hole: A WKB approach}},}\ }\href {\doibase
  10.1103/PhysRevD.81.024031} {\bibfield  {journal} {\bibinfo  {journal}
  {Phys.\ Rev.\ D}\ }\textbf {\bibinfo {volume} {81}},\ \bibinfo {pages}
  {024031} (\bibinfo {year} {2010})},\ \Eprint {http://arxiv.org/abs/0906.2601}
  {arXiv:0906.2601 [gr-qc]} \BibitemShut {NoStop}%
\bibitem [{\citenamefont {Decanini}\ and\ \citenamefont
  {Folacci}(2009)}]{Decanini:2009dn}%
  \BibitemOpen
  \bibfield  {author} {\bibinfo {author} {\bibfnamefont {Y.}~\bibnamefont
  {Decanini}}\ and\ \bibinfo {author} {\bibfnamefont {A.}~\bibnamefont
  {Folacci}},\ }\bibfield  {title} {\enquote {\bibinfo {title} {{Quasinormal
  modes of the BTZ black hole are generated by surface waves supported by its
  boundary at infinity}},}\ }\href {\doibase 10.1103/PhysRevD.79.044021}
  {\bibfield  {journal} {\bibinfo  {journal} {Phys.\ Rev.\ D}\ }\textbf
  {\bibinfo {volume} {79}},\ \bibinfo {pages} {044021} (\bibinfo {year}
  {2009})},\ \Eprint {http://arxiv.org/abs/0901.1642} {arXiv:0901.1642
  [hep-th]} \BibitemShut {NoStop}%
\bibitem [{\citenamefont {Dolan}\ and\ \citenamefont
  {Ottewill}(2009)}]{Dolan:2009nk}%
  \BibitemOpen
  \bibfield  {author} {\bibinfo {author} {\bibfnamefont {S.~R.}\ \bibnamefont
  {Dolan}}\ and\ \bibinfo {author} {\bibfnamefont {A.~C.}\ \bibnamefont
  {Ottewill}},\ }\bibfield  {title} {\enquote {\bibinfo {title} {{On an
  expansion method for black hole quasinormal modes and Regge poles}},}\ }\href
  {\doibase 10.1088/0264-9381/26/22/225003} {\bibfield  {journal} {\bibinfo
  {journal} {Class.\ Quant.\ Grav.}\ }\textbf {\bibinfo {volume} {26}},\
  \bibinfo {pages} {225003} (\bibinfo {year} {2009})},\ \Eprint
  {http://arxiv.org/abs/0908.0329} {arXiv:0908.0329 [gr-qc]} \BibitemShut
  {NoStop}%
\bibitem [{\citenamefont {Decanini}\ \emph {et~al.}(2010)\citenamefont
  {Decanini}, \citenamefont {Folacci},\ and\ \citenamefont
  {Raffaelli}}]{Decanini:2010fz}%
  \BibitemOpen
  \bibfield  {author} {\bibinfo {author} {\bibfnamefont {Y.}~\bibnamefont
  {Decanini}}, \bibinfo {author} {\bibfnamefont {A.}~\bibnamefont {Folacci}}, \
  and\ \bibinfo {author} {\bibfnamefont {B.}~\bibnamefont {Raffaelli}},\
  }\bibfield  {title} {\enquote {\bibinfo {title} {{Unstable circular null
  geodesics of static spherically symmetric black holes, Regge poles and
  quasinormal frequencies}},}\ }\href {\doibase 10.1103/PhysRevD.81.104039}
  {\bibfield  {journal} {\bibinfo  {journal} {Phys.\ Rev.\ D}\ }\textbf
  {\bibinfo {volume} {81}},\ \bibinfo {pages} {104039} (\bibinfo {year}
  {2010})},\ \Eprint {http://arxiv.org/abs/1002.0121} {arXiv:1002.0121 [gr-qc]}
  \BibitemShut {NoStop}%
\bibitem [{\citenamefont {Decanini}\ \emph
  {et~al.}(2011{\natexlab{a}})\citenamefont {Decanini}, \citenamefont
  {Folacci},\ and\ \citenamefont {Raffaelli}}]{Decanini:2011eh}%
  \BibitemOpen
  \bibfield  {author} {\bibinfo {author} {\bibfnamefont {Y.}~\bibnamefont
  {Decanini}}, \bibinfo {author} {\bibfnamefont {A.}~\bibnamefont {Folacci}}, \
  and\ \bibinfo {author} {\bibfnamefont {B.}~\bibnamefont {Raffaelli}},\
  }\bibfield  {title} {\enquote {\bibinfo {title} {{Resonance and absorption
  spectra of the Schwarzschild black hole for massive scalar perturbations: A
  complex angular momentum analysis}},}\ }\href {\doibase
  10.1103/PhysRevD.84.084035} {\bibfield  {journal} {\bibinfo  {journal}
  {Phys.\ Rev.\ D}\ }\textbf {\bibinfo {volume} {84}},\ \bibinfo {pages}
  {084035} (\bibinfo {year} {2011}{\natexlab{a}})},\ \Eprint
  {http://arxiv.org/abs/1108.5076} {arXiv:1108.5076 [gr-qc]} \BibitemShut
  {NoStop}%
\bibitem [{\citenamefont {Decanini}\ \emph
  {et~al.}(2011{\natexlab{b}})\citenamefont {Decanini}, \citenamefont
  {Esposito-Farese},\ and\ \citenamefont {Folacci}}]{Decanini:2011xi}%
  \BibitemOpen
  \bibfield  {author} {\bibinfo {author} {\bibfnamefont {Y.}~\bibnamefont
  {Decanini}}, \bibinfo {author} {\bibfnamefont {G.}~\bibnamefont
  {Esposito-Farese}}, \ and\ \bibinfo {author} {\bibfnamefont {A.}~\bibnamefont
  {Folacci}},\ }\bibfield  {title} {\enquote {\bibinfo {title} {{Universality
  of high-energy absorption cross sections for black holes}},}\ }\href
  {\doibase 10.1103/PhysRevD.83.044032} {\bibfield  {journal} {\bibinfo
  {journal} {Phys.\ Rev.\ D}\ }\textbf {\bibinfo {volume} {83}},\ \bibinfo
  {pages} {044032} (\bibinfo {year} {2011}{\natexlab{b}})},\ \Eprint
  {http://arxiv.org/abs/1101.0781} {arXiv:1101.0781 [gr-qc]} \BibitemShut
  {NoStop}%
\bibitem [{\citenamefont {Decanini}\ \emph
  {et~al.}(2011{\natexlab{c}})\citenamefont {Decanini}, \citenamefont
  {Folacci},\ and\ \citenamefont {Raffaelli}}]{Decanini:2011xw}%
  \BibitemOpen
  \bibfield  {author} {\bibinfo {author} {\bibfnamefont {Y.}~\bibnamefont
  {Decanini}}, \bibinfo {author} {\bibfnamefont {A.}~\bibnamefont {Folacci}}, \
  and\ \bibinfo {author} {\bibfnamefont {B.}~\bibnamefont {Raffaelli}},\
  }\bibfield  {title} {\enquote {\bibinfo {title} {{Fine structure of
  high-energy absorption cross sections for black holes}},}\ }\href {\doibase
  10.1088/0264-9381/28/17/175021} {\bibfield  {journal} {\bibinfo  {journal}
  {Class.\ Quant.\ Grav.}\ }\textbf {\bibinfo {volume} {28}},\ \bibinfo {pages}
  {175021} (\bibinfo {year} {2011}{\natexlab{c}})},\ \Eprint
  {http://arxiv.org/abs/1104.3285} {arXiv:1104.3285 [gr-qc]} \BibitemShut
  {NoStop}%
\bibitem [{\citenamefont {Dolan}\ \emph {et~al.}(2012)\citenamefont {Dolan},
  \citenamefont {Oliveira},\ and\ \citenamefont {Crispino}}]{Dolan:2011ti}%
  \BibitemOpen
  \bibfield  {author} {\bibinfo {author} {\bibfnamefont {S.~R.}\ \bibnamefont
  {Dolan}}, \bibinfo {author} {\bibfnamefont {L.~A.}\ \bibnamefont {Oliveira}},
  \ and\ \bibinfo {author} {\bibfnamefont {L.~C.~B.}\ \bibnamefont
  {Crispino}},\ }\bibfield  {title} {\enquote {\bibinfo {title} {{Resonances of
  a rotating black hole analogue}},}\ }\href {\doibase
  10.1103/PhysRevD.85.044031} {\bibfield  {journal} {\bibinfo  {journal}
  {Phys.\ Rev.\ D}\ }\textbf {\bibinfo {volume} {85}},\ \bibinfo {pages}
  {044031} (\bibinfo {year} {2012})},\ \Eprint {http://arxiv.org/abs/1105.1795}
  {arXiv:1105.1795 [gr-qc]} \BibitemShut {NoStop}%
\bibitem [{\citenamefont {Macedo}\ \emph {et~al.}(2013)\citenamefont {Macedo},
  \citenamefont {Leite}, \citenamefont {Oliveira}, \citenamefont {Dolan},\ and\
  \citenamefont {Crispino}}]{Macedo:2013afa}%
  \BibitemOpen
  \bibfield  {author} {\bibinfo {author} {\bibfnamefont {C.~F.~B.}\
  \bibnamefont {Macedo}}, \bibinfo {author} {\bibfnamefont {L.~C.~S.}\
  \bibnamefont {Leite}}, \bibinfo {author} {\bibfnamefont {E.~S.}\ \bibnamefont
  {Oliveira}}, \bibinfo {author} {\bibfnamefont {S.~R.}\ \bibnamefont {Dolan}},
  \ and\ \bibinfo {author} {\bibfnamefont {L.~C.~B.}\ \bibnamefont
  {Crispino}},\ }\bibfield  {title} {\enquote {\bibinfo {title} {{Absorption of
  planar massless scalar waves by Kerr black holes}},}\ }\href {\doibase
  10.1103/PhysRevD.88.064033} {\bibfield  {journal} {\bibinfo  {journal}
  {Phys.\ Rev.\ D}\ }\textbf {\bibinfo {volume} {88}},\ \bibinfo {pages}
  {064033} (\bibinfo {year} {2013})},\ \Eprint {http://arxiv.org/abs/1308.0018}
  {arXiv:1308.0018 [gr-qc]} \BibitemShut {NoStop}%
\bibitem [{\citenamefont {Nambu}\ and\ \citenamefont
  {Noda}(2016)}]{Nambu:2015aea}%
  \BibitemOpen
  \bibfield  {author} {\bibinfo {author} {\bibfnamefont {Y.}~\bibnamefont
  {Nambu}}\ and\ \bibinfo {author} {\bibfnamefont {S.}~\bibnamefont {Noda}},\
  }\bibfield  {title} {\enquote {\bibinfo {title} {{Wave optics in black hole
  spacetimes: The Schwarzschild case}},}\ }\href {\doibase
  10.1088/0264-9381/33/7/075011} {\bibfield  {journal} {\bibinfo  {journal}
  {Class.\ Quant.\ Grav.}\ }\textbf {\bibinfo {volume} {33}},\ \bibinfo {pages}
  {075011} (\bibinfo {year} {2016})},\ \Eprint
  {http://arxiv.org/abs/1502.05468} {arXiv:1502.05468 [gr-qc]} \BibitemShut
  {NoStop}%
\bibitem [{\citenamefont {Folacci}\ and\ \citenamefont {Ould
  El~Hadj}(2018)}]{Folacci:2018sef}%
  \BibitemOpen
  \bibfield  {author} {\bibinfo {author} {\bibfnamefont {A.}~\bibnamefont
  {Folacci}}\ and\ \bibinfo {author} {\bibfnamefont {M.}~\bibnamefont {Ould
  El~Hadj}},\ }\bibfield  {title} {\enquote {\bibinfo {title} {{Alternative
  description of gravitational radiation from black holes based on the Regge
  poles of the ${\cal S}$-matrix and the associated residues}},}\ }\href
  {\doibase 10.1103/PhysRevD.98.064052} {\bibfield  {journal} {\bibinfo
  {journal} {Phys.\ Rev.\ D}\ }\textbf {\bibinfo {volume} {98}},\ \bibinfo
  {pages} {064052} (\bibinfo {year} {2018})},\ \Eprint
  {http://arxiv.org/abs/1807.09056} {arXiv:1807.09056 [gr-qc]} \BibitemShut
  {NoStop}%
\bibitem [{\citenamefont {Benone}\ \emph {et~al.}(2018)\citenamefont {Benone},
  \citenamefont {Leite}, \citenamefont {Crispino},\ and\ \citenamefont
  {Dolan}}]{Benone:2018rtj}%
  \BibitemOpen
  \bibfield  {author} {\bibinfo {author} {\bibfnamefont {C.~L.}\ \bibnamefont
  {Benone}}, \bibinfo {author} {\bibfnamefont {L.~C.~S.}\ \bibnamefont
  {Leite}}, \bibinfo {author} {\bibfnamefont {L.~C.~B.}\ \bibnamefont
  {Crispino}}, \ and\ \bibinfo {author} {\bibfnamefont {S.~R.}\ \bibnamefont
  {Dolan}},\ }\bibfield  {title} {\enquote {\bibinfo {title} {{On-axis scalar
  absorption cross section of Kerr-Newman black holes: Geodesic analysis, sinc
  and low-frequency approximations}},}\ }\href {\doibase
  10.1142/S0218271818430125} {\bibfield  {journal} {\bibinfo  {journal} {Int.\
  J.\ Mod.\ Phys.\ D}\ }\textbf {\bibinfo {volume} {27}},\ \bibinfo {pages}
  {1843012} (\bibinfo {year} {2018})},\ \Eprint
  {http://arxiv.org/abs/1809.08275} {arXiv:1809.08275 [gr-qc]} \BibitemShut
  {NoStop}%
\bibitem [{\citenamefont {Matzner}(1968)}]{Matzner1968}%
  \BibitemOpen
  \bibfield  {author} {\bibinfo {author} {\bibfnamefont {R.~A.}\ \bibnamefont
  {Matzner}},\ }\bibfield  {title} {\enquote {\bibinfo {title} {{Scattering of
  massless scalar waves by a Schwarzschild ``singularity''}},}\ }\href@noop {}
  {\bibfield  {journal} {\bibinfo  {journal} {J.\ Math.\ Phys.}\ }\textbf
  {\bibinfo {volume} {9}},\ \bibinfo {pages} {163} (\bibinfo {year}
  {1968})}\BibitemShut {NoStop}%
\bibitem [{\citenamefont {Mashhoon}(1973)}]{Mashhoon:1973zz}%
  \BibitemOpen
  \bibfield  {author} {\bibinfo {author} {\bibfnamefont {B.}~\bibnamefont
  {Mashhoon}},\ }\bibfield  {title} {\enquote {\bibinfo {title} {{Scattering of
  electromagnetic radiation from a black hole}},}\ }\href {\doibase
  10.1103/PhysRevD.7.2807} {\bibfield  {journal} {\bibinfo  {journal} {Phys.\
  Rev.\ D}\ }\textbf {\bibinfo {volume} {7}},\ \bibinfo {pages} {2807}
  (\bibinfo {year} {1973})}\BibitemShut {NoStop}%
\bibitem [{\citenamefont {Mashhoon}(1974)}]{Mashhoon:1974cq}%
  \BibitemOpen
  \bibfield  {author} {\bibinfo {author} {\bibfnamefont {B.}~\bibnamefont
  {Mashhoon}},\ }\bibfield  {title} {\enquote {\bibinfo {title}
  {{Electromagnetic scattering from a black hole and the glory effect}},}\
  }\href {\doibase 10.1103/PhysRevD.10.1059} {\bibfield  {journal} {\bibinfo
  {journal} {Phys.\ Rev.\ D}\ }\textbf {\bibinfo {volume} {10}},\ \bibinfo
  {pages} {1059} (\bibinfo {year} {1974})}\BibitemShut {NoStop}%
\bibitem [{\citenamefont {Fabbri}(1975)}]{Fabbri:1975sa}%
  \BibitemOpen
  \bibfield  {author} {\bibinfo {author} {\bibfnamefont {R.}~\bibnamefont
  {Fabbri}},\ }\bibfield  {title} {\enquote {\bibinfo {title} {{Scattering and
  absorption of electromagnetic waves by a Schwarzschild black hole}},}\ }\href
  {\doibase 10.1103/PhysRevD.12.933} {\bibfield  {journal} {\bibinfo  {journal}
  {Phys.\ Rev.\ D}\ }\textbf {\bibinfo {volume} {12}},\ \bibinfo {pages} {933}
  (\bibinfo {year} {1975})}\BibitemShut {NoStop}%
\bibitem [{\citenamefont {Sanchez}(1976)}]{Sanchez:1976fcl}%
  \BibitemOpen
  \bibfield  {author} {\bibinfo {author} {\bibfnamefont {N.~G.}\ \bibnamefont
  {Sanchez}},\ }\bibfield  {title} {\enquote {\bibinfo {title} {{Scattering of
  scalar waves from a Schwarzschild black hole}},}\ }\href {\doibase
  10.1063/1.522949} {\bibfield  {journal} {\bibinfo  {journal} {J.\ Math.\
  Phys.}\ }\textbf {\bibinfo {volume} {17}},\ \bibinfo {pages} {688} (\bibinfo
  {year} {1976})}\BibitemShut {NoStop}%
\bibitem [{\citenamefont {Sanchez}(1978)}]{Sanchez:1977vz}%
  \BibitemOpen
  \bibfield  {author} {\bibinfo {author} {\bibfnamefont {N.~G.}\ \bibnamefont
  {Sanchez}},\ }\bibfield  {title} {\enquote {\bibinfo {title} {{Elastic
  scattering of waves by a black hole}},}\ }\href {\doibase
  10.1103/PhysRevD.18.1798} {\bibfield  {journal} {\bibinfo  {journal} {Phys.\
  Rev.\ D}\ }\textbf {\bibinfo {volume} {18}},\ \bibinfo {pages} {1798}
  (\bibinfo {year} {1978})}\BibitemShut {NoStop}%
\bibitem [{\citenamefont {Matzner}\ \emph {et~al.}(1985)\citenamefont
  {Matzner}, \citenamefont {DeWitt-Morette}, \citenamefont {Nelson},\ and\
  \citenamefont {Zhang}}]{Matzner:1985rjn}%
  \BibitemOpen
  \bibfield  {author} {\bibinfo {author} {\bibfnamefont {R.~A.}\ \bibnamefont
  {Matzner}}, \bibinfo {author} {\bibfnamefont {C.}~\bibnamefont
  {DeWitt-Morette}}, \bibinfo {author} {\bibfnamefont {B.}~\bibnamefont
  {Nelson}}, \ and\ \bibinfo {author} {\bibfnamefont {T.-R.}\ \bibnamefont
  {Zhang}},\ }\bibfield  {title} {\enquote {\bibinfo {title} {{Glory scattering
  by black holes}},}\ }\href {\doibase 10.1103/PhysRevD.31.1869} {\bibfield
  {journal} {\bibinfo  {journal} {Phys.\ Rev.\ D}\ }\textbf {\bibinfo {volume}
  {31}},\ \bibinfo {pages} {1869} (\bibinfo {year} {1985})}\BibitemShut
  {NoStop}%
\bibitem [{\citenamefont {Anninos}\ \emph {et~al.}(1992)\citenamefont
  {Anninos}, \citenamefont {DeWitt-Morette}, \citenamefont {Matzner},
  \citenamefont {Yioutas},\ and\ \citenamefont {Zhang}}]{Anninos:1992ih}%
  \BibitemOpen
  \bibfield  {author} {\bibinfo {author} {\bibfnamefont {P.}~\bibnamefont
  {Anninos}}, \bibinfo {author} {\bibfnamefont {C.}~\bibnamefont
  {DeWitt-Morette}}, \bibinfo {author} {\bibfnamefont {R.~A.}\ \bibnamefont
  {Matzner}}, \bibinfo {author} {\bibfnamefont {P.}~\bibnamefont {Yioutas}}, \
  and\ \bibinfo {author} {\bibfnamefont {T.~R.}\ \bibnamefont {Zhang}},\
  }\bibfield  {title} {\enquote {\bibinfo {title} {{Orbiting cross-sections:
  Application to black hole scattering}},}\ }\href {\doibase
  10.1103/PhysRevD.46.4477} {\bibfield  {journal} {\bibinfo  {journal} {Phys.\
  Rev.\ D}\ }\textbf {\bibinfo {volume} {46}},\ \bibinfo {pages} {4477}
  (\bibinfo {year} {1992})}\BibitemShut {NoStop}%
\bibitem [{\citenamefont {Andersson}(1995)}]{Andersson:1995vi}%
  \BibitemOpen
  \bibfield  {author} {\bibinfo {author} {\bibfnamefont {N.}~\bibnamefont
  {Andersson}},\ }\bibfield  {title} {\enquote {\bibinfo {title} {{Scattering
  of massless scalar waves by a Schwarzschild black hole: A phase integral
  study}},}\ }\href {\doibase 10.1103/PhysRevD.52.1808} {\bibfield  {journal}
  {\bibinfo  {journal} {Phys.\ Rev.\ D}\ }\textbf {\bibinfo {volume} {52}},\
  \bibinfo {pages} {1808} (\bibinfo {year} {1995})}\BibitemShut {NoStop}%
\bibitem [{\citenamefont {Glampedakis}\ and\ \citenamefont
  {Andersson}(2001)}]{Glampedakis:2001cx}%
  \BibitemOpen
  \bibfield  {author} {\bibinfo {author} {\bibfnamefont {K.}~\bibnamefont
  {Glampedakis}}\ and\ \bibinfo {author} {\bibfnamefont {N.}~\bibnamefont
  {Andersson}},\ }\bibfield  {title} {\enquote {\bibinfo {title} {{Scattering
  of scalar waves by rotating black holes}},}\ }\href {\doibase
  10.1088/0264-9381/18/10/309} {\bibfield  {journal} {\bibinfo  {journal}
  {Class.\ Quant.\ Grav.}\ }\textbf {\bibinfo {volume} {18}},\ \bibinfo {pages}
  {1939} (\bibinfo {year} {2001})},\ \Eprint
  {http://arxiv.org/abs/gr-qc/0102100} {arXiv:gr-qc/0102100} \BibitemShut
  {NoStop}%
\bibitem [{\citenamefont {Dolan}(2008{\natexlab{a}})}]{Dolan:2008kf}%
  \BibitemOpen
  \bibfield  {author} {\bibinfo {author} {\bibfnamefont {S.~R.}\ \bibnamefont
  {Dolan}},\ }\bibfield  {title} {\enquote {\bibinfo {title} {{Scattering and
  absorption of gravitational plane waves by rotating black holes}},}\ }\href
  {\doibase 10.1088/0264-9381/25/23/235002} {\bibfield  {journal} {\bibinfo
  {journal} {Class.\ Quant.\ Grav.}\ }\textbf {\bibinfo {volume} {25}},\
  \bibinfo {pages} {235002} (\bibinfo {year} {2008}{\natexlab{a}})},\ \Eprint
  {http://arxiv.org/abs/0801.3805} {arXiv:0801.3805 [gr-qc]} \BibitemShut
  {NoStop}%
\bibitem [{\citenamefont {Crispino}\ \emph
  {et~al.}(2009{\natexlab{a}})\citenamefont {Crispino}, \citenamefont {Dolan},\
  and\ \citenamefont {Oliveira}}]{Crispino:2009xt}%
  \BibitemOpen
  \bibfield  {author} {\bibinfo {author} {\bibfnamefont {L.~C.~B.}\
  \bibnamefont {Crispino}}, \bibinfo {author} {\bibfnamefont {S.~R.}\
  \bibnamefont {Dolan}}, \ and\ \bibinfo {author} {\bibfnamefont {E.~S.}\
  \bibnamefont {Oliveira}},\ }\bibfield  {title} {\enquote {\bibinfo {title}
  {{Electromagnetic wave scattering by Schwarzschild black holes}},}\ }\href
  {\doibase 10.1103/PhysRevLett.102.231103} {\bibfield  {journal} {\bibinfo
  {journal} {Phys.\ Rev.\ Lett.}\ }\textbf {\bibinfo {volume} {102}},\ \bibinfo
  {pages} {231103} (\bibinfo {year} {2009}{\natexlab{a}})},\ \Eprint
  {http://arxiv.org/abs/0905.3339} {arXiv:0905.3339 [gr-qc]} \BibitemShut
  {NoStop}%
\bibitem [{\citenamefont {Falcke}\ \emph {et~al.}(2000)\citenamefont {Falcke},
  \citenamefont {Melia},\ and\ \citenamefont {Agol}}]{Falcke:1999pj}%
  \BibitemOpen
  \bibfield  {author} {\bibinfo {author} {\bibfnamefont {H.}~\bibnamefont
  {Falcke}}, \bibinfo {author} {\bibfnamefont {F.}~\bibnamefont {Melia}}, \
  and\ \bibinfo {author} {\bibfnamefont {E.}~\bibnamefont {Agol}},\ }\bibfield
  {title} {\enquote {\bibinfo {title} {{Viewing the shadow of the black hole at
  the galactic center}},}\ }\href {\doibase 10.1086/312423} {\bibfield
  {journal} {\bibinfo  {journal} {Astrophys.\ J.}\ }\textbf {\bibinfo {volume}
  {528}},\ \bibinfo {pages} {L13} (\bibinfo {year} {2000})},\ \Eprint
  {http://arxiv.org/abs/astro-ph/9912263} {arXiv:astro-ph/9912263} \BibitemShut
  {NoStop}%
\bibitem [{\citenamefont {Castelvecchi}(2017)}]{EHT}%
  \BibitemOpen
  \bibfield  {author} {\bibinfo {author} {\bibfnamefont {D.}~\bibnamefont
  {Castelvecchi}},\ }\bibfield  {title} {\enquote {\bibinfo {title} {{How to
  hunt for a black hole with a telescope the size of Earth}},}\ }\href
  {\doibase 10.1038/543478a} {\bibfield  {journal} {\bibinfo  {journal}
  {Nature}\ }\textbf {\bibinfo {volume} {543}},\ \bibinfo {pages} {478}
  (\bibinfo {year} {2017})}\BibitemShut {NoStop}%
\bibitem [{\citenamefont {Collaboration}(2019)}]{EHTC2019I}%
  \BibitemOpen
  \bibfield  {author} {\bibinfo {author} {\bibfnamefont {The Event
  Horizon~Telescope}\ \bibnamefont {Collaboration}},\ }\bibfield  {title}
  {\enquote {\bibinfo {title} {{First M87 Event Horizon Telescope Results. I.
  The Shadow of the Supermassive Black Hole}},}\ }\href {\doibase
  10.3847/2041-8213/ab0ec7} {\bibfield  {journal} {\bibinfo  {journal}
  {Astrophys.\ J.\ Lett.}\ }\textbf {\bibinfo {volume} {875}},\ \bibinfo
  {pages} {L1} (\bibinfo {year} {2019})}\BibitemShut {NoStop}%
\bibitem [{\citenamefont {Goebel}(1972)}]{Goebel}%
  \BibitemOpen
  \bibfield  {author} {\bibinfo {author} {\bibfnamefont {C.~J.}\ \bibnamefont
  {Goebel}},\ }\bibfield  {title} {\enquote {\bibinfo {title} {{Comments on the
  ``vibrations'' of a black hole}},}\ }\href {\doibase 10.1086/180898}
  {\bibfield  {journal} {\bibinfo  {journal} {Astrophys.\ J.}\ }\textbf
  {\bibinfo {volume} {172}},\ \bibinfo {pages} {L95} (\bibinfo {year}
  {1972})}\BibitemShut {NoStop}%
\bibitem [{\citenamefont {Decanini}\ \emph {et~al.}()\citenamefont {Decanini},
  \citenamefont {Folacci},\ and\ \citenamefont {Ould El~Hadj}}]{DFOEH2019}%
  \BibitemOpen
  \bibfield  {author} {\bibinfo {author} {\bibfnamefont {Y.}~\bibnamefont
  {Decanini}}, \bibinfo {author} {\bibfnamefont {A.}~\bibnamefont {Folacci}}, \
  and\ \bibinfo {author} {\bibfnamefont {M.}~\bibnamefont {Ould El~Hadj}},\
  }\bibfield  {title} {\enquote {\bibinfo {title} {{Regge mode excitation and
  strong gravitational lensing (work in progress)}},}\ }\href@noop {} {\
  }\BibitemShut {NoStop}%
\bibitem [{\citenamefont {Yennie}\ \emph {et~al.}(1954)\citenamefont {Yennie},
  \citenamefont {Ravenhall},\ and\ \citenamefont {Wilson}}]{Yennie:1954zz}%
  \BibitemOpen
  \bibfield  {author} {\bibinfo {author} {\bibfnamefont {D.~R.}\ \bibnamefont
  {Yennie}}, \bibinfo {author} {\bibfnamefont {D.~G.}\ \bibnamefont
  {Ravenhall}}, \ and\ \bibinfo {author} {\bibfnamefont {R.~N.}\ \bibnamefont
  {Wilson}},\ }\bibfield  {title} {\enquote {\bibinfo {title} {{Phase-shift
  calculation of high-energy electron scattering}},}\ }\href {\doibase
  10.1103/PhysRev.95.500} {\bibfield  {journal} {\bibinfo  {journal} {Phys.\
  Rev.}\ }\textbf {\bibinfo {volume} {95}},\ \bibinfo {pages} {500} (\bibinfo
  {year} {1954})}\BibitemShut {NoStop}%
\bibitem [{\citenamefont {Abramowitz}\ and\ \citenamefont
  {Stegun}(1965)}]{AS65}%
  \BibitemOpen
  \bibfield  {author} {\bibinfo {author} {\bibfnamefont {M.}~\bibnamefont
  {Abramowitz}}\ and\ \bibinfo {author} {\bibfnamefont {I.~A.}\ \bibnamefont
  {Stegun}},\ }\href@noop {} {\emph {\bibinfo {title} {Handbook of Mathematical
  Functions}}}\ (\bibinfo  {publisher} {Dover, New York},\ \bibinfo {year}
  {1965})\BibitemShut {NoStop}%
\bibitem [{\citenamefont {Inc.}()}]{Mathematica}%
  \BibitemOpen
  \bibfield  {author} {\bibinfo {author} {\bibfnamefont {Wolfram~Research{,}}\
  \bibnamefont {Inc.}},\ }\href@noop {} {\enquote {\bibinfo {title} {{\it
  Mathematica}, {V}ersion 10.0},}\ }\bibinfo {note} {(Wolfram Research{,} Inc.,
  Champaign, IL, 2014)}\BibitemShut {NoStop}%
\bibitem [{\citenamefont {Handler}\ and\ \citenamefont
  {Matzner}(1980)}]{Handler:1980un}%
  \BibitemOpen
  \bibfield  {author} {\bibinfo {author} {\bibfnamefont {F.~A.}\ \bibnamefont
  {Handler}}\ and\ \bibinfo {author} {\bibfnamefont {R.~A.}\ \bibnamefont
  {Matzner}},\ }\bibfield  {title} {\enquote {\bibinfo {title} {{Gravitational
  wave scattering}},}\ }\href {\doibase 10.1103/PhysRevD.22.2331} {\bibfield
  {journal} {\bibinfo  {journal} {Phys.\ Rev.\ D}\ }\textbf {\bibinfo {volume}
  {22}},\ \bibinfo {pages} {2331} (\bibinfo {year} {1980})}\BibitemShut
  {NoStop}%
\bibitem [{\citenamefont {Ford}\ and\ \citenamefont
  {Wheeler}(1959)}]{Ford1959259}%
  \BibitemOpen
  \bibfield  {author} {\bibinfo {author} {\bibfnamefont {K.~W.}\ \bibnamefont
  {Ford}}\ and\ \bibinfo {author} {\bibfnamefont {J.~A.}\ \bibnamefont
  {Wheeler}},\ }\bibfield  {title} {\enquote {\bibinfo {title} {Semiclassical
  description of scattering},}\ }\href {\doibase 10.1016/0003-4916(59)90026-0}
  {\bibfield  {journal} {\bibinfo  {journal} {Annals of Physics}\ }\textbf
  {\bibinfo {volume} {7}},\ \bibinfo {pages} {259} (\bibinfo {year}
  {1959})}\BibitemShut {NoStop}%
\bibitem [{\citenamefont {Crispino}\ \emph
  {et~al.}(2009{\natexlab{b}})\citenamefont {Crispino}, \citenamefont {Dolan},\
  and\ \citenamefont {Oliveira}}]{Crispino:2009ki}%
  \BibitemOpen
  \bibfield  {author} {\bibinfo {author} {\bibfnamefont {L.~C.~B.}\
  \bibnamefont {Crispino}}, \bibinfo {author} {\bibfnamefont {S.~R.}\
  \bibnamefont {Dolan}}, \ and\ \bibinfo {author} {\bibfnamefont {E.~S.}\
  \bibnamefont {Oliveira}},\ }\bibfield  {title} {\enquote {\bibinfo {title}
  {{Scattering of massless scalar waves by Reissner-Nordstr\"om black
  holes}},}\ }\href {\doibase 10.1103/PhysRevD.79.064022} {\bibfield  {journal}
  {\bibinfo  {journal} {Phys.\ Rev.\ D}\ }\textbf {\bibinfo {volume} {79}},\
  \bibinfo {pages} {064022} (\bibinfo {year} {2009}{\natexlab{b}})},\ \Eprint
  {http://arxiv.org/abs/0904.0999} {arXiv:0904.0999 [gr-qc]} \BibitemShut
  {NoStop}%
\bibitem [{\citenamefont {Schutz}\ and\ \citenamefont
  {Will}(1985)}]{Schutz:1985zz}%
  \BibitemOpen
  \bibfield  {author} {\bibinfo {author} {\bibfnamefont {B.~F.}\ \bibnamefont
  {Schutz}}\ and\ \bibinfo {author} {\bibfnamefont {C.~M.}\ \bibnamefont
  {Will}},\ }\bibfield  {title} {\enquote {\bibinfo {title} {{Black-hole normal
  modes: A semianalytic approach}},}\ }\href {\doibase 10.1086/184453}
  {\bibfield  {journal} {\bibinfo  {journal} {Astrophys.\ J.}\ }\textbf
  {\bibinfo {volume} {291}},\ \bibinfo {pages} {L33} (\bibinfo {year}
  {1985})}\BibitemShut {NoStop}%
\bibitem [{\citenamefont {Iyer}\ and\ \citenamefont
  {Will}(1987)}]{Iyer:1986vv}%
  \BibitemOpen
  \bibfield  {author} {\bibinfo {author} {\bibfnamefont {S.}~\bibnamefont
  {Iyer}}\ and\ \bibinfo {author} {\bibfnamefont {C.~M.}\ \bibnamefont
  {Will}},\ }\bibfield  {title} {\enquote {\bibinfo {title} {{Black-hole normal
  modes: A WKB approach. I. Foundations and application of a higher-order WKB
  analysis of potential-barrier scattering}},}\ }\href {\doibase
  10.1103/PhysRevD.35.3621} {\bibfield  {journal} {\bibinfo  {journal} {Phys.\
  Rev.\ D}\ }\textbf {\bibinfo {volume} {35}},\ \bibinfo {pages} {3621}
  (\bibinfo {year} {1987})}\BibitemShut {NoStop}%
\bibitem [{\citenamefont {Iyer}(1987)}]{Iyer:1986nq}%
  \BibitemOpen
  \bibfield  {author} {\bibinfo {author} {\bibfnamefont {S.}~\bibnamefont
  {Iyer}},\ }\bibfield  {title} {\enquote {\bibinfo {title} {{Black-hole normal
  modes: A WKB approach. II. Schwarzschild black holes}},}\ }\href {\doibase
  10.1103/PhysRevD.35.3632} {\bibfield  {journal} {\bibinfo  {journal} {Phys.\
  Rev.\ D}\ }\textbf {\bibinfo {volume} {35}},\ \bibinfo {pages} {3632}
  (\bibinfo {year} {1987})}\BibitemShut {NoStop}%
\bibitem [{\citenamefont {Dolan}\ and\ \citenamefont
  {Ottewill}(2011)}]{Dolan:2011fh}%
  \BibitemOpen
  \bibfield  {author} {\bibinfo {author} {\bibfnamefont {S.~R.}\ \bibnamefont
  {Dolan}}\ and\ \bibinfo {author} {\bibfnamefont {A.~C.}\ \bibnamefont
  {Ottewill}},\ }\bibfield  {title} {\enquote {\bibinfo {title} {{Wave
  Propagation and Quasinormal Mode Excitation on Schwarzschild Spacetime}},}\
  }\href {\doibase 10.1103/PhysRevD.84.104002} {\bibfield  {journal} {\bibinfo
  {journal} {Phys.\ Rev.\ D}\ }\textbf {\bibinfo {volume} {84}},\ \bibinfo
  {pages} {104002} (\bibinfo {year} {2011})},\ \Eprint
  {http://arxiv.org/abs/1106.4318} {arXiv:1106.4318 [gr-qc]} \BibitemShut
  {NoStop}%
\bibitem [{\citenamefont {Matzner}\ and\ \citenamefont
  {Ryan}(1977)}]{Matzner:1977dn}%
  \BibitemOpen
  \bibfield  {author} {\bibinfo {author} {\bibfnamefont {R.~A.}\ \bibnamefont
  {Matzner}}\ and\ \bibinfo {author} {\bibfnamefont {M.~P.}\ \bibnamefont
  {Ryan}},\ }\bibfield  {title} {\enquote {\bibinfo {title} {{Low frequency
  limit of gravitational scattering}},}\ }\href {\doibase
  10.1103/PhysRevD.16.1636} {\bibfield  {journal} {\bibinfo  {journal} {Phys.\
  Rev.\ D}\ }\textbf {\bibinfo {volume} {16}},\ \bibinfo {pages} {1636}
  (\bibinfo {year} {1977})}\BibitemShut {NoStop}%
\bibitem [{\citenamefont {Matzner}\ and\ \citenamefont
  {Ryan}(1978)}]{MatznerRyan1978}%
  \BibitemOpen
  \bibfield  {author} {\bibinfo {author} {\bibfnamefont {R.~A.}\ \bibnamefont
  {Matzner}}\ and\ \bibinfo {author} {\bibfnamefont {M.~P.}\ \bibnamefont
  {Ryan}},\ }\bibfield  {title} {\enquote {\bibinfo {title} {{Scattering of
  gravitational radiation from vacuum black holes}},}\ }\href@noop {}
  {\bibfield  {journal} {\bibinfo  {journal} {Astrophys.\ J.\ Suppl.}\ }\textbf
  {\bibinfo {volume} {36}},\ \bibinfo {pages} {451} (\bibinfo {year}
  {1978})}\BibitemShut {NoStop}%
\bibitem [{\citenamefont {Dolan}(2008{\natexlab{b}})}]{Dolan:2007ut}%
  \BibitemOpen
  \bibfield  {author} {\bibinfo {author} {\bibfnamefont {S.~R.}\ \bibnamefont
  {Dolan}},\ }\bibfield  {title} {\enquote {\bibinfo {title} {{Scattering of
  long-wavelength gravitational waves}},}\ }\href {\doibase
  10.1103/PhysRevD.77.044004} {\bibfield  {journal} {\bibinfo  {journal}
  {Phys.\ Rev.\ D}\ }\textbf {\bibinfo {volume} {77}},\ \bibinfo {pages}
  {044004} (\bibinfo {year} {2008}{\natexlab{b}})},\ \Eprint
  {http://arxiv.org/abs/0710.4252} {arXiv:0710.4252 [gr-qc]} \BibitemShut
  {NoStop}%
\end{thebibliography}%

\end{document}